\RequirePackage{fix-cm}
\documentclass[smallextended, natbib]{svjour3}
\smartqed 

\usepackage{graphicx}

\usepackage{mathptmx}      
\usepackage[export]{adjustbox}
\usepackage[colorlinks=true,citecolor=blue,urlcolor=blue,breaklinks]{hyperref}
\usepackage{amsmath}
\usepackage{amssymb}
\usepackage{amsfonts}

\newcommand{\pd}{\partial}
\newcommand{\ud}{\mathrm{d}}

\journalname{Living Reviews in Computational Astrophysics}

\begin{document}

\title{Hydrodynamics of core-collapse supernovae and their progenitors}
\author{Bernhard M\"uller}

\institute{B.~M\"uller \at
              School of Physics and Astronomy, Monash University \\
              Tel.: +61-3-99051876\\
              \email{bernhard.mueller@monash.edu}           %  \\
}

\date{Received: : 26 November 2019 / Accepted: : 18 April 2020}

\maketitle

\begin{abstract}
Multi-dimensional fluid flow plays a paramount role in the explosions
of massive stars as core-collapse supernovae. In recent years,
three-dimensional (3D) simulations of these phenomena have matured
significantly. Considerable progress has been made towards identifying
the ingredients for shock revival by the neutrino-driven mechanism,
and successful explosions have already been obtained in a number of
self-consistent 3D models. These advances also bring new challenges,
however. Prompted by a need for increased physical realism and
meaningful model validation, supernova theory is now moving towards a
more integrated view that connects multi-dimensional phenomena in the
late convective burning stages prior to collapse, the explosion
engine, and mixing instabilities in the supernova envelope. Here we
review our current understasnding of multi-D fluid flow in
core-collapse supernovae and their progenitors. We start by outlining
specific challenges faced by hydrodynamic simulations of core-collapse
supernovae and of the late convective burning stages. We then discuss
recent advances and open questions in theory and simulations.
\end{abstract}

\setcounter{tocdepth}{3}
\tableofcontents

\section{Introduction}
\label{sec:intro}
The death of massive stars is invariably spectacular. In the cores of these stars, nuclear fusion proceeds all the way to the iron group through a sequence
of burning stages. At the end of the star's life,
nuclear energy generation has ceased in the degenerate
Fe core (i.e., the core
has become ``inert''),
while nuclear burning
continues in shells composed of progressively
light nuclei further out. Once the core
approaches its effective Chandrasekhar mass
and becomes sufficiently compact, electron
captures on heavy nuclei and partial
nuclear disintegration lead to collapse
on a free-fall time scale, leaving
behind a neutron star or black hole. In most
cases, a small fraction of the potential
energy liberated during collapse is
transferred to the stellar envelope, which
is expelled in a powerful explosion known
as a \emph{core-collapse supernova},
as first recognized by \citet{baade_34a}.

How precisely the envelope is ejected has
remained one of the foremost questions in computational astrophysics ever since
the first modeling attempts in the
1960s \citep{colgate_61,colgate_66}. 
In this
review we  focus on the critical
role of multi-dimensional (multi-D) fluid flow during
the supernova explosion itself and the final pre-collapse stages of their progenitors.

For pedagogical reasons, it is preferable to commence our brief exposition of multi-D hydrodynamic
effects with the supernova explosion mechanism rather than to follow the sequence of events in 
nature, or historical chronology.

\subsection{The multi-dimensional nature of the explosion mechanism}
In principle, the collapse of an iron core to a 
neutron star opens a reservoir of several $10^{53}\, \mathrm{erg}$ of potential energy, which appears
more than sufficient to account for the typical inferred
kinetic energies of observed core-collapse
supernovae of order $10^{51}\, \mathrm{erg}$
\citep[see, e.g.,][]{kasen_09,pejcha_15c}.

Transferring the requisite amount of energy
from the young ``proto-neutron star'' (PNS) is not trivial, however. The simplest idea is that
the energy is delivered when the collapsing
core overshoots nuclear density and ``bounces''
due to the high incompressibility of
matter above nuclear saturation density, which launches a shock wave into the surrounding
shells \citep{colgate_61}. However, the shock wave
 stalls within milliseconds
 as nuclear dissociation of the shocked
material and neutrino losses drains its initial kinetic energy
\citep[e.g.,][]{mazurek_82,burrows_85,bethe_90}.
The shock then turns into an accretion shock, whose radius is essentially determined by the pre-shock ram pressure and
the condition of hydrostatic equilibrium between
the shock and the PNS surface. It typically reaches
a radius between $100$--$200\, \mathrm{km}$
a few tens of milliseconds after bounce and then recedes
again.

\begin{figure*}
    \centering
    \includegraphics[width=\linewidth]{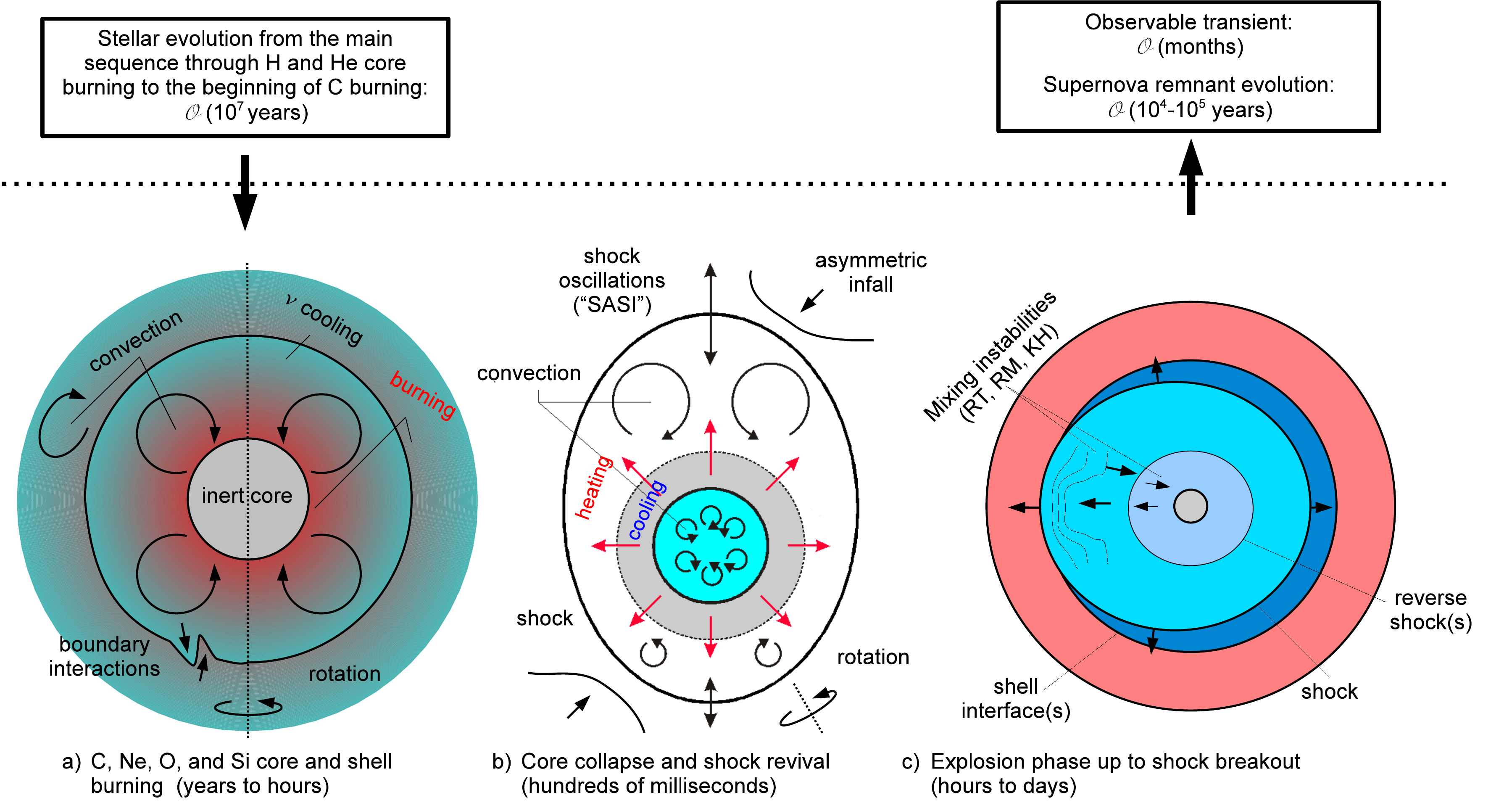}
    \caption{Overview of the multi-D effects operating prior to and
      during a core-collapse supernova as discussed in this review
      (below the dashed line) in the broader context of the evolution
      of a massive star.  \textbf{a)} After millions of years of
      H and He burning, the star enters the neutrino-cooled
      burning stages (C, Ne, O, Si burning).
These  advanced core and shell burning stages tpyically proceed convectively
because burning and neutrino cooling at the top and bottom
of an active shell/core establish an unstable negative entropy gradient.
    The interaction of the flow with convective boundaries
    can lead to mixing and transfer energy and angular
    momentum by wave excitation. Rotation may modify the flow dynamics.
 \textbf{b)} After the star has formed a sufficiently massive
iron core, the core undergoes gravitational collapse, and
a young proto-neutron star is formed. The shock wave
launched by the ``bounce'' of the core quickly stalls, and
is likely revived by neutrino heating in most cases.
In the phase leading up to shock revival,
neutrino heating drives convection in
    the heating or ``gain'' region, and the shock may 
    execute large-scale oscillations due the
    standing accretion shock instability (SASI). Rotation and
    the asymmetric infall of convective burning shells
    can modify the dynamics. There is also a convective
    region below the cooling layer at the proto-neutron
    star (PNS)
    surface. \textbf{c)} After the shock has been revived and
sufficient energy has been pumped into the explosion by
neutrino heating or some other mechanism,
the shock propagates through the outer shells on a time
scale of hours to days. During this phase, the interaction
    of the (deformed) shock with shell interfaces as well as
    reverse shock formation trigger mixing by the
    Rayleigh--Taylor (RT), the Richtmyer--Meshkov (RM)
    instability, and
    (as a secondary process) the Kelvin--Helmholtz (KH) instability.
Once the shock breaks out through the stellar surface,
the explosion becomes visible as an electromagnetic
transient. Mixing instabilities continue to operate
on longer time scales throughout the evolution of the
supernova remnant.
    }
    \label{fig:sketch}
\end{figure*}

Among the various ideas to ``revive'' the shock 
\citep[for a more exhaustive overview see][]{mezzacappa_05,kotake_06,janka_12,burrows_13} the neutrino-driven
mechanism is the most promising scenario and has been
explored most comprehensively
since it was originally conceived --
in a form rather different from the modern paradigm
-- by \citet{colgate_66}.
The modern version of this mechanism is
illustrated in Fig.~\ref{fig:sketch}b:
A fraction of the neutrinos emitted from the PNS
and the cooling layer at its surface are
reabsorbed further out in the ``gain region''.
If the neutrino heating is sufficiently strong, the
 increased thermal pressure drives the shock
out, and the heating powers an outflow of matter in its wake.

However, according to the most sophisticated spherically symmetric (1D)
models using Boltzmann neutrino transport to accurately
model neutrino heating and cooling, this mechanism does not work
in 1D
\citep{liebendoerfer_00,rampp_00} except
for the least massive\footnote{More precisely,
the neutrino-driven mechanism works in 1D
for progenitors with a steeply declining
density profile outside the core \citep{mueller_16b} as in 
the case of electron-capture supernovae
from super-AGB stars \citep{kitaura_06} or the least massive
iron core progenitors \citep{melson_15a}.}
supernova progenitors.
For all other progenitors, it is crucial that
multi-D effects support the neutrino heating.
Convection occurs in the gain region
because neutrino heating establishes a negative entropy
gradient \citep{bethe_90}, and was shown to be
highly beneficial for obtaining neutrino-driven explosions
by the first generation of multi-D models from the 1990s
\citep{herant_92,herant_94,burrows_95,janka_95,janka_96}.
Another instability, 
the standing-accretion shock instability 
\citep[SASI;][]{blondin_03,blondin_06,foglizzo_06,foglizzo_07,laming_07},
arises due to an advective-acoustic amplification cycle and
gives rise to large-scale
($\ell=1,2$) shock oscillations; it plays a similarly
beneficial role in the neutrino-driven mechanism as convection
\citep{scheck_08,mueller_12b}. Rapid rotation could also 
modify the dynamics in the supernova core and support the
development of neutrino-driven explosions \citep{janka_16,summa_18,takiwaki_16}.

Evidently, multi-D effects are also at the heart of the most serious alternative to neutrino-driven explosions, 
the magnetohydrodynamic (MHD) mechanism
\citep[e.g.,][]{akiyama_03,dessart_07a,winteler_12,moesta_14b},
which likely explains unusually energetic 
``hypernovae''. But whether core-collapse supernovae are
driven by neutrinos or magnetic fields, it is pertinent to
ask: How important are  the initial conditions for the
multi-D flow dynamics that leads to shock revival?

\subsection{The multi-D structure of supernova progenitors}
For pragmatic reasons, supernova models have long relied
on 1D stellar evolution models as input, or
at best on ``1.5D'' rotating models using the shellular
approximation \citep{zahn_92}. For non-rotating progenitors, spherical symmetry
was either broken by introducing perturbations 
in supernova simulations by hand, or
due to grid perturbations. For rotating and magnetized
progenitors, spherical symmetry is broken naturally, but on
the other hand the stellar evolution models do not
provide the detailed multi-D angular momentum distribution and magnetic field geometry, which must be specified by hand.

In reality, even non-rotating progenitors are not spherically
symmetric at the onset of collapse. Outside the iron core,
there are typically several active convective burning shells
(Fig.~\ref{fig:sketch}a) that will collapse in
the wake of the iron core within hundreds of milliseconds.
It was realized in recent years that the infall of asymmetric 
shells can be important for shock revival 
\citep{couch_13,mueller_15b,couch_15,mueller_17}.

The multi-D structure of supernova progenitors is thus
directly relevant for the neutrino-driven mechanism, but the
potential ramifications of multi-D effects during the
pre-collapse phase are in fact much broader: How do
they affect mixing at convective boundaries, and hence the
evolution of the shell structure on secular time scales? How
do they affect the angular momentum distribution and
magnetic fields in supernova progenitors?

\subsection{Observational evidence for multi-D effects in core-collapse supernovae}
Observations contain abundant clues about the multi-D nature of
core-collapse supernova explosion. Large birth kicks
of neutron stars \citep{hobbs_05,faucher_06,ng_07} and even black holes
\citep{repetto_12} cannot be explained by stellar dynamics alone
and require asymmetries in the supernova engine.
There is also evidence for mixing processes 
during the explosion and large-scale
asymmetries in the ejecta from the spectra and polarization
signatures of many observed transients \citep[e.g.,][]{wang_08,patat_17},
and from young supernova remnants like Cas~A \citep{grefenstette_14}.

The relation between the asymmetries in the
progenitor and the supernova core, and the
asymmetries in observed transients and gaseous remnants
is not straightforward, however. The
observable symmetries are rather shaped
by  mixing processes that operate
as the shock propagates through the stellar envelope
(Fig.~\ref{fig:sketch}c).
Rayleigh--Taylor instability occurs behind the
shock as it scoops up material and decelerates
\citep{chevalier_76,bandiera_84}, and the interaction
of a non-spherical shock with shell interfaces can
give rise to the Richtmyer--Meshkov instability
\citep{kifonidis_06}. The asymmetries imprinted
during the first seconds of an explosion provide
the seed for these late-time mixing instabilities,
and 3D supernova modellling is now  moving
towards an integrated approach from
the early to the late stages of the to better link
the observations to the physics of the explosion mechanism
\citep[e.g.,][]{hammer_10,wongwathanarat_13,mueller_18,chan_18,chan_20b}, and, in future, even to the multi-D progenitor structure.

\subsection{Scope and structure of this
review}
In this review, we are primarily concerned
with the numerical techniques for modeling
multi-D fluid flow in core-collapse supernovae
and their progenitors, and with
our current understanding of the theory
and phenomenology of the multi-D fluid
instabilities. Although
multi-D effects are relevant to virtually
all aspects of core-collapse supernova
theory, we can only afford cursory attention to
many of them in order to keep this overview focused. 

There are many other reviews to fill the gap, or
provide a different perspective. \citet{janka_12}
provides a very broad, but less technical, overview of
the core-collapse supernova problem. For the explosion
mechanism and a different
take on the role of multi-D effects, the reader may also consult \citet{mezzacappa_05,kotake_06,burrows_13,mueller_16b,janka_16,couch_17}.
The important problem of neutrino transport is
treated in considerable depth by  \citet{mezzacappa_05,mezzacappa_19} and is
therefore not addressed in this review.
A number of reviews address the potential of
neutrinos \citep[e.g.,][]{kotake_06,mueller_19d}
and gravitational waves \citep{ott_08b,kotake_13,kalogera_19} as diagnostics of the multi-dimensional
dynamics in the supernova core.

We shall start by discussing the governing
equations for reactive, self-gravitating flow
and their numerical solution in the context
of core-collapse supernovae and convective
burning in Sect.~\ref{sec:methods}. 
We do not treat numerical methods
for MHD in
supernova simulations, although we occasionally
comment on the role of MHD effects in the
later sections.
In the subsequent sections, we
then review recent simulation results and
progress in the theoretical understanding
of convection during the late burning
stages (Sects.~\ref{sec:prog3d}),
of supernova shock revival
(Sect.~\ref{sec:sn}), and 
the hydrodynamics of the explosion phase
(Sect.~\ref{sec:expl_phase}).

\section{Numerical methods}
\label{sec:methods}
Modeling the late stages of nuclear burning and
the subsequent supernova explosion
involves solving the familiar equations for
mass, momentum, and energy conservation
with source terms that account for nuclear
burning and the exchange of energy and
momentum with neutrinos. Viscosity and
thermal heat conduction mediated by
photons, electron/positrons, and ions can
be neglected, and so we have 
(in the Newtonian
limit and neglecting magnetic fields),
\begin{eqnarray}
\label{eq:hydro1}
\frac{\pd \rho}{\pd t}
+\nabla \cdot (\rho \mathbf{v})
&=&
0,
\\
\label{eq:hydro2}
\frac{\pd \rho \mathbf{v}}{\pd t}
+\nabla \cdot (\rho \mathbf{v} \otimes \mathbf{v})
+ \nabla P
&=&
-\rho \nabla \Phi
+\mathbf{Q}_\mathrm{m},
\\
\label{eq:hydro3}
\frac{\pd \rho (\varepsilon+v^2/2)}{\pd t}
+\nabla \cdot \left[\rho (\varepsilon+v^2/2)\mathbf{v}
+ P\mathbf{v}\right]
&=&
-\rho \mathbf{v} \cdot \nabla \Phi
+ Q_\mathrm{e}
+ \mathbf{Q}_\mathrm{m} \cdot \mathbf{v},
\end{eqnarray}
in terms of the density $\rho$,
the fluid velocity $\mathbf{v}$, the pressure $P$,
internal energy density $\varepsilon$, the gravitational potential $\Phi$, and
the neutrino energy and momentum source
terms $Q_\mathrm{e}$ and $\mathbf{Q}_\mathrm{m}$. If we
take $\varepsilon$ to include nuclear rest-mass
contributions, there is no source term for the nuclear energy generation rate; otherwise
an additional source term $\dot{Q}_\mathrm{nuc}$
appears on the right-hand side (RHS)
of Eq.~(\ref{eq:hydro3}).
These equations are supplemented by 
conservation equations for the mass fractions
$X_i$ of different nuclear species and
the electron fraction $Y_\mathrm{e}$ (net number
of electrons per baryon),
\begin{eqnarray}
\label{eq:xi}
    \frac{\pd \rho X_i}{\pd t}
    +\nabla  \cdot (\rho X_i \mathbf{v})&=&
    \dot{X}_{i,\mathrm{burn}},
\\
\label{eq:ye}
    \frac{\pd \rho Y_\mathrm{e}}{\pd t}
    +\nabla  \cdot (\rho Y_\mathrm{e} \mathbf{v})&=&
    \dot{Q}_{Y_\mathrm{e}},
\end{eqnarray}
where the source terms $\dot{X}_{i,\mathrm{burn}}$
and $\dot{Q}_{Y_\mathrm{e}}$ account for nuclear reactions
and the change of the electron fraction by $\beta$
processes.

In the regime of sufficiently high optical
depth, the effect of the neutrino source terms
could alternatively be expressed by non-ideal
terms for heat conduction, viscosity, and diffusion
of lepton number
\citep[e.g.,][]{imshenik_72,bludman_78,goodwin_82,van_den_horn_83,van_den_horn_84,yudin_08}, but this approach would
break down at low optical depth.
The customary approach to  Eqs.~(\ref{eq:hydro1}--\ref{eq:ye}) is, therefore,
to apply an operator-split approach and combine
a solver for ideal hydrodynamics for
the left-hand side (LHS) and
the gravitational source terms
with separate solvers for the 
source terms due to neutrino interactions and
nuclear reactions. Simulations
of the Kelvin--Helmholtz
cooling phase of the PNS over
time scales of seconds form an exception;
here only the PNS interior
is of interest so that it is possible
and useful to formulate the neutrino 
source terms in the equilibrium diffusion
approximation
\citep{keil_96,pons_99}.

\subsection{Hydrodynamics}
A variety of computational 
methods are employed to solve the 
equations of ideal hydrodynamics
in the context of supernova
explosions or the late stellar burning
stages. Nowadays, the vast majority
of codes use Godunov-based high-resolution
shock capturing (HRSC) schemes with
higher-order reconstruction
(see, e.g., \citealt{leveque,marti_15,balsara_17}
for a thorough introduction).
Examples include 
implementations
of the piecewise
parabolic method
of \citet{colella_85} 
or extensions thereof in
the Newtonian hydroydnamics codes
\textsc{Prometheus}
\citep{fryxell_89,fryxell_91,mueller_91},
which has been integrated
into various neutrino transport
solvers by the Garching group
\citep{rampp_02,buras_06a,scheck_06},
its offshoot \textsc{Prompi}  \citep{meakin_07},
\textsc{PPMstar}
\citep{woodward_18},
\textsc{VH1} 
\citep{blondin_93,hawley_12}
as used within the \textsc{Chimera}
transport code  \citep{bruenn_18},
\textsc{Flash}
\citep{fryxell_00},
\textsc{Castro} 
\citep{almgren_06},
\textsc{Alcar} \citep{just_15},
and \textsc{Fornax}
\citep{skinner_19}.
This approach is also 
used in most general relativistic (GR)
hydrodynamics codes
for core-collapse supernovae
like
\textsc{CoCoNuT} \citep{dimmelmeier_02_a,mueller_10},
\textsc{Zelmani} \citep{reisswig_13,roberts_16},
and \textsc{GRHydro}
\citep{moesta_14a}.  
Godunov-types
scheme with piecewise-linear 
total-variation diminishing (TVD) reconstruction are still used in the
\textsc{Fish} code
\citep{kaeppeli_11}
and in the relativistic  \textsc{Fugra} code of
\citep{kuroda_12,kuroda_16}. The 3DnSNe code of the Fukuoka group
\citep[e.g.,][]{takiwaki_12}, which is
based on the \textsc{ZEUS} code of \citet{stone_92},
has also switched from  an artificial viscosity scheme to a Godunov-based finite-volume
approach with TVD reconstruction
\citep{yoshida_19}.

Alternative strategies are less frequently employed.
The \textsc{Vulcan} code \citep{livne_93,livne_04}
uses a staggered grid and von Neumann--Richtmyer
artificial viscosity
\citep{neumann_50}.
The \textsc{SNSPH} code  \citep{fryer_06}
uses smoothed-particle hydrodynamics
(\citealp{gingold_77,lucy_77};
for modern reviews see
    \citealp{price_12,rosswog_15}).
Although less widely used in supernova
modeling today,
the SPH approach has been utilized for some
of the early studies of Rayleigh--Taylor mixing 
\citep{herant_91} and convectively-driven
explosions \citep{herant_92} in 2D
and later for the first 3D supernova
simulations with gray neutrino transport
\citep{fryer_02}. Multi-dimensional moving mesh schemes
have been occasionally employed to simulate
magnetorotational supernovae
\citep{ardeljan_05} and reactive-convective flow
in stellar interiors \citep{dearborn_06,stancliffe_11}.
More recently ``second-generation'' moving-mesh
codes based on Voronoi tessellation, such 
as \textsc{Tess} \citep{duffell_11} and
\textsc{Arepo} \citep{springel_10}, have been
developed and employed for simulations of
jet outflows \citep{duffell_13,duffell_15}, and
fallback supernovae \citep{chan_18}.
Spectral solvers for the equations
of hydrodynamics, while popular for solar convection,
have so far been applied only once for simulating
oxygen burning \citep{kuhlen_03}, but never for
the core-collapse supernova problem.

Since Godunov-based finite-volume solvers are now most commonly
employed for simulating core-collapse supernovae and the
late burning stages, we shall focus on the problem-specific
challenges for this approach in this section.

\subsubsection{Problem geometry and choice of grids}
The physical problem geometry in global simulations of core-collapse
supernovae and the late convective burning stages is characterized by 
approximate spherical symmetry, and one frequently needs to deal
with strong radial stratification and a large range of radial scales.
For example, during the pre-explosion and early explosion phase,
the PNS develops 
a ``density cliff'' at its surface
that is approximately in radiative equilibrium and
can be approximated as an exponential isothermal atmosphere
with a scale height $H$ of
\begin{equation}
    H=\frac{k T R^2}{G M m_\mathrm{b} },
\end{equation}
in terms of the PNS mass $M$, radius $R$, and surface temperature $T$, and the
baryon mass $m_\mathrm{b}$. With typical values of $M\sim 1.5\,M_\odot$,
$R$ shrinking down to a final value of $\mathord{\sim}12\, \mathrm{km}$,
and a temperature of a few MeV, the scale height soon shrinks to a few $100 \, \mathrm{m}$.
Later on during the explosion, the scales of interest shift to the radius
of the entire star, which is of order $\mathord{\sim} 10^8 \, \mathrm{km}$
for red supergiants.

The spherical problem geometry and the multi-scale nature of the flow
is a critical element in the choice of the numerical grid for
``star-in-a-box'' simulations.\footnote{Of course, some problems can
or need to be studied using simplified geometries (planar
or cylindrical) or local simulations.} Cartesian grids, various
spherical grids, and, on occasion, unstructured grids have been
used in the field for global simulations and face different challenges. 

\paragraph{Grid-induced perturbations.}
Cartesian grids have the virtue of algorithmic simplicity and do not
suffer from coordinate singularities, but also
come with disadvantages as they are not adapted to the approximate
symmetry of the physical problem. The unavoidable non-spherical
perturbations from the grid make it impossible to reproduce
the spherically symmetric limit in multi-D even for
perfectly spherical initial conditions, or to study the
growth of non-spherical perturbations
in a fully controlled manner. The former deficiency is arguably an acceptable
sacrifice, though it can limit the possibilities for code testing and
verification, but the latter can introduce visible artifacts in simulations. For example, Cartesian codes 
sometimes  produce
non-vanishing gravitational wave signals from the bounce
of non-rotating cores \citep{scheidegger_10}, and often show
dominant $\ell=4$ modes during the growth phase
of non-radial instabilities
\citep{ott_12}.

\paragraph{Handling the multi-scale problem.}
Furthermore, a single Cartesian grid cannot easily handle
the multiple scales encountered in the supernova problem. 
Even with $\mathcal{O}(1000^3)$ zones, such a grid
can at best cover the region inside $1000 \, \mathrm{km}$
with acceptable resolution, but
following the infall of matter for several $100 \, \mathrm{ms}$
without boundary artifacts
and the development of an explosion requires covering a region
of at least $10,000 \, \mathrm{km}$. This problem is often dealt with
by using adaptive mesh refinement \citep[AMR; see, e.g., ][]{berger_89,fryxell_00}, which is
usually implemented as ``fixed mesh refinment'' 
for pre-defined nested cubic patches \citep[e.g.,][]{schnetter_04}.
Other codes have opted to combine a single central
Cartesian patch or nested patches with a spherically
symmetric region
\citep{scheidegger_10} or multiple spherical polar patches
\citep{ott_12} outside. For long-time simulations of Rayleigh--Taylor
mixing in the envelope, standard adaptive or
pre-defined mesh refinement may not be sufficiently efficient for
covering the range of changing scales and necessitate manual
remapping to a coarser grids (``homographic expansion'') as the simulation
proceeds \citep{chen_13}.

In spherical coordinates, the multi-scale nature
of the problem can be accommodated to a large
degree by employing a non-uniform radial grid that transitions
to roughly equal spacing in $\log r$ at large radii.
Radial resolution can be added selectively in strongly
stratified regions like the PNS surface
\citep{buras_06a},
or one can use an adaptive moving
radial grid \citep{liebendoerfer_04,bruenn_18}.
However, some care must be exercised in using
non-uniform radial grids. Rapid variations
in the radial grid resolution $\Delta r/r$ can produce
artifacts such as artificial waves and disturbances
of hydrostatic equilibrium. 

It is also straightforward to implement a moving radial
grid to adapt to changing resolution requirements or
the bulk contraction/expansion of the region of interest;
see  \citet{winkler_84,mueller_94} for an explanation
of this technique. The MPA/Monash group routinely apply
such a moving radial grid in quasi-Lagrangian mode
during the collapse phase \citep{rampp_00},
and, with a prescribed grid function,
in parameterized multi-D simulations of 
neutrino-driven explosions \citep{janka_96,scheck_06}
and in simulations of convective
burning \citep{mueller_16c}. 
The Oak Ridge group
uses a truly adaptive
radial grid in their supernova simulations with the
\textsc{Chimera} code \citep{bruenn_18}. 
A moving radial mesh might also appear useful
for following the expansion of the ejecta
and the formation of a strongly diluted central region
in simulations of mixing instabilities in the envelope,
but the definition of an appropriate grid function
is non-trivial. Most simulations of mixing instabilities
in spherical polar coordinates have therefore relied
on simply removing zones continuously from the evacuated
region of the blast wave to increase
the time step \citep{hammer_10} rather than
 implementing a moving radial mesh \citep{mueller_18}.

Both fixed mesh refinement and spherical grids
with non-uniform radial mesh spacing only provide
limited adaptability to the structure of the flow.
Truly adaptive mesh refinement can provide superior
resolution in cases where very tenuous, non volume-filling flow structures emerge. Mixing instabilities in the
envelope are a prime example for such a situation,
and have often been studied using AMR in spherical
polar coordinates \citep{kifonidis_00,kifonidis_03,kifonidis_06} in 2D and Cartesian
coordinates \citep{chen_17}.

\paragraph{The time step constraint in spherical polar
coordinates.}
While spherical polar coordinates are well-adapted
to the problem geometry, they also suffer from drawbacks.
One of these drawbacks  --- among others
that we discuss further below --- is that the converging cell geometry
imposes stringent constraints on the time step near
the grid axis and the origin. The 
Courant--Friedrichs--Lewy limit
\citep{courant_28} for the time step $\Delta t$
requires that 
$\Delta t < r \,\Delta \theta/(|\mathbf{v}|+c_\mathrm{s})$
in 2D
and $\Delta t < r \sin \theta \,\Delta \varphi
/(|\mathbf{v}|+c_\mathrm{s})$ in 3D in terms of the grid spacing
$\Delta \theta$ and $\Delta \varphi$ in latitude
and longitude, and the fluid velocity $\mathbf{v}$ and sound speed $c_\mathrm{s}$. If $\Delta \theta =\Delta \varphi$, this
is worse than a Cartesian code with grid spacing comparable
to $\Delta r$
by a factor $\Delta \theta \ll 1$ in 2D
and  $\Delta \theta^2 \ll 1$ in 3D near the origin.

Various workarounds have been developed to tame this
time-step constraint. Some
core-collapse supernova codes
(\textsc{Prometheus-Vertex}, \textsc{Chimera}, \textsc{CoCoNuT})
simulate
the innermost region of the grid assuming spherical symmetry. The approximation of
spherical symmetry is well justified in the core, since the
innermost region of the PNS is convectively stable during
the first seconds after collapse and explosion until the late Kelvin--Helmholtz cooling phase. Even more savings can
be achieved by treating the PNS convection zone using
mixing-length theory \citep{mueller_15b}, but this is a
more severe approximation that significantly affects
the predicted gravitational wave signals and certain
features of the neutrino emission and nucleosynthesis.
Concerns have also been voiced that the imposition of
a spherical core region creates an immobile obstacle
to the flow that leads to the violation of momentum
conservation, which might have repercussions on
neutron star kicks \citep{nordhaus_10b}. While it is true that
the PNS tends not to move in 
simulations with a spherical core region,
\citet{scheck_06} found (using a careful
analysis based on hydro simulations in the
accelerated frame comoving with the PNS)
that the assumption of an immobile PNS
does not gravely affect the dynamics in the supernova interior
and the PNS kick in particular.

Even with a 1D treatment for the innermost grid zones, one
is still left with a severe time-step constraint at the
grid axis in 3D. A number of alternatives to
spherical polar grids with uniform spacing in
latitude $\theta$ and
longitude $\varphi$ can help to remedy this.
The simplest workaround is to adopt uniform
spacing in $\mu=\cos \theta$ instead of $\theta$.
In this case, one has
$\sin \theta = (2N_\theta-1)^{1/2}/N_\theta\approx\sqrt{2} N_\theta^{-1/2}$
in the zones adjacent to the axis 
for $N_\theta$ zones in latitude
instead of
$\sin \theta \approx N_\theta^{-1}/2$,
so the time step limit scales as
$\Delta t \propto \sqrt{2} N_\theta^{-1/2} N_\phi^{-1}$
instead of 
$\Delta t \propto N_\theta^{-1} N_\phi^{-1}/2$, where $N_\varphi$ is the number of zones in longitude. Alternatively, one can selectively increase the
$\theta$-grid spacing in the zones close to the axis.
However, the time step constraint at the axis is still
more restrictive than at the equator in this approach,
and the aspect ratio of the grid cells becomes extreme
near the pole, which can create problems with numerical
stability and accuracy.

\begin{figure}
    \centering
    a)
        \includegraphics[width=0.24\textwidth]{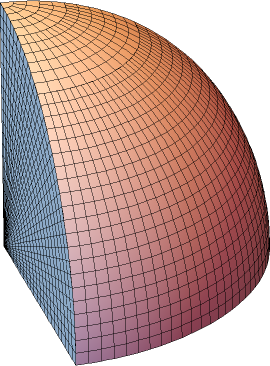}
        \hfill
        b)
    \includegraphics[width=0.32 \textwidth]{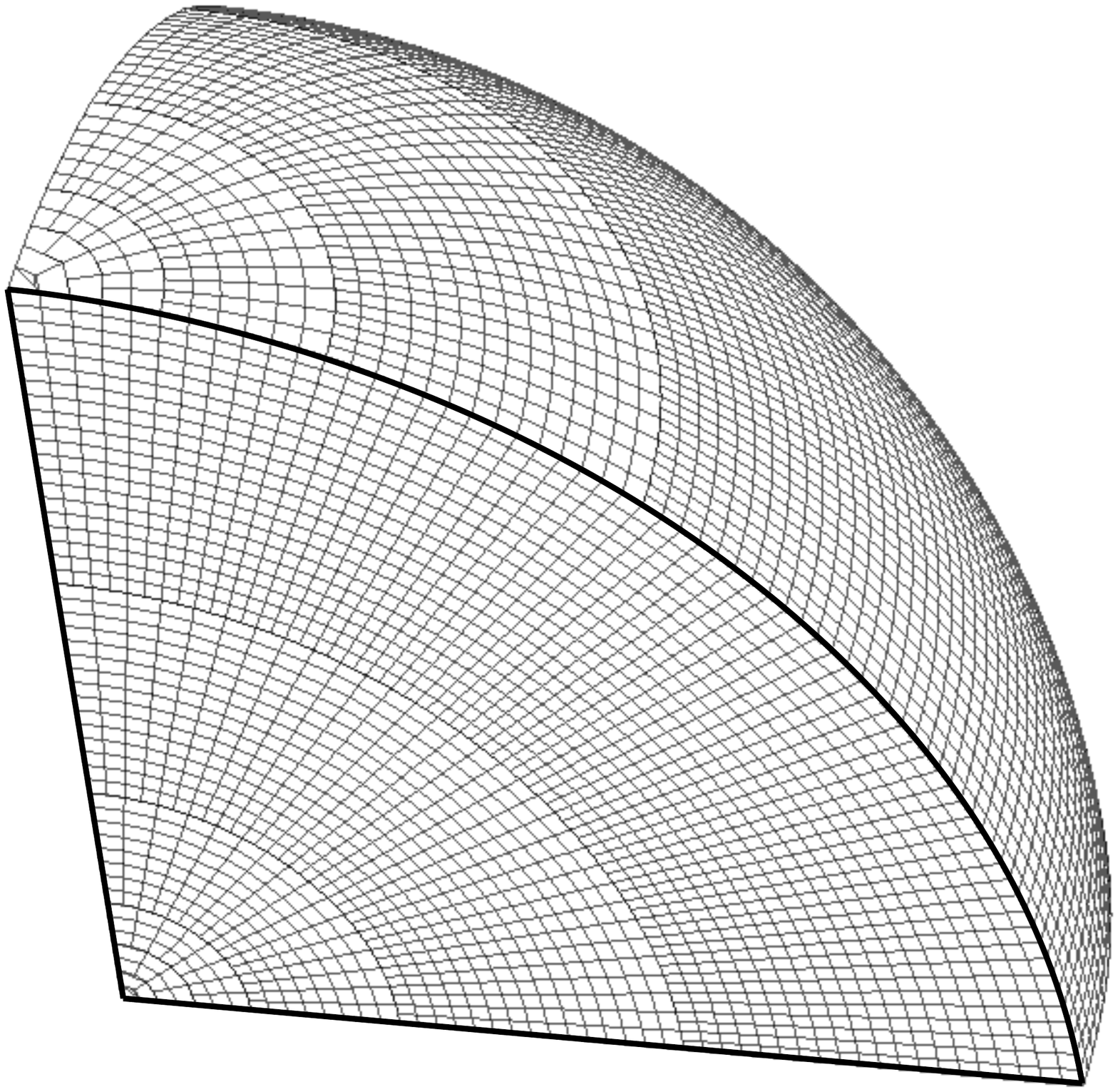}
    \hfill
    c)
    \includegraphics[width=0.32\textwidth]{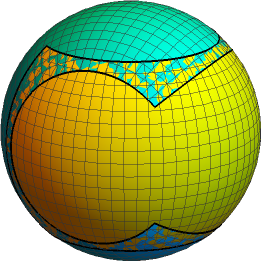}
    \caption{Alternative spherical grids that avoid the tight time step constraint at the axis of standard spherical polar grids:
\textbf{a)} Grid with mesh coarsening in the $\varphi$-direction
    only. Only an octant of the
    entire grid is shown.
    \textbf{b)} Dendritic grid with
    coarsening in the $\theta$- and $\varphi$-direction
    Image reproduced with permission from \citep{skinner_19}, copyright by AAS.
    \textbf{c)} Overset Yin-Yang grid \citep{kageyama_04,wongwathanarat_10a} with two overlapping spherical
    polar patches in yellow and cyan.
    }
    \label{fig:grids}
\end{figure}

One approach to fully cure the time step problem,
which was first proposed for simulations of
compact objects by \citet{cerda_09},
consists in abandoning the logically Cartesian grid
in $r$, $\theta$, and $\varphi$ and selectively
coarsening the grid spacing in $\varphi$ (and possibly
$\theta$) near the axis (and optionally at small $r$)
as illustrated in Fig.~\ref{fig:grids}.
Such a mesh coarsening scheme has been included
in the \textsc{CoCoNuT-FMT} code \citep{mueller_15b}
with coarsening in the $\varphi$-direction, 
and as a ``dendritic grid'' with coarsening in
the $\theta$- and $\varphi$-direction in the
\textsc{Fornax} code \citep{skinner_19} and
 the 3DnSNe code \citep{nakamura_19}. 
Mesh coarsening can be implemented
 following  standard AMR
practice by prolongating
from the coarser grids to the finer grids in the
reconstruction step. Alternatively, one can
continue using the hydro solver on a fine
uniform grid in $\theta$ and $\varphi$, 
and average the solution over coarse ``supercells'' 
after each time step, followed by a conservative prolongation
or ``pre-reconstruction''
step back onto the fine grid to ensure higher-order
convergence. This has the advantage of retaining the data layout and algorithmic
structure of a spherical polar code, but care is required
to to ensure that the prolongation of the conserved
variables does not introduce non-monotonicities in
the primitive variables, which limits the pre-reconstruction
step to second-order accuracy in practice \citep{mueller_19a}.
A possible concern with mesh coarsening
on standard spherical polar grids is that it  may
favor the emergence of axis-aligned bipolar flow
structures during the explosion phase
in supernova simulations \citep{mueller_15b,nakamura_19}. In practice, however, strong physical seed perturbations
easily break any grid-induced alignment of the flow with the axis.
As the more simulations with mesh coarsening
become available \citep{mueller_19a,burrows_19b},
it does not appear that axis alignment is a 
recurring problem.

Filtering in Fourier space, which has long
been used in the meteorology community \citep{boyd},
provides another means
of curing the restrictive time step constraint
near the axis and has been implemented in the
\textsc{CoCoNuT-FMT} code \citep{mueller_19a}.
This can also be implemented with minimal interventions
in a solver for spherical polar grids, and is
attractive because the amount of smoothing that
is applied to the solution increases more gradually
towards the axis than with mesh coarsening schemes.
\citet{mueller_19a} suggests that this eliminates the problem
of axis-aligned flows. On the downside, simulations with
Fourier filtering may occasionally encounter problems
with the Gibbs phenomenon at the shock.

More radical solutions to the axis problem include overset
grids
and non-orthogonal spherical grids.
An overset Yin-Yang grid \citep{kageyama_04}
has been implemented in the \textsc{Prometheus} code
\citep{wongwathanarat_10a,melson_msc} and used successfully
for simulations of supernovae and convective burning.
The Yin-Yang grid provides near-uniform resolution
in all directions, solves the time step problem, and
also eliminates the delicate problem of boundary conditions
at the axis of a spherical polar grid. The added
algorithmic complexity is limited to interpolation routines
that provide boundary conditions; since each patch
is part of a spherical polar grid, no modifications of
the hydro solver for non-orthogonal grids are required.
As a downside, it is more complicated --
but possible \citep{peng_06} -- to implement
overset grids in a strictly conservative manner.
In future, non-orthogonal grids spherical grids
\citep{ronchi_96,calhoun_08,wongwathanarat_16}
may provide
another solution that avoids the axis problem
and ensures conservation in a straightforward manner,
but applications are so far limited to other
astrophysical problems
\citep{koldoba_02,fragile_09,shiota_10}.

\paragraph{Boundary conditions.}
The definition of the outer boundary conditions for Cartesian
grids can be more delicate and less flexible than
for spherical grids. Simulations of supernova shock revival
and the late convective burning stages usually do not
cover the entire star for efficiency reasons, and sometimes
it can even be desirable to excise an inner core
region, e.g.\ the PNS interior in supernova simulations
\citep[e.g.,][]{janka_96,scheck_08,ugliano_12,ertl_15,sukhbold_16}
 or the Fe/Si core in the O shell burning
models of \citet{mueller_17}. 
To minimize artifacts near the outer and
(for annular domains) the inner boundary, the
best strategy is often to impose hydrostatic
boundary conditions assuming constant entropy,
so that the pressure $P$, density $\rho$,
and radial velocity $v_r$
in the ghost cells are obtained as
\begin{eqnarray}
\ud P&=&\int \rho g \,\ud r,\\
\ud \rho&=& \int c_\mathrm{s}^{-2} \,\ud P,\\
v_r &=&0.
\end{eqnarray}
in terms of the radial gravitational acceleration
$g$ and the sound speed $c_\mathrm{s}$
(cf.\ \citealp{zingale_02} for hydrostatic
extrapolation in the plane-parallel case).
This can be readily implemented for spherical
grid, and the same is true for inflow, outflow,
or wall boundary conditions for excised
outer shells or an excised core.

In Cartesian coordinates, however,
defining boundary conditions as
a function of radius is at odds with the usual
strategy of enforcing the boundary conditions
by populating ghost zones along individual
grid lines separately. 
For pragmatic reasons, 
one often enforces standard boundary conditions
(reflecting/inflow/outflow) on the faces of
the cubical domain instead \citep[e.g.,][]{couch_15},
which is viable as long as the domain boundaries
are sufficiently distant from the region of interest.
Alternatively, one can impose fixed boundary conditions \emph{inside} the cubical domain,
but  outside a smaller spherical region of
interest \citep[e.g.,][]{woodward_18b}. 
Outflow conditions on an interior boundary,
e.g.\ for fallback onto a compact remnant,
can also be implemented relatively easily
\citep{joggerst_09}. 

On the other hand, the boundary conditions
at the axis and the origin require careful consideration
in case of a spherical polar in order
to minimize artifacts from the grid singularities.
Conventionally, one uses reflecting boundary 
conditions to populate the ghost zones before
performing the reconstruction in the $r$-
and $\theta$- direction, i.e., one assumes
odd parity for the velocity components
$v_r$ and $v_\theta$ respectively, and 
even parity for scalar quantities and the
transverse velocity components. This usually ensures that
$v_r$ and $v_\theta$ do not blow up near the grid singularities,
but in some cases stronger measures are required; e.g., 
one can enforce zero $v_r$ or $v_\theta$ in the cell next
to the origin/grid axis, or switch to step function reconstruction
in the first cell. One may also need to impose odd parity
for $v_\varphi$ or for better stability, or reconstruct
the angular velocity component $\omega_\varphi=v_\varphi/r$
instead of $v_\varphi$. 

No hard-and-fast rules for such fixes
at the axis and the origin can be given, except perhaps that 
one should also consider treating the geometric source terms
in spherical polar coordinates differently (see below) before
applying fixes to the boundary conditions that
reduce the order of reconstruction, or 
before manually damping or zeroing
velocity components. In fact, the symmetry assumptions 
behind reflecting boundary conditions 
(i.e., $v_r\rightarrow 0$ at the origin and
$v_\theta \rightarrow 0$) are actually too strong. Strictly
speaking, one should only impose the condition that
the \emph{Cartesian} velocity components $v_x$, $v_y$, and
$v_z$ are continuous across the singularity for smooth flow.
In principle, this can be accommodated during the
reconstruction by populating the ghost zones for 
$r<0$, $\theta<0$, and $\theta>\pi$ with values from
the corresponding grid lines \emph{across} the origin
or the axis, bearing in mind any flip
in direction of the basis vectors $\mathbf{e}_r$,
$\mathbf{e}_\theta$, and $\mathbf{e}_\varphi$ across
the coordinate singularity. For the reconstruction along the radial
grid line with constant $\theta$ and $\varphi$, this comes
down to defining
\begin{eqnarray}
    v_r (r) &=& \left\{
    \begin{array}{ll}
    \phantom{-}v_r(r,\theta,\varphi),& r>0\\
    -v_r(r,\pi-\theta,\varphi+2\pi),& r<0
    \end{array}
    \right.\\
    v_\theta (r) &=& \left\{
    \begin{array}{ll}
    \phantom{-}v_\theta(r,\theta,\varphi),& r>0\\
    \phantom{-}v_\theta(r,\pi-\theta,\varphi+2\pi),& r<0
    \end{array}
    \right. \\
    v_\varphi (r) &=& \left\{
    \begin{array}{ll}
    \phantom{-}v_\varphi(r,\theta,\varphi),& r>0\\
    -v_\varphi(r,\pi-\theta,\varphi+2\pi),& r<0
    \end{array}
    \right.,
\end{eqnarray}
and, similarly, for reconstruction in the $\theta$-direction
along a grid line with constant $r$ and $\varphi$:
\begin{eqnarray}
    v_r (\theta) &=& \left\{
    \begin{array}{ll}
        \phantom{-}v_r(r,-\theta,\varphi+2\pi),& \theta<0\\
       \phantom{-} v_r(r,\theta,\varphi),& 0<\theta<\pi\\
        \phantom{-}v_r(r,2\pi-\theta,\varphi+2\pi),& \pi<\theta
    \end{array}
    \right.
    \\
    v_\theta (\theta) &=& \left\{
    \begin{array}{ll}
        -v_\theta(r,-\theta,\varphi+2\pi),& \theta<0\\
        \phantom{-}v_\theta(r,\theta,\varphi),& 0<\theta<\pi\\
        -v_\theta(r,2\pi-\theta,\varphi+2\pi),& \pi<\theta
    \end{array}
    \right.
    \\
    v_\varphi (\theta) &=& \left\{
    \begin{array}{ll}
        -v_\varphi(r,-\theta,\varphi+2\pi),& \theta<0\\
        \phantom{-}v_\varphi(r,\theta,\varphi),& 0<\theta<\pi\\
        -v_\varphi(r,2\pi-\theta,\varphi+2\pi),& \pi<\theta
    \end{array}
    \right..
\end{eqnarray}    
This allows for non-zero values of $v_r$ and $v_\theta$ at the origin
to reflect that matter can flow across the origin and the axis.
Such special polar boundary conditions have been implemented
for 3D light-bulb\footnote{Broadly speaking,
light-bulb simulations manually fix the neutrino
luminosities and spectral properties, or compute
them based on simple analytic considerations,
and also use simplified neutrino source terms.} simulations of SASI and convection
with \textsc{Flash} \citep{fernandez_15},
and are also used in the \textsc{Fornax} code 
\citep{skinner_19}.
In practice, however, reflecting boundary conditions do not appear
to pose a major obstacle for flows across the axis or the origin if the
diverging
fictitious force terms are treated appropriately (see below).
The reason is that reflecting boundaries merely slightly degrade
the accuracy of the first cell interfaces away from the origin
and the axis; the fact that velocity components \emph{at} the coordinate
singularity are (incorrectly) forced to zero on the cell interfaces
at $r=0$, $\theta=0$, and $\theta=\pi$ 
does not matter much because these interfaces
have a vanishing surface area, so that the interface \emph{fluxes} must vanish anyway.

\paragraph{Geometric source terms.}
Another obstacle in spherical polar coordinates is
the occurrence of fictitious force terms in the momentum
equation. In terms of the density $\rho$ and
the orthonormal components $v_i$ and $g_i$ of the velocity 
and gravitational acceleration, the equations read,
\begin{align}
\frac{\pd \rho v_r}{\pd t}
&+\frac{1}{r^2}\frac{\pd  r^2 \rho v_r^2}{\pd r}
+\frac{1}{r^2}\frac{\pd  r^2 \rho v_r v_\theta}{\pd r}
+\frac{1}{r^2}\frac{\pd  r^2 \rho v_r v_\varphi}{\pd r}
+\frac{\pd P}{\pd r}
=
\rho g_r
+
\rho\frac{v_\theta^2+v_\varphi^2}{r},
\\
\frac{\pd \rho v_\theta}{\pd t}
\nonumber
&+\frac{1}{r \sin \theta}\frac{\pd  \sin \theta \rho v_r v_\theta}{\pd \theta}
+\frac{1}{r \sin \theta}\frac{\pd  \sin \theta \rho v_\theta^2}{\pd \theta}
+\frac{1}{r \sin \theta}\frac{\pd  \sin \theta \rho v_\theta v_\varphi}{\pd \theta}
\\
&+\frac{1}{r}\frac{\pd P}{\pd \theta}
=
\label{eq:v2}
\rho g_\theta
+
\rho\frac{\cot \theta v_\varphi^2-v_r v_\theta}{r},
\\
\nonumber
\frac{\pd \rho v_\varphi}{\pd t}
&+\frac{1}{r \sin \theta}\frac{\pd  \rho v_r v_\varphi}{\pd \varphi}
+\frac{1}{r \sin \theta}\frac{\pd \rho v_\theta v_\varphi}{\pd \varphi}
+\frac{1}{r \sin \theta}\frac{\pd  \rho v_\varphi^2}{\pd \varphi}
+\frac{1}{r \sin \theta}\frac{\pd P}{\pd \varphi}
\nonumber
\\
\label{eq:v3}
&=
\rho g_\varphi
-\rho\frac{v_r v_\varphi+v_\theta v_\varphi \cot \theta}{r},
\end{align}
where the fictitious force terms are singular at the origin
and at the axis. Often  straightforward time-explicit
discretization is sufficient for these source terms, especially
in unsplit codes with Runge-Kutta time integration.
 In a dimensionally split
implementation it can be advantageous to include  
a characteristic state correction in the Riemann problem due to
 (some of the) fictitious force terms \citep{colella_84}. Time-centering of the geometric source terms
can also lead to minor differences \citep{fernandez_15}.

When stability problems or pronounced axis artifacts
are encountered, one can adopt a more radical solution
and transport the Cartesian momentum density 
$\mathbf{\rho v}=(\rho v_x,\rho v_y,\rho v_z)$
while still using the components $v_\alpha$ 
in the spherical
polar basis as advection velocities, so that the fictitious
force terms disappear entirely,
\begin{equation}
    \frac{\pd \rho \mathbf{v}}{\pd t}
    +
    \frac{1}{\sqrt{\gamma}}\frac{\pd \sqrt{\gamma} \rho \mathbf{v} v^\alpha  }{\pd x^\alpha}
    +
    \nabla P
    =\rho \mathbf{g},
\end{equation}
where $\alpha \in \{r,\theta,\varphi\}$ and
$\gamma=r^4 \sin^2 \theta$ is the determinant
of the metric. This has been implemented in the
\textsc{CoCoNuT-FMT} code as one of several options
for the solution of the momentum equation.
Transforming back and forth between
the spherical polar basis for the reconstruction and
solution of the Riemann problem and the Cartesian
components for the update of the conserved quantities
might appear cumbersome, but in fact one need not
explicitly transform to Cartesian components at all.
Instead one only needs to rotate vectorial
quantities from the interface to the cell center
when updating the
momentum components in the spherical polar basis.
For example, for uniform
grid spacing $\Delta \theta$ in the 
$\theta$-direction,
the flux difference terms from the
$\theta$-interfaces $j$ and $j+1$ for updating
$\rho v_r$ and $\rho v_\theta$ in zone $j+1/2$ become
\begin{eqnarray}
\nonumber
    \left(\frac{\pd \rho v_{r,j+1/2}}{\pd t}\right)_\theta
    &=&
     \phantom{-}\cos \Delta \theta/2
    [
    \rho v_r v_\theta\Delta A]_{j\phantom{+1}}
    +\sin \Delta \theta/2 [
    (\rho v_\theta^2+P) \Delta A]_{j\phantom{+1}}
    \\
    &&-\cos \Delta \theta/2
    [
    \rho v_r v_\theta \Delta A]_{j+1}
    +\sin \Delta \theta/2 [
    (\rho v_\theta^2+P) \Delta A]_{j+1},
    \\
    \nonumber
    \left(\frac{\pd \rho v_{\theta,j+1/2}}{\pd t}\right)_\theta
    &=&
    \phantom{-} \cos \Delta \theta/2
    [
    (\rho v_\theta^2+P) \Delta A]_{j\phantom{+1}}
    -\sin \Delta \theta/2
    [\rho v_r v_\theta \Delta A]_{j\phantom{+1}}
    \\
    &&-\cos \Delta \theta/2
    [(\rho v_\theta^2+P)\Delta A]_{j+1}
    -\sin \Delta \theta/2 [
    \rho v_r v_\theta \Delta A]_{j+1},
\end{eqnarray}
where $\Delta A$ is the interface area and $\Delta \theta$
is the grid spacing in the $\theta$-direction,
which is assumed to be uniform here.
The term for $\rho v_\varphi$ is not modified at all.
Apart from eliminating the fictitious force terms in 
favor of flux flux terms, this alternative discretization
of the momentum advection and pressure terms also complies
with the conservation of total momentum (although
discretization of the gravitational source term may still 
violate momentum conservation). 

The singularities at the origin and the pole
constitute
a more severe problem for relativistic codes
using free evolution schemes for the metric, where
they can jeopardize the stability of the metric solver.
We refer
to \citet{baumgarte_13,baumgarte_15} for a robust
solution to this problem that has been
implemented in their \textsc{nada} code; they employ a reference metric
formulation both for the field equations 
and the fluid equations that factors out
metric terms that become singular and use a partially implicit
Runge--Kutta scheme to evolve the problematic terms.

\paragraph{Angular momentum conservation.}
A somewhat related issue concerns the violation
of angular momentum conservation in standard
finite-volume codes (both with Eulerian
grids and moving meshes). This is a concern
especially for problems such as convection in
rotating stars and  magnetorotational explosions
where rotation plays
a major dynamical role and the evolution of
the flow needs to be followed over long time scales. It is also
an issue for question such as pulsar spin-up
by asymmetric accretion, although
a post-processing of the numerical angular
momentum flux can help to obtain meaningful
results even when there is a substantial
violation of angular momentum conservation
\citep{wongwathanarat_13}.

\begin{figure}
    \centering
    \includegraphics[width=0.75\linewidth]{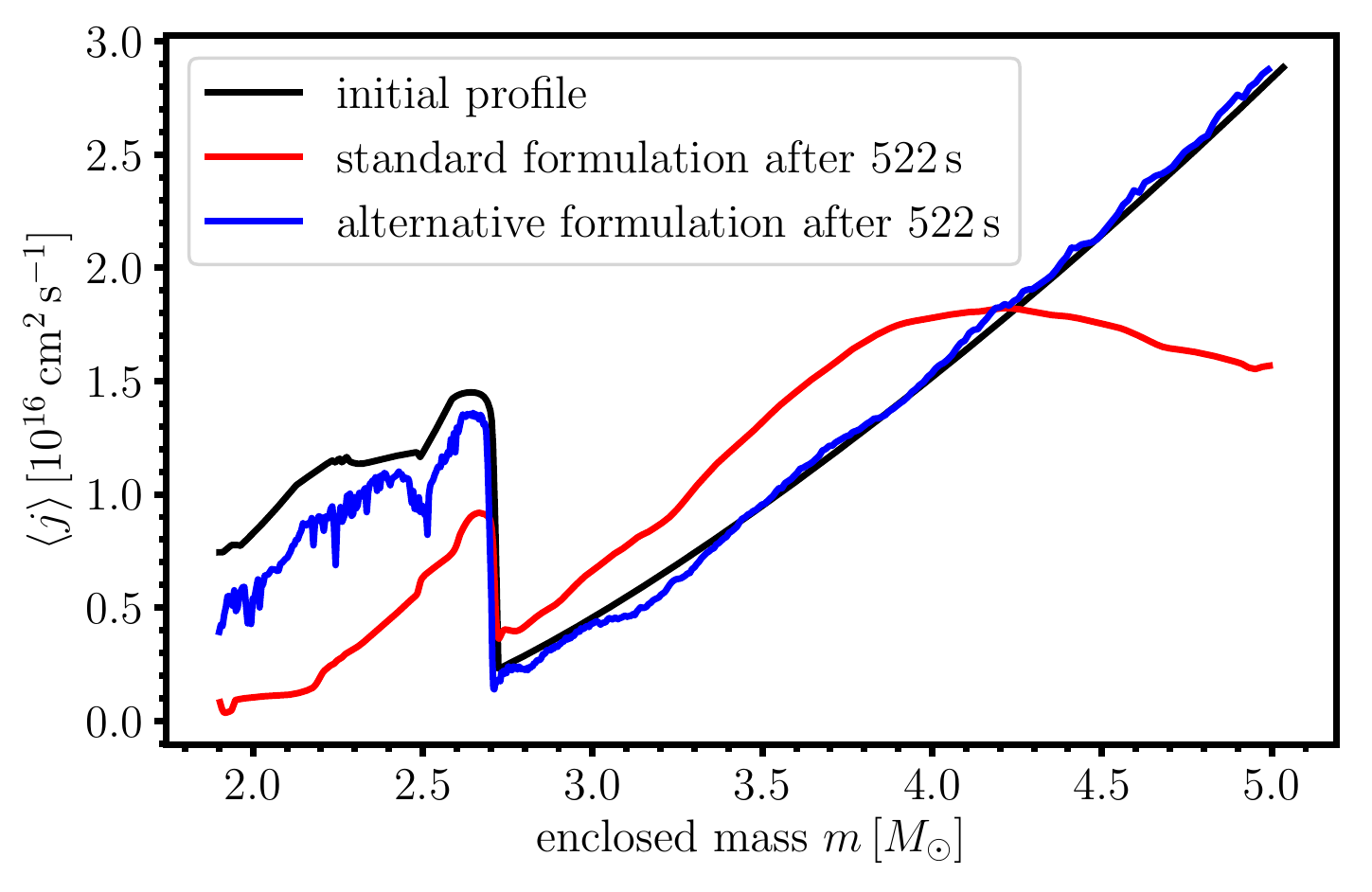}
    \caption{Profiles of the mass-weighted, spherically-averaged
    specific angular momentum
    $\langle j\rangle$ in 3D simulations of convective oxygen shell
    burning in a rapidly rotating gamma-ray
    burst progenitor with an initial helium core mass
    of $16\,M_\odot$ from \citet{woosley_06}. The red
    and blue curves show the angular momentum distribution
    after a simulation time of
    $522 \, \mathrm{s}$ (about 15 convective turnovers)
    for the standard formulation (red)
    of the fictitious force terms
    in Eqs.~(\ref{eq:v2},\ref{eq:v3})
    and for the alternative formulation (blue)
    in Eqs.~(\ref{eq:jnew1},\ref{eq:jnew2}).
    The initial angular momentum profile is shown 
    in black. The alternative formulation reduces
    the violation of global angular momentum conservation
    from 20\% to 7\%. However, for the standard formulation
    the effects of
    angular momentum non-conservation are much bigger
    locally than suggested by the global conservation error.
    The fictitious force terms proportional to
    $v_r$ lead to a considerable loss of angular momentum near
    the reflecting boundaries, even though the angular momentum
    \emph{flux} through the boundaries is exactly zero,
    and there is a spurious increase of angular momentum
    at the bottom of the convective oxygen burning shell outside
    the mass coordinate  $m=2.7\,M_\odot$.
    }
    \label{fig:ang_cons}
\end{figure}

The problem of 
angular momentum non-conservation can be
solved, or at least mitigated, using Discontinuous
Galerkin methods  \citep[][]{despres_15,mocz_14,schaal_15},
which are not currently used in this field however,
and can be avoided entirely in SPH \citep{price_12}.
For a given numerical scheme, increasing the resolution
is usually the only solution to minimize the
conservation error, but in 3D spherical polar coordinates
(and in 2D cylindrical coordinates), one can still
ensure exact conservation of the angular momentum
component $L_z$ along the grid axis by conservatively
discretizing
the conservation equation for
$\rho v_\varphi r \sin\theta$,
\begin{align}
\frac{\pd \rho v_\varphi r \sin \theta}{\pd t}
&+\frac{1}{r \sin \theta}\frac{\pd  \rho v_r v_\varphi r \sin \theta}{\pd \varphi}
+\frac{1}{r \sin \theta}\frac{\pd \rho v_\theta v_\varphi r \sin \theta}{\pd \varphi}
\nonumber
\\
&
+\frac{1}{r \sin \theta}\frac{\pd  \rho v_\varphi^2 r \sin \theta}{\pd \varphi}
+\frac{\pd P}{\pd \varphi}
=
\rho g_\varphi r \sin \theta,
\end{align}
instead of Eq.~(\ref{eq:v3}). Incidentally,
this angular-momentum conserving formulation
emerges automatically in GR hydrodynamics in spherical
polar coordinates if one solves for
the covariant momentum density components
as in the \textsc{CoCoNuT} code
\citep{dimmelmeier_02_a}.
However, as a price for exact conservation of
$L_z$, one occasionally encounters very rapid
rotational flow around the axis, and enforcing
conservation of only \emph{one} angular component
may add to artificial flow anisotropies due to the spherical
polar grid geometry. Moreover, this recipe
cannot be used for Yin-Yang-type overset spherical grids
or for non-orthogonal spherical grids. If angular momentum
conservation is a concern, one can, however, resort
to a compromise  by conservatively discretizing
the equations for $\rho v_\theta r$ and $\rho v_\varphi r$,
\begin{align}
\label{eq:jnew1}
\frac{\pd \rho v_\theta r}{\pd t}
&+\frac{1}{r \sin \theta}\frac{\pd  \sin \theta \rho v_r v_\theta r}{\pd \theta}
+\frac{1}{r \sin \theta}\frac{\pd  \sin \theta \rho v_\theta^2 r}{\pd \theta}
+\frac{1}{r \sin \theta}\frac{\pd  \sin \theta \rho v_\theta v_\varphi r}{\pd \theta}
\nonumber
\\
&+\frac{\pd P}{\pd \theta}
=
\rho g_\theta r
+
\rho\frac{\cot \theta v_\varphi^2}{r}
\\
\label{eq:jnew2}
\frac{\pd \rho v_\varphi r}{\pd t}
&+\frac{1}{r \sin \theta}\frac{\pd  \rho v_r v_\varphi r}{\pd \varphi}
+\frac{1}{r \sin \theta}\frac{\pd \rho v_\theta v_\varphi r}{\pd \varphi}
+\frac{1}{r \sin \theta}\frac{\pd  \rho v_\varphi^2  r}{\pd \varphi}
+\frac{1}{ \sin \theta}\frac{\pd P}{\pd \varphi}
\nonumber
\\
&=
\rho g_\varphi r
-\rho\frac{v_\theta v_\varphi \cot \theta}{r},
\end{align}
which eliminates some of the fictitious force terms.
This sometimes considerably
improves angular momentum conservation for
\emph{all} angular momentum components and works
for any spherical grid. Figure~\ref{fig:ang_cons} illustrates
the difference between the standard form of the
fictitious force term and the alternative
form in Eqs.~(\ref{eq:jnew1},\ref{eq:jnew2}) for
a simulation of oxygen shell convection in a
rapidly rotating gamma-ray burst progenitor.

\subsubsection{Challenges of subsonic turbulent
flow}
\label{sec:subsonic}
In the final stages of massive stars one encounters
a broad range of flow regimes. The convective Mach number, i.e., the ratio of the typical
convective velocity to the sound speed, during
the advanced convective burning stages is
fairly low,
ranging
from $\mathord{\sim} 10^{-3}$
or less \citep{cristini_17}
to  $\mathord{\sim}0.1$--$0.2$ in the innermost shells
at the onset of collapse \citep{collins_18}.
The flow is highly turbulent with 
Reynolds numbers of order
$10^{13}$--$10^{16}$.
During the supernova, one finds
convective Mach numbers
 from a few $10^{-2}$ in PNS convection to
$\mathord{\sim} 0.4$ \citep{mueller_15a,summa_16}
in the gain region around shock revival, and as
the shock propagates through the envelope,
the flow becomes extremely supersonic
with Mach numbers of up to several $10^2$.
The flow in unstable regions is typically highly turbulent. In the gain region, one obtains
a nominal Reynolds number 
of order  $10^{17}$ based on the neutron viscosity \citep{abdikamalov_15},
but non-ideal effects of \emph{neutrino} viscosity and drag play a role in the environment of the PNS
\citep{burrows_88b,guilet_15,melson_19}. In the
outer regions of the PNS convection zone
neutrino viscosity keeps 
the Reynolds number as low as $\mathord{\sim}100$
during some phases \citep{burrows_88b},
and in the gain region drag effects are
still so large that the flow cannot be assumed
to behave like ordinary high-Reynolds number
flow \citep{melson_19}. Later on, as
the shock propagates through the envelope
and mixing instabilities develop, neutrino
drag becomes unimportant, and the
flow is again in the regime of very high
Reynolds numbers.

Both the vast range in Mach number and the turbulent
nature of the flow present challenges for the accuracy
and robustness of numerical simulations. While even
relatively simple HRSC schemes can deal with 
Mach numbers of $\mathrm{Ma}\gtrsim 1$ with aplomb at reasonably
high resolution, they
can become excessively diffusive at low Mach number
because of spurious acoustic waves arising from
the discontinuities in the reconstruction
\citep{guillard_04,miczek_15}. Moreover, the acoustic
time step constraint in explicit finite-volume codes becomes excessively restrictive at low Mach number compared
to the physical time scales of interest. These
problems can be dealt with by using various low-Mach
number approximations,
e.g., the anelastic
approximation as in \citet{glatzmaier_84},
or more general formulations as
in the \textsc{Maestro} code of
\citealt{nonaka_10}, or by a time-implicit
discretization of the full compressible Euler equations
\citep{viallet_11,miczek_15}. However, few studies \citep{kuhlen_03,michel_19} have used such methods
to deal with low-Mach number flow in the late
stages of convective burning massive stars as yet;
they have been employed more widely to study, e.g., 
the progenitors of thermonuclear supernovae \citep{zingale_11,nonaka_12}

Part of the reason is that advanced HRSC schemes
remain accurate and competitive down to
Mach numbers of $10^{-2}$ and below depending
on the reconstruction method and the Riemann solver.
Although it is impossible to decide a priori
whether a particular choice of methods is
adequate for a given physical problem, or
what its resolution requirements are, it is useful to be
aware of strengths and weaknesses of different schemes 
in the context of subsonic turbulent flow. 
Unfortunately, our discussion of these strengths
and weaknesses must remain rather qualitative because
very few studies in the field have compared
the performance of different Riemann solvers and reconstruction
schemes in full-scale simulations and not only
for idealized test problems.

\begin{figure*}
  \includegraphics[width=0.49\textwidth]{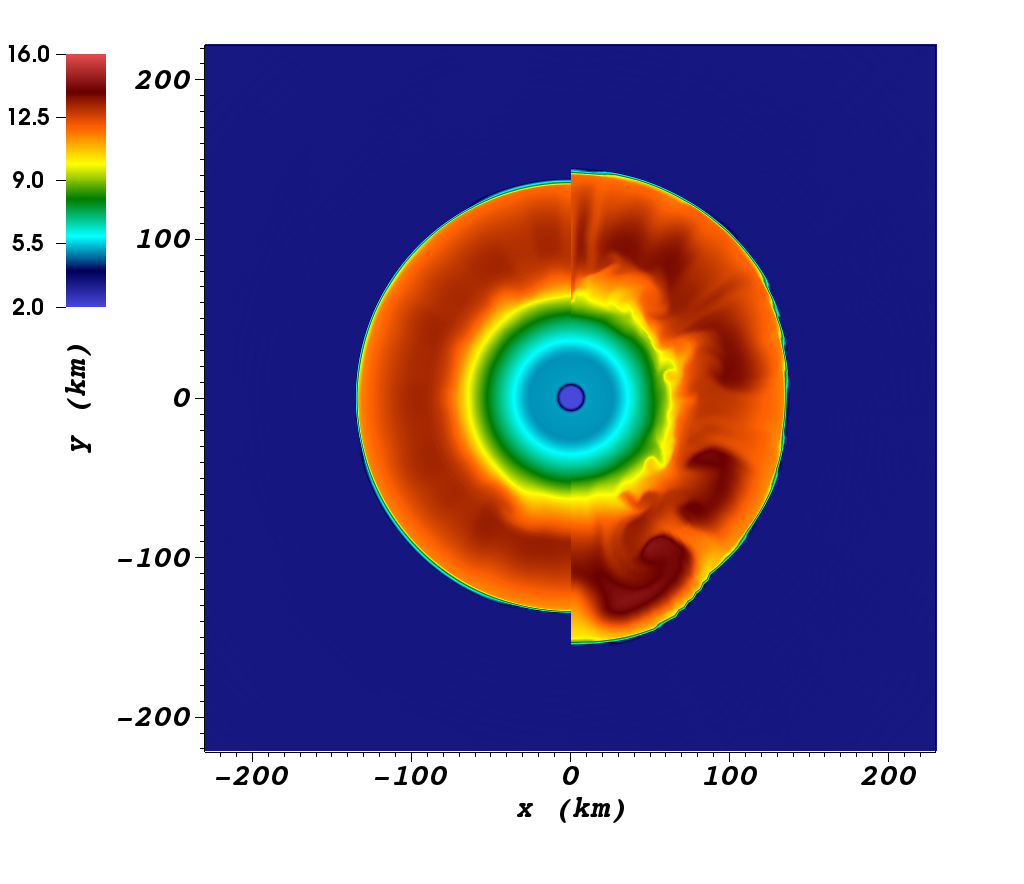}
  \hfill
  \includegraphics[width=0.49\textwidth]{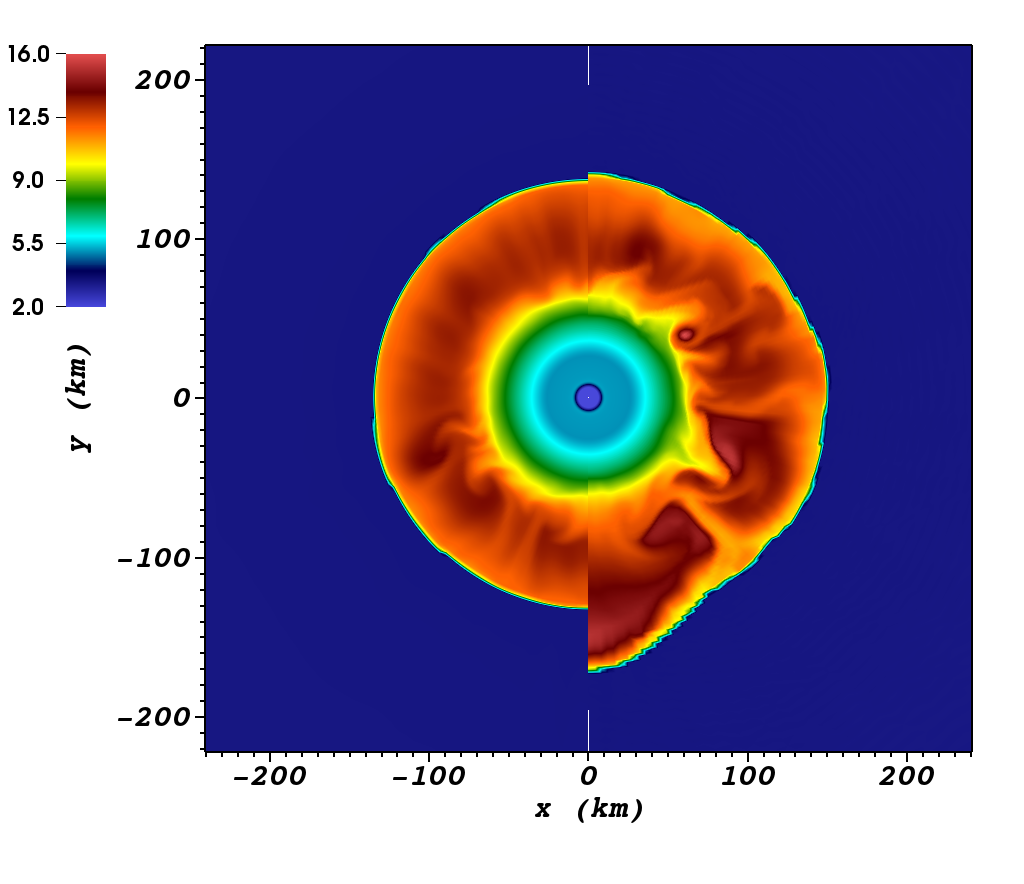}
  \\
  \includegraphics[width=0.49\textwidth]{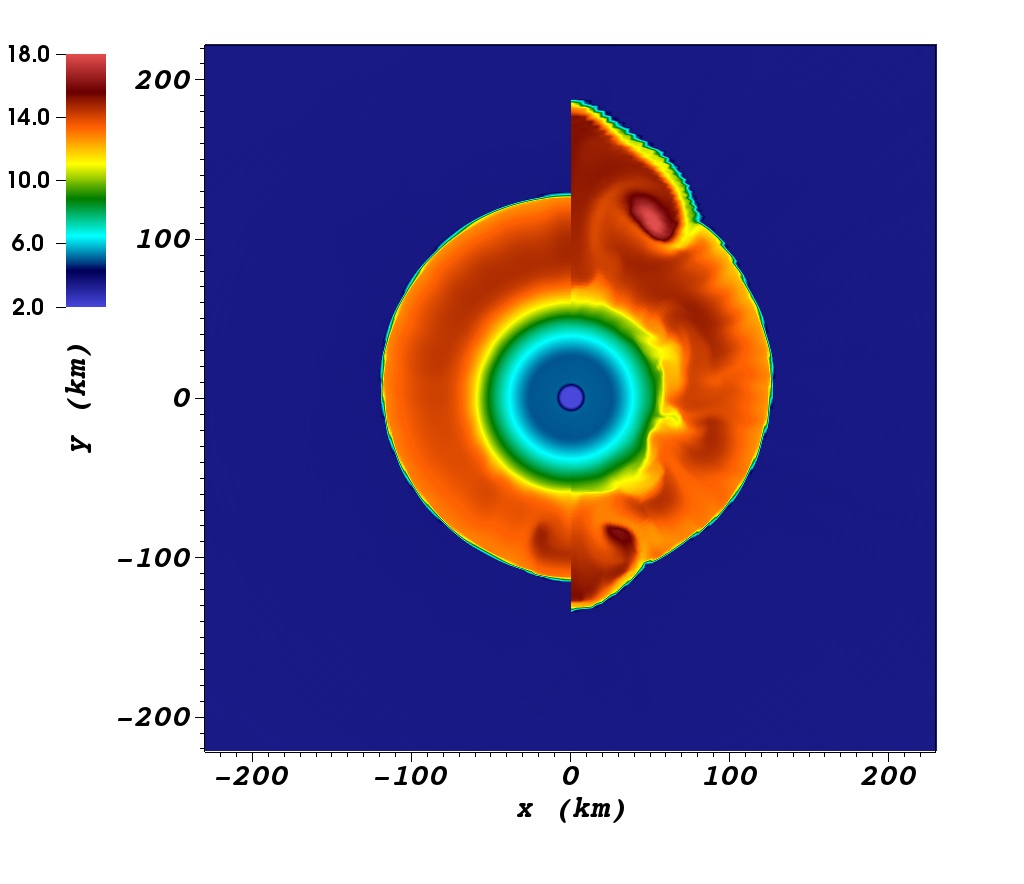}
  \hfill
    \includegraphics[width=0.49\textwidth]{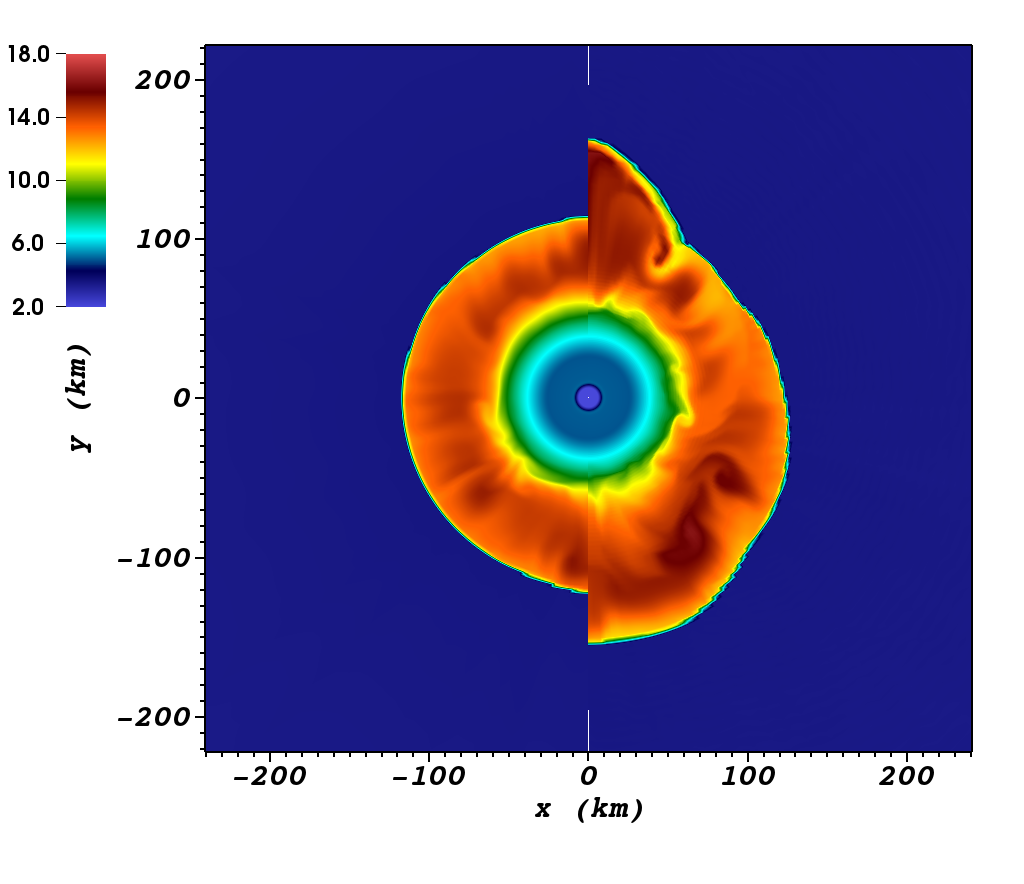}
  \\
  \includegraphics[width=0.49\textwidth]{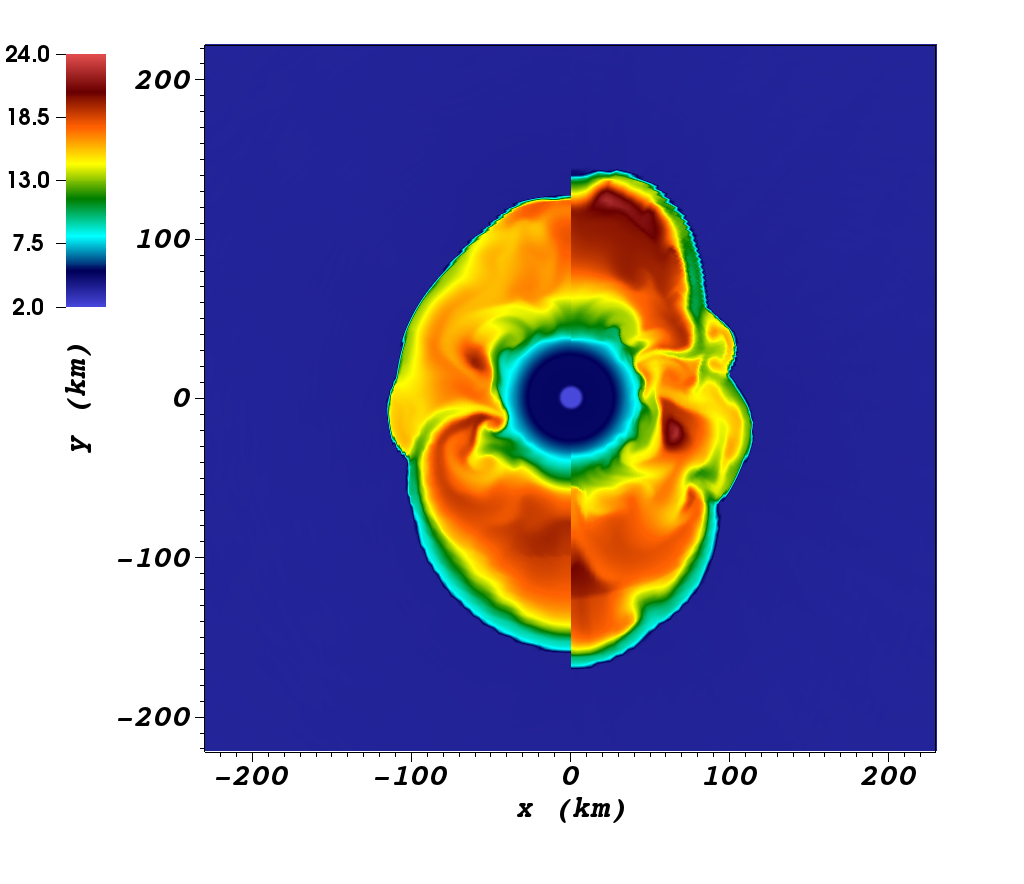}
  \hfill
  \includegraphics[width=0.49\textwidth]{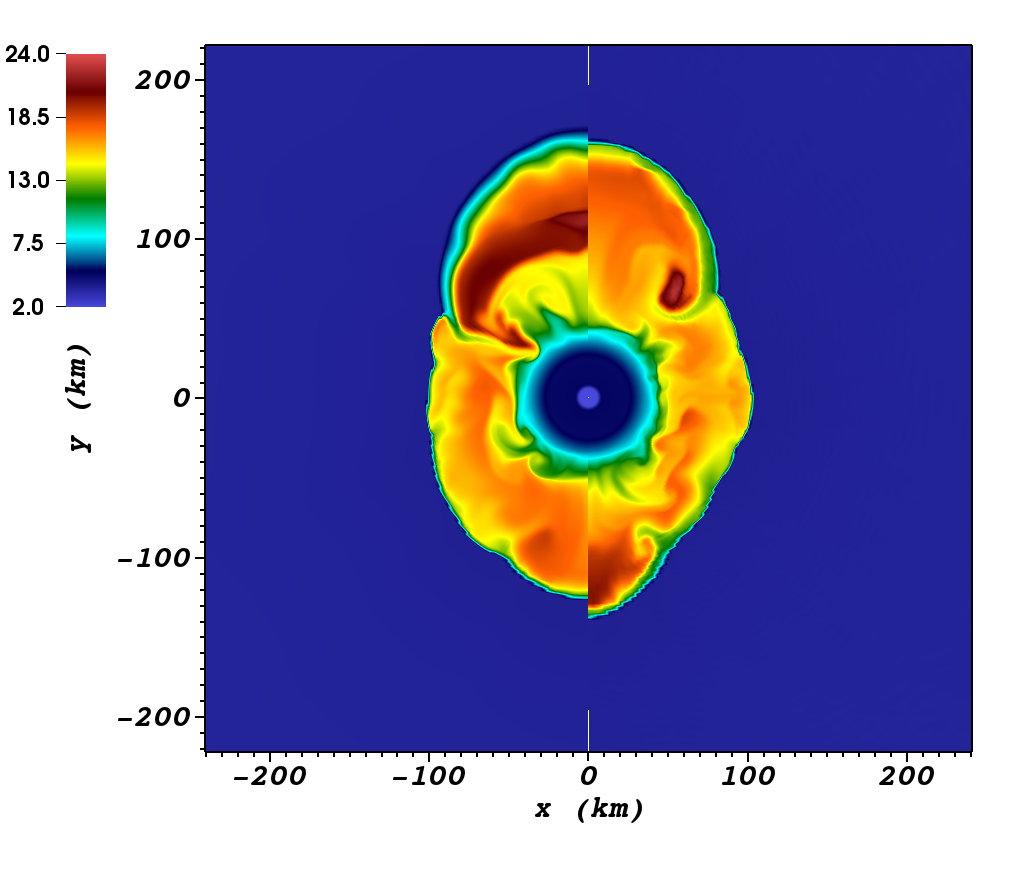}
\caption{Snapshots of the entropy from 2D simulations
of the $20\,M_\odot$ progenitor
of \citet{woosley_07} using the \textsc{CoCoNuT-FMT} code at post-bounce
time of $137 \, \mathrm{ms}$ (top row),
 $163 \, \mathrm{ms}$ (middle row),
 and $226 \, \mathrm{ms}$ (bottom row).
 The left and rights halves of the panels
in the left column show the results
for the HLLE and HLLC solver with standard
PPM reconstruction.
The left and rights halves of the  panels 
in the right column show the results
for second-order reconstruction using the MC limiter
and 6th-order extremum-preserving PPM reconstruction
using the HLLC solver.
}
\label{fig:hlle_hllc1}       
\end{figure*}

\paragraph{Riemann solvers.}
For supernova simulations with Godunov-based codes,
a variety of Riemann solvers are currently used. Newtonian codes
typically opt either for a (nearly) exact solution of
the Riemann problem following \citet{colella_85}\footnote{The solver of \citet{colella_85} is not exact in the strict sense because
it involves a local linearization of the equation of state.}
or for approximate solvers that
at least take the full wave structure of the Riemann problem
into account such as the HLLC solver \citep{toro_94}. 
On the other hand, the majority of relativistic simulations still resort to the HLLE solver \citep{einfeldt_88} because of the added
complexity of full-wave approximate Riemann solvers for
GR hydrodynamics; exceptions include the \textsc{CoCoNuT} code which
routinely uses the relativistic HLLC solver
of \citet{mignone_05_a},
the \textsc{CoCoNuT} simulations of \citet{cerda_05}
using the Marquina solver \citep{marquina_96},
and the convection simulations with
the \textsc{WhiskeyTHC} code \citep{radice_16},
which uses a Roe-based flux-split scheme.

The use of simpler solvers in GR simulations
is  a concern because one-wave schemes
behave significantly worse than full-wave solvers
in the subsonic regime. 
The problem of excessive acoustic noise from the
discontinuities introduced by the reconstruction is
exacerbated because solvers like HLLE
essentially have these discontinuities decay
 \emph{only} into acoustic waves. 
 This results in stronger numerical diffusion,
 but can also create artificial numerical noise
 because the diffusive terms in the HLLE or
 Rusanov flux can generate spurious pressure
 perturbations from isobaric conditions.
 While higher-order reconstruction can beat down
 the numerical diffusion for smooth flows, a strong degradation
 in accuracy is unavoidable in turbulent flow with structure
 on all scales.
 
Few attempts have as yet been made to quantify the impact 
of the more diffusive one-wave solvers in supernova
simulations. In an idealized setup, the
problem was addressed by 
\citet{radice_15}, who conducted simulations
of stirred isotropic turbulence with solenoidal forcing
with a turbulent Mach number
of $\mathrm{Ma}\sim 0.35$, two different
Riemann solvers (HLLE vs.\ HLLC), and
different reconstruction methods and grid resolutions.
Even when using PPM reconstruction, they still found
substantial differences between the HLLE and HLLC solver
in the spectral properties of the turbulence with the HLLE
runs requiring about $50\%$ more zones to achieve
an equivalent resolution of the turbulent cascade to HLLC.
Although it is not easy to extrapolate from the idealized setup of \citet{radice_15} to core-collapse supernova simulations, one must
clearly expect resolution requirements to depend sensitively on the Riemann solver. This is also illustrated by 
2D supernova simulations
comparing the HLLE and HLLC solver
using the \textsc{CoCoNuT-FMT} code as shown in 
Fig.~\ref{fig:hlle_hllc1}: Starting from the initial
seed perturbations, the HLLC model shows a faster growth
of large-scale SASI shock oscillations
during its linear phase and earlier
emergence of parasitic instabilities
(see Sect.~\ref{sec:sasi} for the physics behind the
SASI) due to the smaller amount of numerical dissipation. 
The evolution
of the shock differs significantly during the first
$100 \, \mathrm{ms}$ of SASI activity, although the
models become similar in terms of shock radius
and shock asphericity later on. However,
even then HLLC run consistently shows a higher entropy
contrast and higher non-radial velocities within
the gain region.

\paragraph{Reconstruction methods.}
Similar concerns (not restricted to
the low-Mach number regime) as for the simpler Riemann solvers can
be raised about the order of the reconstruction scheme.
There is certainly a clear 
divide between second-order piecewise linear reconstruction and higher-order methods like the PPM,
WENO \citep[weighted essentially non-oscillatory ][]{shu_97}, and 
higher-order monotonicity-preserving
\citep[MP][]{suresh_97}
schemes.  
In their simulations of forced subsonic turbulence,
\citet{radice_15} found similar differences between second-order
reconstruction using the monotonized central (MC;
\citealp{van_leer_77}) limiter ---
one of the shaper second-order limiters --- and
runs using PPM or WENO as between the HLLC and HLLE solver.
Again,
the lower accuracy of second-order schemes
is often clearly visible in full supernova simulations,
which is again illustrated in
 Fig.~\ref{fig:hlle_hllc1}
for the same setup as above.
Similar to the
HLLE run, the simulations
using the MC limiter shows a delayed growth
of the SASI and less small-scale structure.

Comparing the more modern higher-order reconstruction
methods is much more difficult. 
For smooth problems like single-mode linear
waves solutions, going beyond the original 4th-order PPM method of \citet{colella_84} to methods of 5th order or higher
can  substantially reduce numerical dissipation;
in the optimal case, the dissipation
decreases with  the grid spacing $\Delta x$
as $\Delta x^{-q}$
for a $q$-th order method
\citep{rembiasz_17}. However, the higher-order scaling of
the numerical dissipation cannot be generalised
to turbulent flow, because the dissipation
of the shortest realizable modes
at the grid scale does not increase as a higher power
of $q$. Based on similarity arguments, one can
work out that the effective Reynolds number
of turbulent flow  increases only as $\mathrm{Re}\sim \Delta x ^{-4/3}$ \citep{mueller_16b} and not
as $\Delta x^{q}$ as one might hope. The reason behind
this limitation is that increasing the order of
reconstruction
does not increase the maximum wavenumber $k_{\max}$ of modes that can be
represented on the grid, it merely limits numerical dissipation
to a narrow band of wavenumbers below $k_{\max}$.

For the moderately
subsonic turbulent flow in core-collapse supernovae
during the accretion phase,  higher-order
reconstruction often does not bring any
tangible improvements for this reason. 
Fig.~\ref{fig:hlle_hllc1} 
again shows this by comparing runs using standard
4th-order PPM and the 6th-order extremum-preserving 
PPM method of \citet{colella_08}. In both cases,
the evolution of the shock is very similar, even
though the phases of the SASI oscillations eventually falls out of sync. It is not
obvious by visual inspection that the higher-order
method allows smaller structures to develop. Only upon  deeper
analysis can small differences between the two methods
be found, for example the model with
extremum-preserving reconstruction
maintains a measurably higher entropy contrast in the
gain region and a slightly higher turbulent kinetic energy in
the gain region.

There are nonetheless situations where it is useful
to adopt extremum-preserving methods of very high order in global simulations of turbulent flow.
First, such methods open up the regime of low Mach numbers
to explicit Godunv-based codes. Using their \textsc{Apsara} code,
\citet{wongwathanarat_16} were able to solve the 
Gresho vortex problem
\citep{gresho_90} with little dissipation down to a
Mach number of $10^{-4}$ with the extremum-preserving
PPM method of \citet{colella_08}, which is 
about two orders of magnitude better than for the
MC limiter \citep{miczek_15}, and
about
one order of magnitude better than for standard
PPM. 

Modern higher-order methods can also
be crucial in certain 
simulations of mixing at convective
boundaries and nucleosynthesis.
In the case of convective
boundary mixing, 
this has been stressed 
and investigated by \citet{woodward_10,woodward_14},
who achieve higher accuracy for the
advection of mass fractions in their
\textsc{PPMstar} code by evolving
moments of the concentration variables
within each cell (which is somewhat
reminiscent of the Discontinuous
Galerkin method). They found that this
Piecewise-Parabolic Boltzmann method
only requires half the resolution
of standard PPM  to achieve the
same accuracy \citep{woodward_10}. 
Higher-order extremum-preserving
methods may also
prove particularly useful for minimizing the
numerical diffusion of mass fractions in models of Rayleigh--Taylor mixing during the supernova explosion phase, but this is yet to be investigated.

\subsubsection{High-Mach number flow}
Some of the considerations for subsonic flow
carry over to the supersonic and transsonic flow encountered
during the supernova explosion phase where mixing instabilities 
also lead to turbulence, but there are also problems specific
to the  supersonic regime.

\paragraph{Sonic points.} It is well known that
the original Roe solver produces spurious
expansion shocks in transsonic rarefaction fans,
which needs to be remedied by some form of
entropy fix \citep{laney_98,toro_09}. While other
full-wave solvers -- like the
exact solver and HLLC -- never fail as spectacularly as
Roe's, they are still prone to mild instabilities
at sonic points. Under adverse conditions, these instabilities
can be amplified and turn into a serious numerical problem.
In this case, it is advisable to switch to a more dissipative
solver such as HLLE in the vicinity of the sonic point.
In supernova simulations, this problem is sometimes
encountered in the neutrino-driven wind that develops
once accretion onto the PNS has ceased.
It can also occur prior to shock revival in the infall
region and severely affect the infall downstream
of the instability, especially when nuclear burning is included.
In this case, the problem can be easily overlooked or
misidentified because it usually manifests itself
as an unusually strong stationary burning front, which may
seem perfectly physical at first glance.

\begin{figure}
    \centering
    \includegraphics[width=0.5 \linewidth]{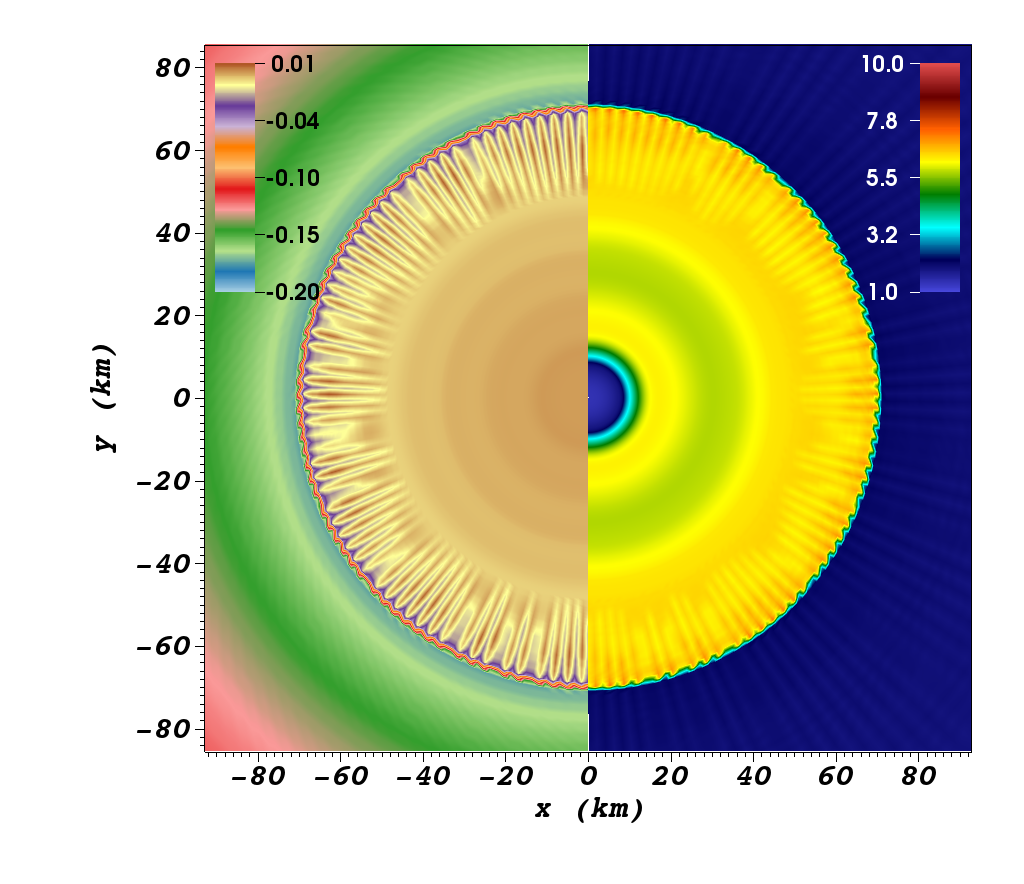}
    \caption{Odd-even decoupling in a 
     2D core-collapse supernova simulation  of a $20\,M_\odot$ star with 
     the \textsc{CoCoNuT} that uses the HLLC solver everywhere
     instead of switching to HLLE at shocks. The left and
     right panels show the radial velocity in units of the speed of light
    and the entropy $s$ in units of $k_\mathrm{b}/\mathrm{nucleon}$ 
    about $10 \, \mathrm{ms}$ after bounce.
    The characteristic radial streaks from odd-even decoupling
    are clearly visible behind the shock.}
    \label{fig:odd_even}
\end{figure}

\paragraph{Odd-even decoupling and the carbuncle phenomenon.}
Full-wave solvers like the exact solver
and HLLC are subject to an instability
at shock fronts \citep{quirk_94,liou_00b}: For grid-aligned shocks,
insufficient dissipation in the direction
parallel to the shock can cause odd-even decoupling
in the solution, which manifests
itself in artificial stripe-like patterns downstream of the 
shock. When the shock is only locally tangential
to a grid line, this instability can give rise
to protrusions, which is known as the carbuncle
phenomenon. In supernova simulations, odd-even decoupling
was first recognized  as a problem
by \citet{kifonidis_00}, and since
then the majority of supernova codes
(e.g., \textsc{Prometheus}, \textsc{Flash}, \textsc{CoCoNuT},
\textsc{Fornax}) have
opted to handle this problem by adaptively switching
to the more dissipative HLLE solver at strong shocks following
the suggestion of \citet{quirk_94}.
The \textsc{Chimera} code \citep{bruenn_18} adopts the alternative
approach of a local oscillation filter
\citep{sutherland_03}, which has the advantage of not 
degrading the resolution in the direction perpendicular to
the shock, but has the drawback of allowing
the instability to grow to a minute level (which
may be undetectable in practice) before smoothing
is applied to the solution. 
The carbuncle phenomenon can also occur in 
Richtmyer-type artificial viscosity schemes and be
cured by modifying the artificial viscosity
\citep{iwakami_08}.
The carbuncle
instability remains a subject of active research
in computational fluid dynamics, and a number
of papers \citep[e.g.,][]{nishikawa_08,huang_11,rodinov_17,simon_19}
have attempted to construct Riemann solvers
or artificial viscosity schemes that avoid
the instability without sacrificing accuracy away from
shocks, and may eventually prove useful for supernova simulations.

\paragraph{Kinetically-dominated flow.}
In HRSC codes that solve the total energy equation, one
obtains the mass-specific internal energy $\varepsilon$ by subtracting
the kinetic energy $v^2/2$ from the total energy $e$.
In high Mach-number flow, one has
$e\gg \varepsilon$ and $v^2/2 \gg \varepsilon$, and hence subtracting
these two large terms can introduce large errors
in  the internal energy density and the pressure and
sometimes leads to severe stability problems. A similar
problem can occur in magnetically-dominated regions
in MHD. Sometimes the resulting stability problems can
be remedied by evolving the internal
energy equation
\begin{equation}
\frac{\pd \rho \varepsilon}{\pd t}
+\nabla \cdot \left(\rho \varepsilon \mathbf{v}
+ P\mathbf{v}\right)
-\mathbf{v}\cdot \nabla P
=
-\rho \mathbf{v} \cdot \nabla \Phi
+ Q_\mathrm{e}
+ \mathbf{Q}_\mathrm{m} \cdot \mathbf{v},
\end{equation}
instead of Eq.~(\ref{eq:hydro3}) in
regions of high Mach number or low plasma-$\beta$.
However, in doing so one sacrifices strict
energy conservation, and hence one should
apply this recipe as
parsimoniously as possible.

\subsection{Treatment of gravity}
Convective burning and core-collapse
supernovae introduce specific challenges in the
treatment of gravity. In the subsonic flow regimes,
one needs to be wary of introducing undue artifical
perturbations from hydrostatic equilibrium and
take care to avoid secular conservation errors.
Moreover, in the core-collapse supernova problem,
general relativistic effects become important
in the vicinity of the PNS. 

\subsubsection{Hydrostatic balance and
conservation properties}
For nearly hydrostatic flow, one has 
$\nabla P \approx - \rho \nabla \Phi$, but this
near cancellation is not automatically reflected
in the numerical solution when
using a Godunov-based scheme. Instead, the stationary
numerical solution may be one with
\emph{non-zero} advection terms that are exactly 
(but incorrectly) balanced by the gravitational
source term \citep{greenberg_96,leveque_98}.
Schemes that avoid this pathology are
called \emph{well-balanced}.  The proper
cancellation between the pressure gradient
and the gravitational source term is
particularly delicate if those two terms
are treated in operator-split steps.
Different methods have been proposed to incoroprate
well-balancing into Godunov-based schemes.
One approach is to use piecewise hydrostatic
reconstruction \citep[e.g.,][]{kastaun_06,kaeppeli_14}.
A related technique suggested by
\citet{leveque_98} introduces discrete jumps in
the middle of cells to obtain modified interface
states for the Riemann problem and absorb the
source terms altogether.

In practice, these special techniques are not
used widely in the field for two reasons.
First, it is not trivial to general these schemes
to achieve higher-order accuracy.
Second, one already obtains a very well-balanced
scheme by combing higher-order reconstruction,
an accurate Riemann solver, and unsplit time
integration. For split schemes, one can ensure
a quite accurate cancellation of the pressure
gradient and the source term by including
a characteristic state correction as described
by \citet{colella_84} for the original PPM method.

Nevertheless, the  cancellation of the pressure gradient and
the gravitational source term in hydrostatic
equilibrium is usually not perfect and typically
leads to minute odd-even noise in the velocity field
that is almost undetectable by eye. Computing
the gravitational source term
$\rho \mathbf{v}\cdot \nabla \Phi$
in the energy equation using such a noisy
velocity field $\mathbf{v}$ can lead to an appreciable secular drift of the total energy. For example,
spurious energy generation can stop  proto-neutron
star cooling on simulation time scales longer than a second \citep{mueller_phd}. This problem can be
circumvented by discretizing the energy equation
starting from the form \citep{mueller_10}
\begin{equation}
\label{eq:energy_modified}
\frac{\pd \rho (\varepsilon + v^2/2+\Phi)}{\pd t}
 \nabla \cdot \left[\rho\mathbf{v} (\varepsilon + v^2/2+\Phi)
+P\mathbf{v}\right]
= \rho \frac{\pd \Phi}{\pd t}.
\end{equation}
This guarantees exact total energy conservation
if the time derivative of the
gravitational potential is zero. Under certain
conditions, exact total energy conservation can
be achieved for a time-dependent self-gravitating
configuration as well, and the method
can also be generalized to the relativistic case.

In principle, one can also implement the 
gravitational source term (in the Newtonian
approximation) in the momentum
equation in a conservative form by 
writing $\rho \mathbf{g}$ as the divergence
of a gravitational stress tensor \citep{shu}.
Such a scheme has been implemented
by \citet{livne_04} in the \textsc{Vulcan} code.
However, this procedure involves a more delicate
modification of the equations than in case of
the energy source term, because it essentially
amounts to replacing $\rho$ by the finite-difference
representation of the Laplacian
$(4\pi G)^{-1} \Delta \Phi$ in the momentum source term.
Unless the solution for the gravitational potential
is extremely accurate, large acceleration errors
may thus arise. Moreover, this approach does
not work for effective relativistic potentials
(see Sect.~\ref{sec:gr}). For these reasons,
the conservative form of the gravitational source term
has not been used in practice in other codes.
Even though the issue of momentum conservation is
of relevance in the context of neutron star kicks,
conservation errors do not seem to affect supernova
simulation results qualitatively in practice, and
post-processing techniques can be used to
infer neutron star velocities from simulations with good
accuracy \citep{scheck_06}.

\subsubsection{Treatment of general relativity}
\label{sec:gr}
In core-collapse supernova simulations,
the relativistic compaction of the proto-neutron
star reaches $GM/Rc^2=0.1$--$0.2$ even
for a normal PNS mass $M$ and a
somewhat extended radius $R$ of the warm PNS.
Infall velocities of $0.15$--$0.3c$
are encountered. Hence general relativistic
(GR) and special relativistic effects are
no longer negligible, though the latter is more critical
for the treatment of the neutrino transport
than for the hydrodynamics. For very massive neutron stars,
cases of black hole formation, or jet-driven explosions,
relativistic effects can be more pronounced.

A variety of approaches is used in supernova modelling
to deal with relativistic effects. Purely Newtonian models
have now largely been superseded. Using Newtonian gravity
results in unphysically large PNS radii, and, as a consequence,
lower neutrino luminosities and mean energies and
worse heating heating conditions than in the relativistic
case, even though the stalled accretion shock radius is larger than in 
GR before explosion \citep{mueller_12a,kuroda_12,lentz_12a,oconnor_18a}.
As an economical alternative, one can retain the
framework of Newtonian hydrodynamics but incoroprate
relativistic corrections in the gravitational potential
based on the TOV equation \citep{rampp_02}. This approach
was subsequently refined by \citet{marek_05,mueller_08}
to account for some inconsistencies between
the use of Newtonian hydrodynamics and a potential
based on a relativstic stellar structure equation, but
full consistency can never be achieved in the
pseudo-Newtonian approach. In the multi-D case,
the relativistic potential replaces the monopole
of the Newtonian potential, while higher multipoles
are left unchanged.
From a purist point of view,
this pseudo-Newtonian approach is delicate because one sacrifices
global conservation laws for energy and momentum
(which would still hold in a more complicated form
in an asymptotically flat space in full GR).
In practice, this is less critical; in
PNS cooling simulations by \citet{huedepohl_10}
the total emitted neutrino energy was found
to agree with the neutron star binding energy (computed
from the correct TOV solution) to within $1\%$ for
the modified TOV potential (Case~A) of \citet{marek_05}.

If the framework of Newtonian hydrodyanmics is abandoned,
one may still opt for an approximate method to solve
for the space-time metric as in the
\textsc{CoCoNuT} code \citep{dimmelmeier_04,mueller_10,mueller_15a}. Elliptic
formulations such as CFC
(conformal flatness conditions, \citealp{isenberg_78})
and xCFC (a modification of CFC for
improved numerical stability; \citealp{cordero_09})
can be cheaper and more stable than free-evolution
schemes based on the 3+1 decomposition
\citep[for reviews of these techniques, see][]{baumgarte_10,lehner_14}
and maximally constrained schemes \citep{bonazzola_04,cordero_12}.
However, full GR supernova simulations
without the CFC approximation and with multi-group
transport have also become possible recently
\citep{roberts_16,ott_18,kuroda_16,kuroda_18}. Although CFC remains an approximation, it is exact in spherical symmetry, and
comparisons with free-evolution schemes have shown 
excellent agreement in the context of rotational
collapse have shown excellent agreement
even for rapidly spinning progenitors \citep{ott_07b}.

Comparisons of pseudo-Newtonian and GR simulations
have demonstrated that using an effective potential is at least sufficient
to reproduce the PNS contraction, the shock evolution, 
and the neutrino emission in GR very well \citep{liebendoerfer_05,mueller_10,mueller_12a}.
While \citet{mueller_12a} still found better heating
conditions in the GR case than with an effective
potential in their 2D models, this comparison
was not fully controlled in the sense
that two different hydro solvers were used,
and the effect was related
to subtle differences in the PNS convection zone,
which may well be related to factors other
than the GR treatment (cf.\ Sect.~\ref{sec:pns_convection}).
Further code comparisons are desirable to resolve this.
The pseudo-Newtonian approach, does, however,
systematically distort the eigenfrequencies
of neutron star oscillation modes
\citep{mueller_08}. In particular, the
frequency of the dominant f-/g-mode seen
in the gravitational wave spectrum is shifted up
by 15--20\% compared to the correct
relativistic value \citep{mueller_13}.

\subsubsection{Poisson solvers}
In the Newtonian approximation, the gravitational
field is obtained by solving the Poisson equation
\begin{equation}
    \Delta \Phi=4\pi G\rho.
\end{equation}
In constrained formulations of the Einstein equations
like (x)CFC, one encounters non-linear Poisson
equations.

In simulations of supernovae and the
late convective burning stages, the
density field usually only deviates modestly from spherical
symmetry and is not exceedingly clumpy
(except in the case of mixing instabilities
in the envelope during the explosion phase when
self-gravity is less important to begin with). For this reason,
the usual method of choice for solving the Poisson equation
(even in Cartesian geometry) is to use the multipole expansion of the Green's function \citep{mueller_95}. Typically,
no more than 10--20 multipoles
are needed for good accuracy,
and very often only
the monopole component is retained.
Other methods have been used occasionally,
though, such as pseudospectral methods \citep{dimmelmeier_04}
and finite-difference solvers \citep[e.g.,][]{burrows_07},
and the FFT \citep{hockney_65,eastwood_79} is a viable option for Cartesian simulations.

Although it yields accurate results at fairly cheap
cost, some subtle issues can arise with the multipole expansion.
When projecting the source density onto spherical harmonics
$Y_{\ell m}$
to obtain multipole components $\hat{\rho}_{\ell m}=\int Y_{\ell m}^* \rho \,\ud \Omega$, a naive step-function integration can lead to
a self-potential error \citep{couch_13b} and destroy convergence
with increasing mulitpole number $N_\ell$. This can
be avoided either by performing the
integrals over spherical harmonics $Y_{\ell m}^*$
analytically \citep{mueller_94}, or by using a staggered grid for the
potential \citep{couch_13b}. The accuracy of the solution
can also be degraded if the central mass concentration
moves away from the center of the grid, which can be 
cured by off-centering the multipole expansion
\citep{couch_13b}. Problems with off-centred
or clumpy mass distributions can be cured completely
if an exact solver is used. In Cartesian geometry,
this can be accomplished econmically using the FFT,
and an exact solver for spherical polar grid 
using a discrete eigenfunction expansion has
recently been developed as well \citep{mueller_19c}.
On spherical multi-patch grids, the efficient parallelization
and computation of integration weights requires some
thought and has been addressed by \citep{almanstoetter_18,wongwathanarat_19}.

\subsection{Reactive flow}
Nuclear burning is the principal driver of
the flow for core and shell convection in the
late, neutrino-cooled evolutionary stages
of supernova progenitors. In core-collapse
supernovae nuclear dissociation and recombination
play a critical role for the dynamics and energetics,
and one of the key observables, the mass
of ${}^{56}\mathrm{Ni}$, is determined by nuclear
burning.

Approaches to nuclear transmutations differ
widely between simulation codes,
and range from the assumption of nuclear statistical
equilbrium (NSE) everywhere in some core-collapse supernova
models to rather large reaction networks.
Naturally,the appropriate level of sophistication
 depends on the regime and the observables
of interest. The theory of nuclear reaction
networks is too vast to cover in detail here, and
we can only touch a few salient points related
to their integration into hydrodynamics codes.
For a more extensive coverage, we refer to textbooks
and reviews on the subject
\citep{clayton_68,arnett_96,mueller_97_b,timmes_99,hix_06,iliadis}.

\paragraph{Burning regimes.}
As stellar evolution proceeds towards collapse,
the ratio of the nuclear time scale to both
the sound crossing time scale and convective time
scale decreases, and the nuclear reaction flow
involves an increasing number of reactions.
The burning of  C, Ne, and to some extent of O  is dominated by an overseeable number of main reaction channels,
and the relevant reaction rates are slow compared
to the relevant hydrodynamical time scales.
During oxygen burning,
quasi-equilibrium clusters begin to appear
and eventually merge into  one or two big clusters
during Si burning \citep{bodansky_68,woosley_73} that
are linked by slow ``bottleneck'' reactions.
For sufficiently high temperatures,
NSE is established and the composition
only depends on density $\rho$, temperature $T$, and
the electron fraction $Y_\mathrm{e}$ and
is given the Saha equation. At higher
densities during core-collapse, the assumption
of non-interacting nuclei break down,
and a high-density equation of state is
required  \citep[see][for recent reviews]{lattimer_13,oertel_17,fischer_17};
this regime is not of concern here because
the flow can be treated as non-reactive.

\paragraph{Simple approaches.}
In core-collapse supernova simulations,
one sometimes simply assumes NSE everywhere, which
amounts to an implicit release of energy at the start of
a simulation. Although the Si and O shell
will still collapse in the wake of the Fe core,
this is somewhat problematic, especially for
long-time simulations where the effect on the
infall is bound to be more pronounced. For mitigating
potential artifacts from the inconsistency
of the composition and equation of state with
the underlying stellar evolution model, it can 
be useful to initialise supernova simulations
using the pressure rather than the temperature
of progenitor model.

A considerably
better and very cheap approach, known as ``flashing'',
is to use a few key $\alpha$-elements
and non-symmetric iron group nuclei in addition
to protons, neutrons, and $\alpha$-particles and
burn them instantly into their
reaction products and eventually into NSE upon reaching certain
threshold temperatures \citep{rampp_02}. Such an approach
can capture the energetics of explosive burning in the shock
and the freeze-out from NSE in neutrino-driven outflows
reasonably well, but only gives indicative results
on the composition of the ejected matter. The
choice of the proper NSE threshold temperature $T_\mathrm{NSE}\gtrsim 5\mathrm{GK}$ can be particularly
delicate, since this depends on the entropy
and expansion time scale of the outflow and can
critically affect the degree of recombination in
the neutrino-heated ejecta.

For simulations of convective burning, a
smooth behavior of the nuclear source terms is
required, but for C, Ne, or O burning, one can
still resort to simple fit formulae
and only track the composition of the main
fuel and ash if the goal is merely to understand
the dynamics of the flow
\citep[e.g.,][]{kuhlen_03,jones_17,cristini_16,cristini_17,andrassy_18}.
Often these source terms (or also rates in calculations with veritable network) are rescaled if
low convective Mach numbers make
simulations in the physical regime unfeasible. This can be useful for exploring 
the parameter dependence of flow phenomena, but
caution is required because safe extrapolation
to the physical regime may also require a rescaling
of other terms (e.g., thermal diffusivity, neutrino losses).

\paragraph{Reaction networks.}
In 2D, \citet{bazan_97} already conducted
simulations of convective burning with 123 species,
but the use of large networks in 3D simulations
is still prohibitively expensive. Modern
3D simulations of convective burning 
with the \textsc{Prompi}
\citep[e.g.,][]{meakin_06,mocak_18},
\textsc{Prometheus} \citep{mueller_16c,yadav_19},
 \textsc{Flash} \citep{couch_15} 
 and \textsc{3DnSNe} \citep{yoshida_19}
 codes have therefore
only use networks of 19--25 species consisting
of $\alpha$-elements, light particles, and
at most a few extra iron-group elements.
In multi-D supernova simulations with neutrino transport
the use of such networks is feasible
\citep{groote_phd,bruenn_13,bruenn_16,wongwathanarat_17},
though they have not been used widely yet. It is critical
that such reduced reaction networks appropriately
account for side chains and the effective
reaction flow between light particles
\citep{weaver_78,timmes_00}. Their use
is problematic for Si burning which
requires networks of more than a hundred
species to accurately capture the quasi-equilibrium
clusters and the effects of deleptonization 
\citep{weaver_78}
and for freeze-out from NSE with considerable
neutron excess. Larger networks or special
methods for quasi-equilibrium \citep{weaver_78,hix_07,guidry_13}
will be required for reliable multi-D simulations
of convective Si burning.

\paragraph{Coupling to the hydrodynamics.}
Some numerical issues arise when a nuclear network
is coupled to a Eulerian hydrodynamics
solver, or even
if the composition is just
tracked as a passive tracer.
One such problem concerns the conservation
of partial masses, which is guaranteed
analytically by a
conservation equation
(\ref{eq:xi}) for each species $i$,
\begin{equation}
\label{eq:xi2}
    \frac{\pd \rho X_i}{\pd t}
    +\nabla  \cdot (\rho X_i \mathbf{v})=
    \dot{X}_{i,\mathrm{burn}}.
\end{equation}
This equation can be solved using standard, higher-order
finite-volume techniques. However, the solution
also has to obey the constraint
\begin{equation}
\sum X_i=1,
\end{equation}
which is not fulfilled automatically by
the numerical solution, unless flat reconstruction
for the mass fractions is employed. One
could enforce this constraint by rescaling the mass fractions to sum up to unity, but this would violate
the conservation of partial masses.
\citet{plewa_99} developed the Consistent Multi-fluid
Advection (CMA) method as
the standard treatment to ensure both minimal
numerical diffusion of mass fractions and
enforce conservation of partial masses.
This method
involves a rescaling and coupling of the interpolated
\emph{interface} values of the various mass fractions.
 \citet{plewa_99} demonstrated that simple methods
for the advection of mass fractions can easily
result in wrong yields by a factor of a few
for some isotopes in supernova explosions.

Another class of problems is related to advection errors
and numerical diffusion, especially at contact
discontinuities and shocks, which can lead to
artificial detonations or an incorrect propagation
of physical detonations 
\citep{colella_86,fryxell_89,mueller_97_b}.
To ensure that detonations propagate
at the correct physical velocity, nuclear burning should be 
switched off in shocked zones \citep{mueller_97_b}.\footnote{Note also the use of front-tracking methods for
unresolvable burning fronts \citep{reinecke_02,leung_19a}, which
are commonly used for modelling Type~Ia supernovae and the O deflagration in electron-capture supernova progenitors.}
Due to the extreme temperature dependence
of nuclear reaction rates, similar problems can arise 
away from discontinuities due to
advection errors that produce a small level of noise in the 
temperature. Artificial detonations can easily develop
in highly degenerate regions and around sonic points.
Eliminating such artifacts may require
appropriate switches for pathological zones or
very high spatial resolution
\citep[e.g.,][]{kitaura_06}.

\section{Late-stage convective burning in supernova progenitors}
\label{sec:prog3d}
In the
Introduction, we already outlined the motivation
for multi-D simulations of supernova progenitors in broad 
terms. On the most basic level, multi-D models are
needed to properly intialize supernova simulations
and provide physically correct seed perturbations
for the instabilities that develop after collapse
and in the explosion phase. This does not, in fact,
presuppose that 1D stellar evolution models incorrectly predict
the overall spherical structure of pre-supernova progenitors;
in principle such  an initialization might involve nothing but
adding some degrees of freedom to 1D stellar
evolution models without any noticeable change of the
spherically averaged stratification. 
Historically, however, simulations of
late-stage convection have focused on deviations
of the multi-D flow from the
predictions of traditional  mixing-length theory
(MLT; \citealp{biermann_32,boehm_58,cox}) and not evolved
progenitor models up to core collapse, whereas
the initialization problem has only been tackled
recently by \citet{couch_15,mueller_16c,mueller_19a,yadav_19}. In this section, we therefore address
the interior flow in convective regions and boundary
effects first before specifically discussing
multi-D pre-supernova models.

\subsection{Interior flow}
\label{sec:interior_flow}
Let us first consider the flow within convectively unstable
regions. In MLT as implemented
in modern stellar evolution codes
such as \textsc{Kepler} \citep{weaver_78,heger_10}
and \textsc{Mesa} \citep{paxton_11}, the convective velocity
$v_\mathrm{conv}$ in
such regions is tied to the superadiabaticity
of the density gradient as encoded by the Brunt-V\"ais\"al\"a
$\omega_\mathrm{BV}$
frequency and the local pressure scale height $\Lambda$,
\begin{equation}
    \label{eq:vmlt}
    v_\mathrm{conv}=\alpha \sqrt{\Lambda  \delta \rho/\rho\, g}=
    \alpha \Lambda \omega_\mathrm{BV},
\end{equation}
where $\alpha$ is a tuneable parameter of order
unity, and the  MLT density contrast
$\delta \rho$ is obtained from the 
the difference between the actual
and density gradient $\pd\rho/\pd r$
and the adiabatic density gradient
$(\pd\rho/\pd P)_s (\pd P/\pd r)$,
\begin{equation}
\label{eq:drho_mlt}
    \delta \rho
    =
    \frac{\Lambda \rho \omega_\mathrm{BV}^2}{g}
    =
    \Lambda\left[\frac{\pd \rho }{\pd r}-
    \left(\frac{\pd \rho}{\pd P}\right)_s
    \frac{\pd P}{\pd r}\right]
    =
    \Lambda\left(\frac{\pd \rho }{\pd r}-\frac{1}{c_\mathrm{s}^2}
    \frac{\pd P}{\pd r}\right).
\end{equation}
Note that stellar evolution textbooks usually express
the convective velocity and density contrast in terms
of the difference between the actual and adiabatic
temperature gradient \citep{clayton_68,cox,kippenhahn},
but Eqs.~(\ref{eq:vmlt}) and
(\ref{eq:drho_mlt}) are  fully equivalent formulations
that often prove less cumbersome.

Using Eq.~(\ref{eq:vmlt})
for the convective velocity,
Eq.~(\ref{eq:drho_mlt}) for the MLT density
contrast, and the temperature contrast
$\delta T=(\pd T/\pd \ln \rho)_P (\delta \rho/\rho)$,
we then obtain the convective energy flux $F_\mathrm{conv}$
\citep{kippenhahn,cox},
\begin{eqnarray}
\label{eq:mlt_e}
    F_\mathrm{conv}&=&
      \alpha_e
  \rho c_P \, \delta T \, v_\mathrm{conv}
\\
\nonumber
  &=&
  -\alpha \alpha_e
  \rho c_P \left(\frac{\pd T}{ \pd \ln \rho}\right)_P
  \frac{\Lambda^2 \omega_\mathrm{BV}^3}{g},
\end{eqnarray}
where $c_P$ is the specific heat at constant pressure,
and $\alpha_e$ is another tunable non-dimensional parameter.
Similarly, by estimating the composition contrast
$\delta X_i$ using the local gradient as
$\delta X_i=\alpha_X \Lambda \, \pd X_i/\pd r$, we obtain
the partial mass flux for species $i$
\begin{equation}
\label{eq:mlt_x}
    F_{X_i}=
     \rho v_\mathrm{conv} \delta X_i
    =
    \alpha_X
    \Lambda \rho v_\mathrm{conv} \frac{\pd X_i}{\pd r},
\end{equation}
where $\alpha_X$ is again a dimensionless parameter. When comparing 1D stellar evolution models to each other or to  multi-D simulations, one
must bear in mind that slightly different normalization
conventions for Eqs.~(\ref{eq:mlt_e})
and (\ref{eq:mlt_x}) are in use. Regardless of these
ambiguities, these coefficients are  of order
unity, for example the \textsc{Kepler} code
uses $\alpha \alpha_e = 1/2$ and $\alpha \alpha_X = 1/6$
\citep{weaver_78},
which can be conveniently interpreted as
$\alpha= 1$, $\alpha_e=1/2$ and $\alpha_X = 1/6$
\citep{mueller_16c},

In order to connect more easily to multi-D simulations, it is useful restate the
assumptions and consequences of MLT (without radiative diffusion) in
a slightly different language.
Equation~(\ref{eq:vmlt}) can also be written as
\begin{equation}
\label{eq:balance1}
        \frac{v_\mathrm{conv}^3}{\Lambda}=
        \alpha^2 \, \delta \rho/\rho\,g 
        v_\mathrm{conv},
\end{equation}
which we can interpret as a balance between
the rate of buoyant energy generation  ($\alpha^2\, \delta \rho/\rho\, 
        v_\mathrm{conv}$) and
turbulent dissipation ($\epsilon\sim v_\mathrm{conv}^3/\Lambda$).
Furthermore the work done by bouyancy must ultimately be supplied
by nuclear burning. Using thermodynamic relations, we find that the potential energy
$\Lambda\, \delta \rho/\rho\, g$ liberated by bubbles
rising or sinking by one mixing length is of the order
of the enthalpy contrast $\delta h$ of the bubbles,
which roughly equals the integral of the nuclear energy generation
rate $\dot{q}_\mathrm{nuc}$ over one turnover time $\tau=\Lambda/v_\mathrm{conv}$,
\begin{equation}
\label{eq:balance2}
\Lambda\, \delta \rho/\rho\, g \sim \delta h
\sim \dot{q}_\mathrm{nuc} \Lambda/v_\mathrm{conv}.
\end{equation}
Together, Eqs.~(\ref{eq:balance1}) and (\ref{eq:balance2}) lead
to a scaling law
$v_\mathrm{conv} \sim (\dot{q}_\mathrm{nuc} \Lambda)^{1/3}$
for the typical value of $v_\mathrm{conv}$
in a convective shell. 

\begin{figure}
    \centering
    \includegraphics[width=0.7\textwidth]{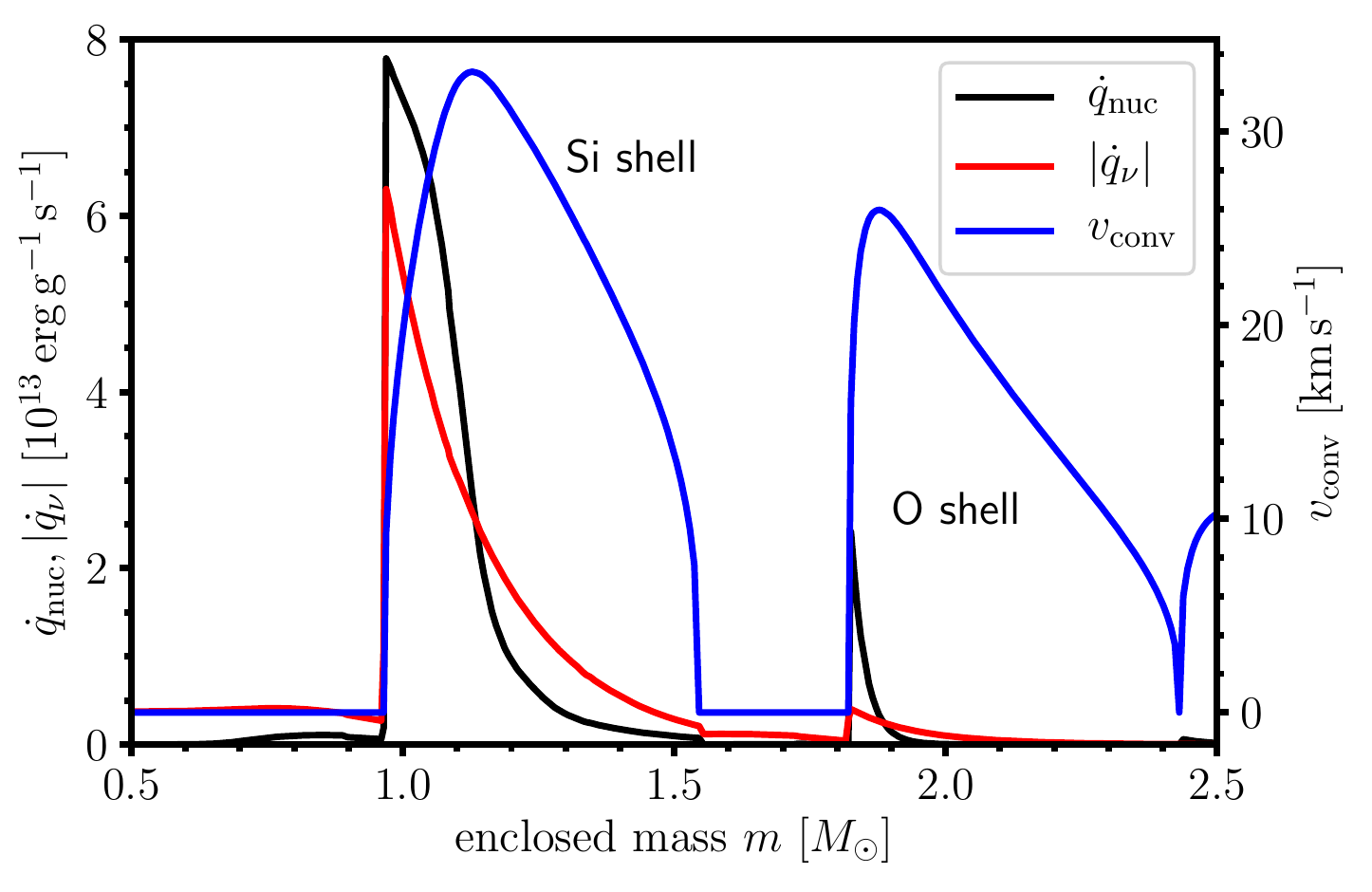}
    \caption{Energy generation rate
    $\dot{q}_\mathrm{nuc}$ (black),
    neutrino cooling rate 
    $|\dot{q}_\nu|$ (red),
    and convective velocity $v_\mathrm{conv}$ (blue)
    in an $18.88\,M_\odot$
    progenitor (1D model, discussed
    in \citealt{yadav_19}
    in
    the innermost shells about $1\, \mathrm{hr}$
    before collapse. At this stage,
    balanced power still obtains. Nuclear
    energy generation dominates at the bottom
    of the shells, while neutrino cooling dominates
    in the outer layers. The integrated 
    energy generation and cooling rate for
    the entire shell, which are given by
    the areas under the black and red curve, nearly balance each other.}
    \label{fig:balanced_power}
\end{figure}

In nature,  balance between nuclear energy generation, buoyant energy generation,
and turbulent dissipation is usually established over a few turnover times.
On longer time scales, active burning shells also adjust by
expansion or contraction until
 the total nuclear energy generation rate and neutrino cooling rate
 balance each other \citep{woosley_72}, with the 
nuclear burning dominating in the inner
 region and neutrino cooling dominating in the outer region of the shell (Fig.~\ref{fig:balanced_power}).
 
Because of the extremely strong temperature sensitivity of the
burning rates, this state of balanced power is difficult to maintain
when setting up multi-D simulations and will only be reestablished
over a long, thermal time scale.\footnote{Note that this
thermal time scale is more difficult to define than during early
burning stages where radiative diffusion is important.}
In fact, the problem of thermal adjustment has not yet been
rigorously analyzed for any multi-D model yet, and insufficient simulation
time for thermal adjustment is a concern that needs to be addressed
in future. However, the problem of thermal adjustment is mitigated
during the latest phases of shell convection prior to collapse:
As the core and the surrounding shells contract, the
nuclear burning rates accelerate to a point where neutrino cooling
and shell expansion by $P\, \ud V$ work can no longer re-establish
thermal balance on the contraction time scale of the core,
and the state of balanced power is physically broken.

\paragraph{Two-dimensional simulations of
convective burning.}
The first attempts to simulate late-stage convection
in massive stars by \citet{arnett_94,bazan_94,bazan_98}
targeted oxygen burning in a $20\,M_\odot$ star
in a 2D shellular domain 
with the \textsc{Prometheus} code
using a small, 12-species reaction network
and neutrino cooling by the pair process. Starting
from a simulation on a small wedge of
$18^\circ$ in  \citet{arnett_94},
\citet{bazan_94,bazan_98} subsequently considered
broader wedges in cylindrical symmetry 
of up to $135^\circ$ 
 with a resolution of up to $460 \times 128$
grid cells,
as well as  cases with meridional
symmetry on a 2D grid $(r,\varphi)$ in radius and longitude.
The simulations were invariably limited to 
short periods (up to $400 \, \mathrm{s}$ in \citealt{bazan_98}) and only a 
few convective turnover times.
One simulation \citep{bazan_97} also tackled
Si burning in 2D with a large network of 123 nuclei.
These first-generation 2D models invariably found
violent convective motions with
Mach numbers of 0.1--0.2 and velocities about
an order of magnitude above the MLT predictions,
which cannot be accounted for by the aforementioned 
ambiguities in the definition of the dimensionless 
coefficients. The convective structures invariably
tended to grow to the largest angular scale allowed
by the chosen wedge geometry, and large density
perturbations were found at the convective boundaries.
\citet{bazan_98} also stressed the high temporal
variability of the convective flow, going so far as
to question whether a steady state is ever established
before collapse. Longer simulations of
the same $20\,M_\odot$ model over $1200 \, \mathrm{ms}$
 by the same group using the
\textsc{Vulcan} code of \citep{livne_93} showed
the emergence of a steady state, albeit quite
different from the 1D stellar evolution model
due to convective boundary mixing \citep{asida_00},

To a large extent, the pronounced differences between these 
first-generation simulations and MLT predictions stem
from the assumption of 2D flow. In 2D turbulence, the energy
cascade is artificially inverted and goes from
small to large scales \citep{kraichnan_67}. As a result,
the flow tends to organise itself into large vortices,
and dissipation occurs primarily in boundary
layers \citep{falkovich_17,clercx_17}. 

\paragraph{Three-dimensional simulations of
convective burning.}
Consequently, 3D simulations of convective 
burning obtained considerably smaller convective
velocities. The first 3D, full $4\pi$ solid angle
models of O shell convection 
(along
with models of core hydrogen burning)
were
presented by \citet{kuhlen_03}  for a $25\,M_\odot$
star with and without rotation. Their simulations
used the anelastic pseudospectral code of 
\citet{glatzmaier_84} to follow convection for
about 90 turnovers in the non-rotating case,
and approximated the burning and neutrino cooling
rates by power-law fits.  Different
from the earlier 2D models, they found convective
velocities in good agreement with the 1D MLT
prediction in the underlying stellar evolution model,
but the still observed the emergence of large-scale
flow patterns.

The use of simplified burning and neutrino loss rates,
the anelastic approximation, and an explicit turbulent
diffusivity in \citet{kuhlen_03} still posed  a concern,
which was subsequently addressed by a series of
2D and 3D simulations 
of O and C burning \citep{meakin_06,meakin_07,meakin_07_b,arnett_09}
in wedge-shaped domains
using the compressible \textsc{Prompi} code and a larger reaction network (25 species) than in 
the first generation of 2D models.
These simulations confirmed the significantly less
violent nature of 3D convection compared to 2D
\citep{meakin_06,meakin_07}, and 
established good agreement between elastic and anelastic
simulations
on the convective
velocities and fluctuations of thermodynamic
quantities 
in the interior of convective zones,
though anelastic codes cannot model fluctuations at convective
boundaries very well \citep{meakin_07_b}. They
also found balance between buoyant driving
and turbulent dissipation (which is essentially
a restatement of the basic assumption of
MLT) and observed rough equipartiton between
the radial and non-radial contributions
to the turbulent kinetic energy
\citep{arnett_09}. Their models
still revealed differences from MLT in detail,
such as different correlation lengths for
velocity and temperature and a non-vanishing
kinetic energy flux \citep{meakin_07}. Moreover,
\citet{meakin_10} suggested that the implicit
identification of the pressure scale height with
the dissipation length in MLT might lead to an
underestimation of the convective velocities.
More recent work by the same group has stressed
the time variability of the convective flow
\citep{arnett_11,arnett_11b} and criticized the
MLT assumption of quasi-stationary convective velocities.
Specifically \citet{arnett_11b} pointed to
strong fluctuations in the turbulent kinetic energy
in the 3D oxygen shell burning simulation in 
a $23\,M_\odot$ star by  \citet{meakin_07}, which they
attempted to motivate by recourse to the Lorenz model for
convection in the Boussinesq approximation. The connection
between the simulations of convective burning and the 
Lorenz model for a viscous-conductive convection problem
remains rather opaque, however.

More recent work on 3D convection by other groups has vindicated
rather than undermined MLT as an approximation for the interior
of convective zones. \citet{mueller_16c} conducted
 a $4\pi$-simulation of O burning in an $18\,M_\odot$ star
up to the point of collapse and found that convection reaches
a quasi-stationary state after a few turnovers with only
small fluctuations in the turbulent kinetic energy.
In line with MLT and as in \citet{arnett_09}, 
the average convective velocity is 
well described by a balance of turbulent dissipation
and buoyant driving in their model,  and is in turn
related to the average nuclear energy generation
rate $\dot{q}_\mathrm{nuc}$ as
\begin{equation}
\label{eq:vconv}
\frac{v_\mathrm{conv}^3}{\Lambda}
\approx 0.7 \dot{q}_\mathrm{nuc},
\end{equation}
and even the profiles of the radial component of the
turbulent velocity perturbation are in good agreement
with the corresponding 1D stellar evolution model.
A similar scaling was reported by \citet{jones_17}
based on idealized high-resolution simulations of O burning
with a simple EoS and parameterized nuclear source terms
and by \citet{cristini_17} based on simulations of
C burning in planar 3D geometry, also with a parameterized
(and artificially boosted) nuclear source term. 
\citet{jones_17}  verified this
scaling over a wider range of
convective luminosities and Mach numbers
by applying different boost factors to the
nuclear generation rate.

\begin{figure}
    \centering
    \includegraphics[width=0.48\linewidth]{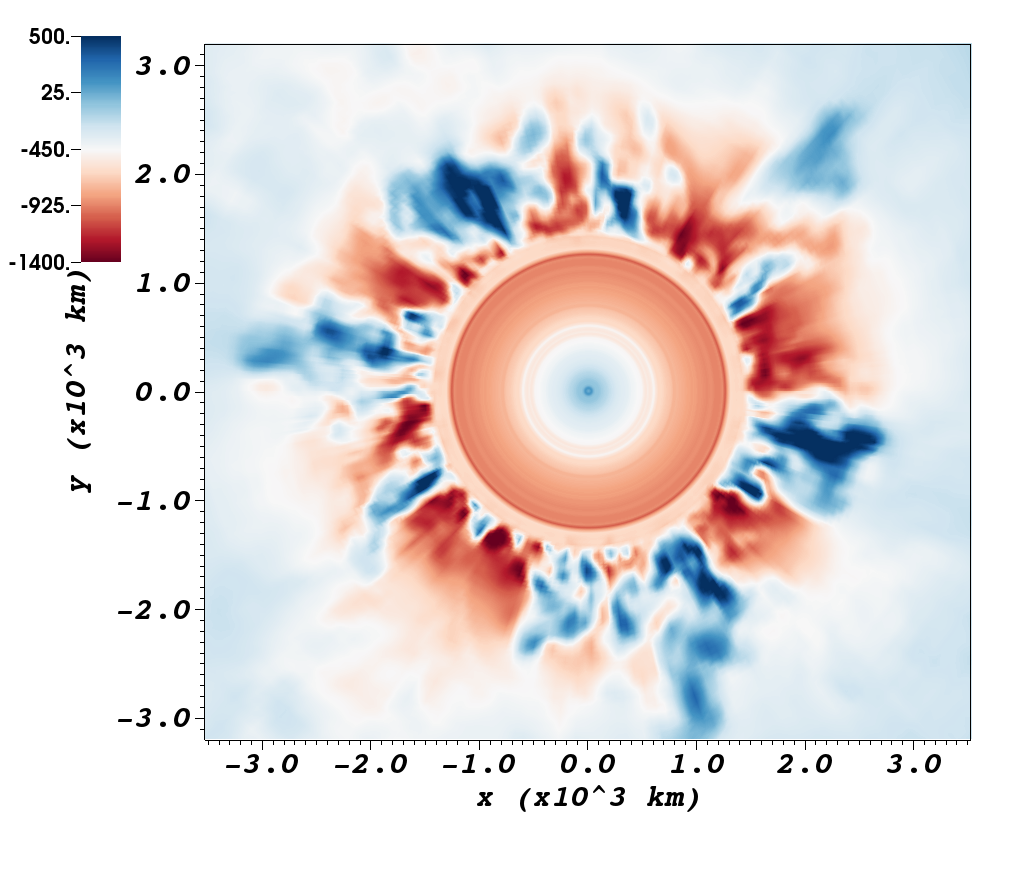}
    \includegraphics[width=0.48\linewidth]{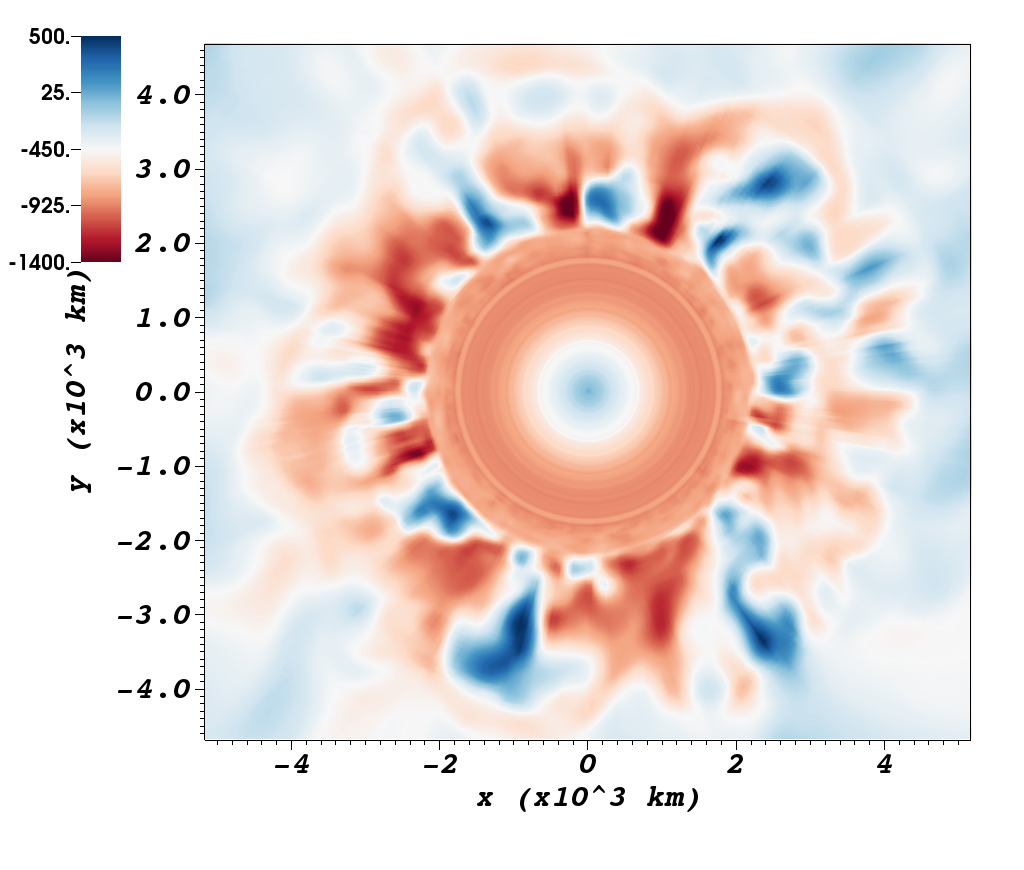}\\
    \includegraphics[width=0.48\linewidth]{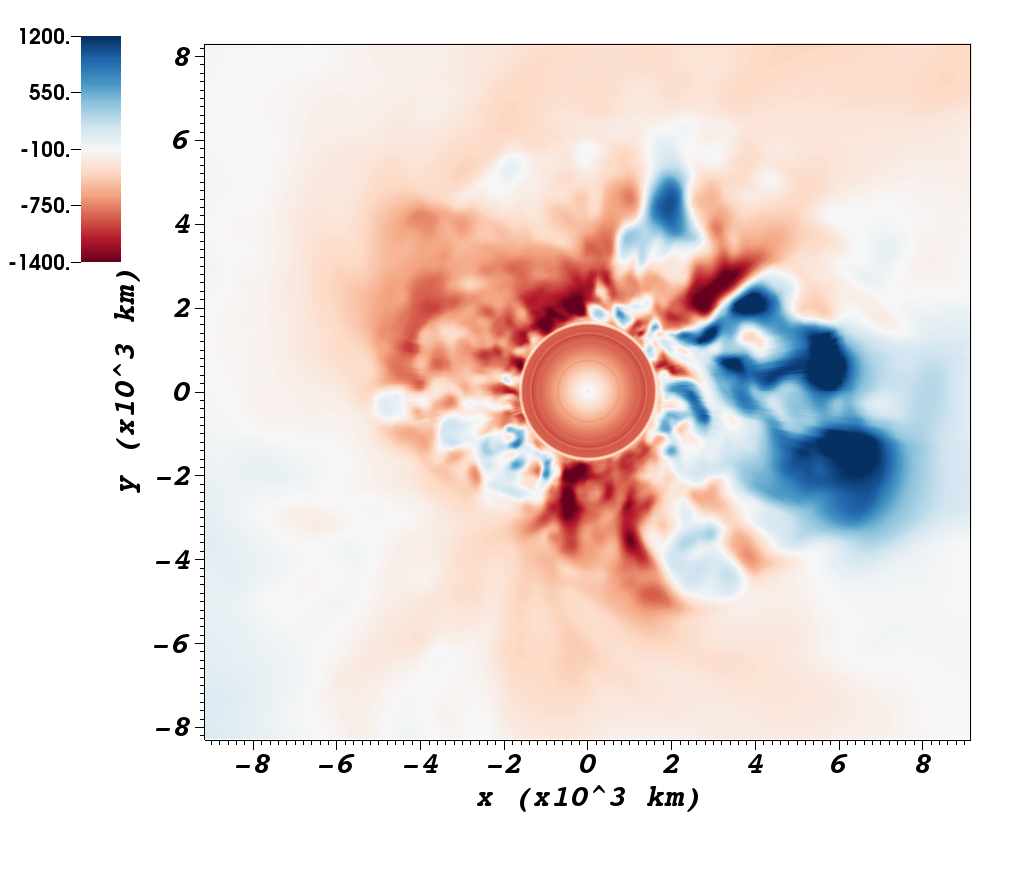}
    \includegraphics[width=0.48\linewidth]{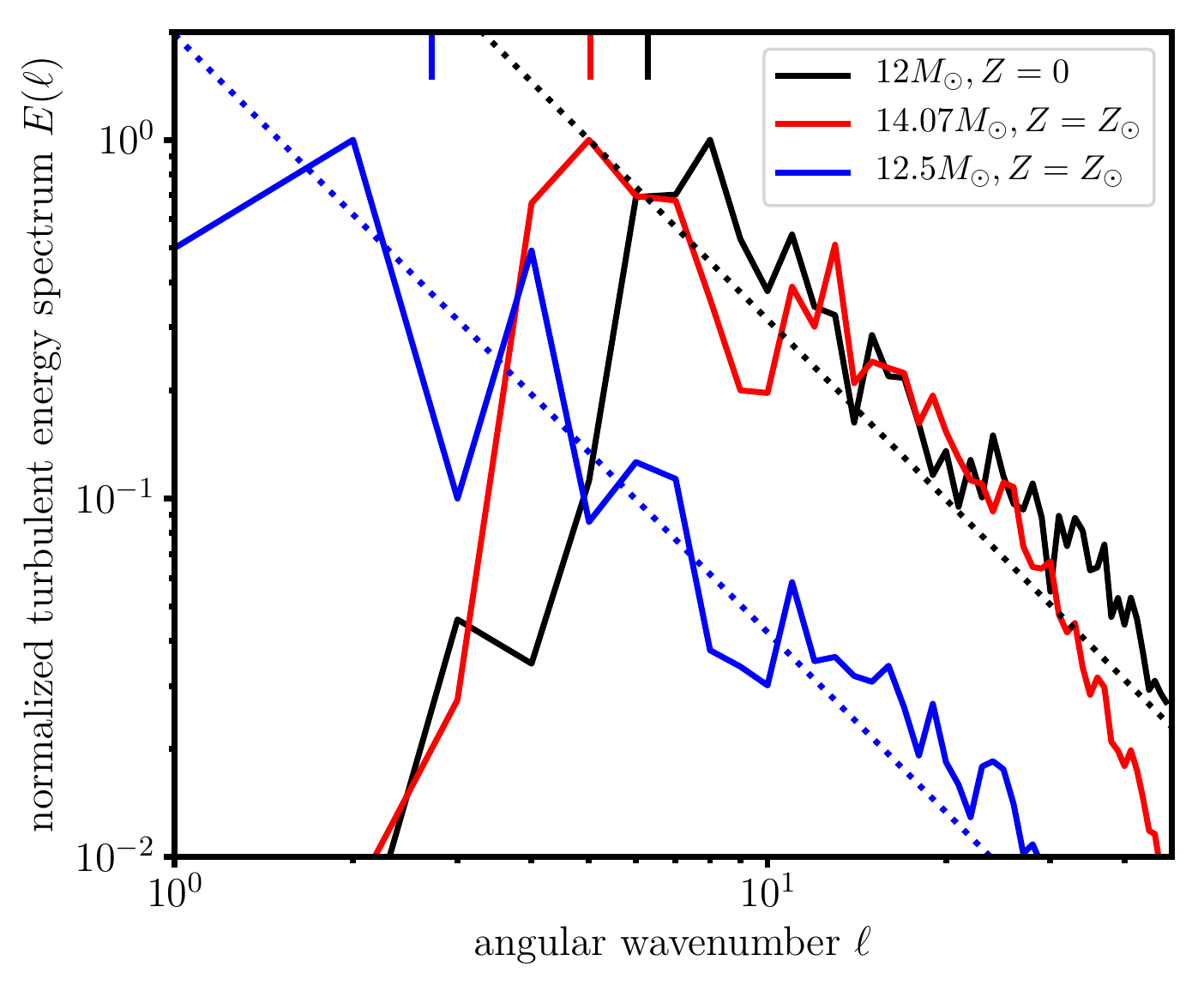}
    \caption{Dependence of the dominant eddy scale
    on the shell geometry illustrated by
    slices through 3D supernova progenitor
    models with convective burning and their
    turbulent energy spectra. The 2D slices
    show the radial velocity at the onset
    of collapse in progenitors of
    $12\,M_\odot$ with metallicity $Z=0$ (top right),
    $14.07\,M_\odot$ with $Z=Z_\odot$ (top left) ,  and $12.5\,M_\odot$
    with $Z=Z_\odot$ (bottom left) with active convective O shells. The bottom right panel shows turbulent
    energy spectra $E(\ell)$ computed from the radial velocity
    around the center of the convective zone. The
    dominant wavenumber expected from
    Eq.~(\ref{eq:ell_0}) is indicated
    at the top; note that there is
    an uncertainty because the outer
    boundaries of the convective zones are
    fuzzy. The dotted lines
    show the slope of a Kolmogorov spectrum.
    The plots for the $12\,M_\odot$ and $12.5M_\odot$
    models have been adapted from \citet{mueller_19a}.
    Image reproduced with permission, copyright by the authors.
    }
    \label{fig:geometries}
\end{figure}

Regarding the dominant scales of the convective
flow, the  recent global 3D shell burning
simulations \citep{chatzopoulos_14,couch_15,mueller_16c,jones_16,yadav_19} confirm the emergence of large eddies
with low angular wavenumber $\ell$ that stretch
across the entire convective zone. \citet{mueller_16c}
verified quantitatively that the peak of the turbulent
energy spectrum in $\ell$ agrees
well with the wavenumber of the first unstable
convective mode at the critical Rayleigh number
\citep{chandrasekhar_61,foglizzo_06},
\begin{equation}
\label{eq:ell_0}
    \ell=
    \frac{\pi (R_+ + R_-)}{2 (R_+ - R_-)}.
\end{equation}
Further simulations that also explored thinner shells
\citep{mueller_19a} show a shift towards higher $\ell$
and corroborate this scaling as illustrated
in Fig.~\ref{fig:geometries}.
Beyond this dominant wavenumber, the turbulence
exhibits a Kolmogorov spectrum \citep{chatzopoulos_14,mueller_16c}.

Naturally, the modern 3D models still exhibit differences to MLT in
detail even within convective zones. For example, 
$\omega_\mathrm{BV}^2$ often changes sign in the outer
parts of a convective layer in 3D, indicating that the 
spherically-averaged stratification is nominally stable
\citep{mocak_09,mueller_16c}. \citet{mueller_16c} also remark
that the spherically-averaged mass fraction profiles
tend to be flatter in 3D than in 1D, due to the usual asymmetric 
choice $\alpha_X = \alpha_e/3$ for the MLT parameters for
material diffusion and energy transport, which probably
ought to be replaced by $\alpha_X = \alpha_e$.
A rigorous approach to quantify the structure of
the convective flow and the differences between 3D and 1D
models is available in the form of spherical 
Reynolds decomposition, which has been pursued
systematically by \citet{viallet_13,mocak_14,arnett_15}.
The mere form of the Reynolds-averaged equations
for bulk (i.e., spherically-averaged) and fluctuating
quantities dictates that such an analysis invariably finds
dozens of terms that are implicitly set to zero
in MLT.

\paragraph{Assessment.} How are we to evaluate these
commonalities and differences between 3D simulations
and 1D stellar evolution flow? For most purposes, the
question is not whether effects \emph{are} missing
in MLT-based 1D models (since the very purpose
of an approximation like MLT is to retain only
the leading effects), but whether those missing
effects matter over secular time scales or have
an impact during the supernova explosion.
As we shall discuss in detail in
Sect.~\ref{sec:perturbations}, the presence of
asymmetries in convective shells indeed matters
during the supernova, but the fact is also
that MLT and linear perturbation theory appear
to predict the relevant parameters -- the velocities and dominant 
scales of  convective eddies -- quite well.
As far as the secular evolution of convective burning shells is 
concerned,  there is little evidence that MLT does not
adequately describe the flow \emph{within} convective shells.
There is typically good agreement in
critical parameters for the shell evolution like
the total nuclear burning rate. Many effects that
MLT captures inaccurately and matter critically
in models of convective envelopes and stellar atmospheres -- such as 
the precise deviation of the stratification from superadiabticity --  
are of minor importance for the bulk evolution of massive stars during
the late burning stages.
For more tangible consequence of multi-D effects on
secular time scales, we need to consider convective boundaries
in Sect.~\ref{sec:cbm}.

\subsection{Supernova progenitor models}
\label{sec:presn}
\paragraph{Simulations to the presupernova stage.}
Only a few models of convective burning have yet
been carried up to the point of core collapse
\citep{couch_15,mueller_16b,mueller_16c,mueller_19a,yadav_19,yoshida_19} because of several obstacles.
In order to accurately follow the composition changes
and the deleptonization in the Fe core
and Si shell (i.e., in the NSE and QSE regime)
that drive the evolution towards collapse, 
reaction networks with well over
a hundred nuclei are required \citep{weaver_78}.
This is feasible in principle, but yet impractical for 
well-resolved 3D simulations up to collapse. Furthermore, the 
initial transient phase and imperfect hydrostatic
equilibrium  after the mapping from 1D to multi-D may
artificially delay the collapse.

Two different strategies have been employed to circumvent
these problems. In their
simulation of Si burning in a $15\,M_\odot$ star
for $160 \, \mathrm{s}$, \citet{couch_15} used
an extended 21-species $\alpha$-network with some
iron group nuclei added to model core deleptonization
\citep{paxton_11}. In order to force the core to collapse,
they increased the electron capture rate on
${}^{56}\mathrm{Fe}$ by a factor of 50, and
their 3D model in fact reaches collapse more than six
times faster than the corresponding 1D stellar
evolution model. This approach
is problematic because any modification of the contraction
time scale of the core also affects the burning in the
outer shells \citep{mueller_16c}. 
Using the same 21-species network,  \citet{yoshida_19}
managed to evolve a 3D
simulation of a $25\,M_\odot$ star and several 2D simulations 
of different progenitors
for the  last $\mathord{\sim} 100 \, \mathrm{s}$ without 
modifying the deleptonization rate. This suggests
that multi-D models can be evolved somewhat self-consistently
to collapse even though the short simulation time is
a concern in this particular case, since it remains unclear to 
what extent the results are affected by the initial 
transient.

The 3D studies of O shell burning in various
progenitors by \citet{mueller_16b,mueller_16c,mueller_19a,yadav_19} have followed a different approach and circumvented
the problems of QSE and deleptonization by excising
the major part of the Fe core and Si shell and replacing
them with an inner boundary condition that is contracted
according to a mass shell trajectory from the
corresponding 1D stellar evolution model. This
approach can be justified for many progenitors, which
have no active convective Si shell, or only weak convection
in the Si shell.

\paragraph{Evolution towards collapse.}
The convective flow in the contracting burning shells
shortly before collapse exhibits few noteworthy
differences to the burning in quasi-hydrostatic shells described in Sect.~\ref{sec:interior_flow}.
The 3D simulations of the different groups
\citep{couch_15,mueller_16c,yoshida_19}
all show the emergence of modes with a
dominant wavelength of the order of the shell
width according to  Eq.~(\ref{eq:ell_0}),
and as far as comparisons have been performed,
the convective velocities remain in good
agreement with MLT until shortly before collapse.
It is noteworthy, however, that the convective velocities
and Mach numbers tend to increase significantly
during the last minutes before collapse because
the temperature at the base of the inner shells,
and hence the burning rate, increase as they contract
in the wake of the core. The convective velocities
then freeze out shortly before collapse once the
burning rate changes on a time scale shorter
than the turnover time scale. This freeze-out
seems to be captured adequately by time-dependent
MLT so that 1D stellar evolution models provide good
estimates for the convective velocities at the
onset of collapse. Bigger differences between
1D and 3D progenitor models can occur in case
of small buoyancy barriers between the O, Ne, and C
shell, in which case 3D models are more likely to
undergo a shell merger \citep{yadav_19}.

The evolution of the convective
shells during collapse will be discussed
in Sect.~\ref{sec:perturbations}.

\begin{figure}
    \centering
    \includegraphics[width=0.48\textwidth]{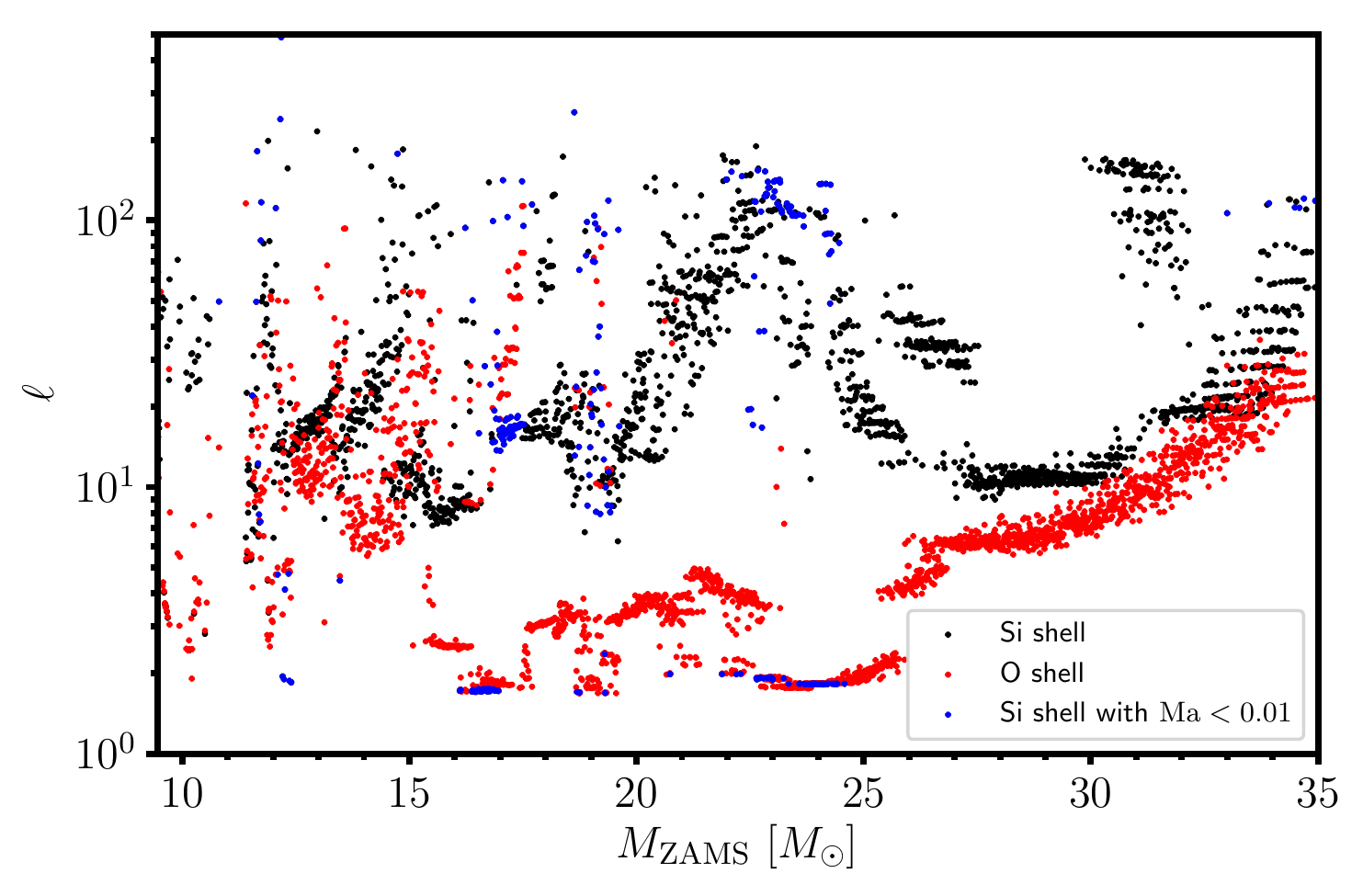}
    \hfill
    \includegraphics[width=0.48\textwidth]{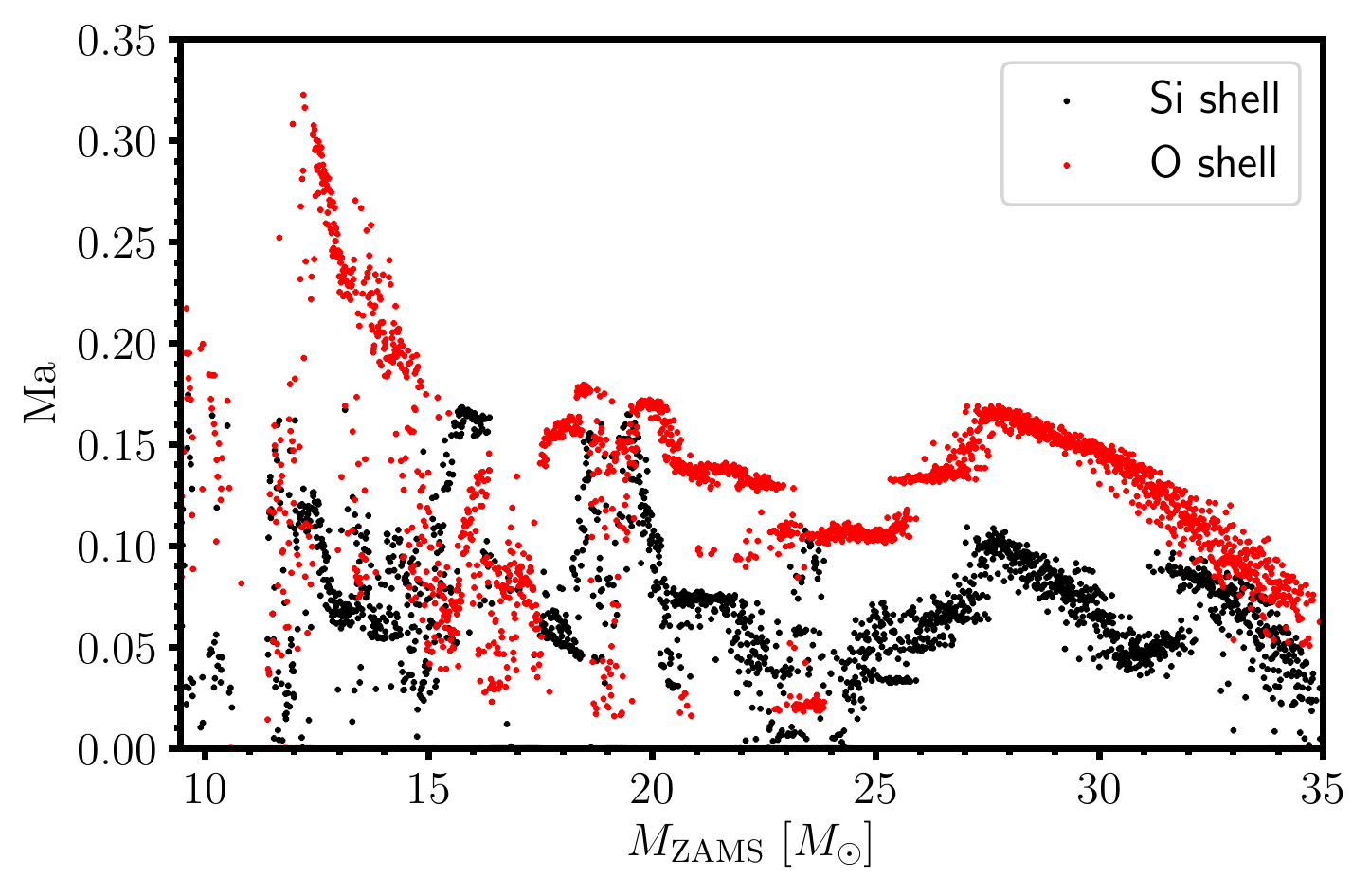}
    \caption{Convective Mach number (left) and dominant
    angular wave number (right) in the Si shell (black) and O
    shell (red) predicted from 1D single-star
    evolution models from the
    study of \citet{collins_18}.
    Image reproduced with permission, copyright by the authors.}
    \label{fig:conv_prog}
\end{figure}

\paragraph{Progenitor dependence.}
Since 3D simulations indicate that convective
velocities and eddy scales can be estimated
fairly well  from 1D stellar evolution models,
one can already roughly outline the progenitor
dependence of convective shell properties as shown
by \citet{collins_18}. Considering the active
Si and O shell burning shells at the onset
of collapse in over 2000 progenitor models,
they find a number of systematic trends
(Fig.~\ref{fig:conv_prog}):
The O shell typically has a higher
convective Mach number
(0.1--0.3) than the Si
shell, where  usually $\mathrm{Ma}<0.1$,
but there are islands around $16\,M_\odot$
and $19\,M_\odot$ in ZAMS mass where
the convective Mach number in the Si shell
reaches about 0.15 and is higher than
in the O shell. The highest convective Mach
numbers of up to 0.3 are reached in the
O shell of low-mass progenitors with small cores
as O burns deeper in the gravitational potential
at higher temperatures. The general
trend towards lower convective velocities
in the O shell with higher progenitor and core
mass is modified by variations in shell entropy
and the residual O mass fraction at the onset of collapse.
Deviations from this general trend also come about
because the various C, Ne, O, and Si 
shell burning episodes do not always occur in the same
order, and because of shell mergers.

The O shell is usually thicker
and therefore allows large-scale modes
with wave numbers $\ell<10$ to dominate.
Large-scale modes are more prevalent in 
progenitors above $16\,M_\odot$ with their
more massive O shells.
The first, thick Si shell is no longer active
at collapse in most cases, and there is
typically only a thin convective Si shell (if any)
between the Fe core and O shell at collapse, which will dominated by small-scale motions.

\citet{collins_18} also find a high prevalence
of late shell O-Ne shell mergers among high-mass
progenitors. In about 40\% of their models
between $16\,M_\odot$ and $26\,M_\odot$ such 
a merger was initiated within the last minutes
of collapse.

Although some of these trends follow from 
robust structural features and trends in the 
progenitor evolution, these findings will need to be 
examined with different stellar evolution codes
and may be modified in detail, especially
when better prescriptions for convective boundary
mixing on secular time scales become available.

\subsection{Convective boundaries}
\label{sec:cbm}
\paragraph{Mixing by entrainment.}
As one of the most conspicuous 
features in their first 2D models of
O shell burning, \citet{bazan_94,bazan_97} noted
the mixing of considerable amounts of C from the 
overlying layer into the active burning region.
Although mixing across convective
boundaries (sometimes indistinctly called ``overshooting'') had already been a long-standing topic
in stellar evolution by then, these results were noteworthy
because \citet{bazan_94,bazan_97} found much stronger
convective boundary mixing (CBM) than compatible
with overshoot prescriptions in 1D stellar
evolution models of massive stars. Second, they observed
that the mixed material can burn vigorously and thereby
in turn dramatically affect the convective flow, i.e.,
there is the possibility of a feedback mechanism
that cannot occur in the case of envelope convection
or surface convection.
\citet{meakin_06} investigated this problem further
in a situation with active and interacting O and C
shells and observed strong excitation of
p- and g-modes at convective-radiative boundaries, which,
as they suggested, might also contribute to compositional
mixing.

Critical steps beyond a mere descriptive analysis
of CBM during the late burning stages were finally
taken by \citet{meakin_07_b}, who established
i) the presence of CBM also in 3D (albeit
weaker than in 2D), ii) identified the dominant
process as entrainment driven by shear (Kelvin--Helmholtz and Holmb\"oe) instabilities
at the convective boundary, and iii) verified
that the mass entrainment $\dot{M}$ rate obeys a power
law that can be motivated theoretically and has
been verified in laboratory experiments of shear-driven
entrainment \citep{fernando_91,strang_01}:
\begin{equation}
\label{eq:entrainment}
    \dot{M}=4\pi A \rho r^2 v_\mathrm{conv} \mathrm{Ri}_\mathrm{b}^{-n},
\end{equation}
Here, $A$ and $n$ are dimensionless constants and
$\mathrm{Ri}_\mathrm{b}$ is the bulk Richardson number,
which can be expressed in terms of the integral scale $L$
of the turbulent flow and the buoyancy jump $\Delta b$
across the boundary,
\begin{equation}
\label{eq:rib}
    \mathrm{Ri}_\mathrm{B}=
    \frac{\Delta b\, L}{v_\mathrm{conv}^2}.
\end{equation}
The buoyancy jump can be obtained by integrating
the square of the Brunt-V\"aisala over the extent
of the boundary layer from $r_1$ to $r_2$,
\begin{equation}
    \Delta b =  \int\limits_{r_1}^{r_2}
    -\omega_\mathrm{BV}^2 \,\ud r.
\end{equation}
In the case of a thin boundary layer, this reduces
to $\Delta b =  g \, \delta \rho/\rho$, where
$\delta \rho/\rho$ is the density contrast across
the convective interface. From
their simulations, \citet{meakin_07} determined
values of $A=1.06$ and $n = 1.05$ for the power-law coefficients.
Since the work expended to entrain material against 
buoyancy the force of buoyancy must be supplied by a fraction
of the convective energy flux (an argument
which was independently redeveloped by \citealt{spruit_15}), one expects
a value of $n=1$ for sufficiently high $\mathrm{Ri}_\mathrm{B}$. 

Several subsequent 3D simulations \citep{mueller_16c,jones_17} have 
confirmed  a value of $n\approx 1$ for the scaling law
(\ref{eq:entrainment}).
\citet{mueller_16c} found a significantly smaller value
of $A\approx 0.1$, however, but this may simply be due
to ambiguities in the definition and measurement of the integral
length scale $L$, which \citet{mueller_16c} identify
with the pressure scale height $\Lambda$, and of the convective
velocity $v_\mathrm{conv}$ that enters Eq.~(\ref{eq:rib})
for the bulk Richardson number. \citet{jones_17}  expressed the 
entrainment law slightly differently by a proportionality
$\dot{M}\propto \dot{Q}_\mathrm{nuc}$ to the total nuclear energy generation rate $\dot{Q}_\mathrm{nuc}$, which is equivalent
to Eq.~(\ref{eq:entrainment}) with $n=1$. Their simulations
are particularly noteworthy because they employed sufficiently
high resolution to establish the entrainment law up to very
high  $\mathrm{Ri}_\mathrm{b}$. Although they do not explicitly
state values of $\mathrm{Ri}_\mathrm{b}$, one can estimate
that their models reach up to $\mathrm{Ri}_\mathrm{b}=700$--$1000$.

The simulations of \citet{cristini_17,cristini_19} are a notable
exception as they find a significantly shallower power
law with $n=0.74$. This different power-law slope has yet
to be accounted for, but it is important to note that
despite the shallower power law, \citet{cristini_17,cristini_19}
generally find \emph{lower} entrainment rates than
\citet{meakin_07} for the same value of $\mathrm{Ri}_\mathrm{b}$
with a much smaller value of $A=0.05$. At 
$\mathrm{Ri}_\mathrm{b}\approx 20$, their entrainment
rate is actually in very good agreement with
\citet{mueller_16c}, and in the region of
$\mathrm{Ri}_\mathrm{b}=40$--$300$ their data
are consistent with a steeper power law of $n\approx1$. Since
\citet{cristini_17,cristini_19} also explore a much broader
range in bulk Richardson number than the
aforementioned studies, one possible interpretation could be that i) the value of $A$ was overestimated in \citet{meakin_07}, and that ii) the low
value of $n=0.74$ may be due to a flattening of the entrainment
law below $\mathrm{Ri}_\mathrm{b}=20$--$30$
for some physical reasons, and perhaps a slight flattening
at $\mathrm{Ri}_\mathrm{b}>200$ because of numerical resolution effects.

\paragraph{Shell mergers.}
Convective boundary mixing can take on a dramatic
form when the buoyancy jump between two shells is 
sufficiently small for the neighboring shells to merge
entirely within a few convective turnover times. Such shell
mergers have long been known to occur in 1D stellar evolution
models, in particular between O, Ne, and C shells
\citep[e.g.,][]{sukhbold_14,collins_18}. This is because
balanced power leads to very similar entropies in
the O, Ne, and C shells, and hence small buoyancy
jumps between the shells. When nuclear energy generation and neutrino cooling
finally fall out of balance due to shell contraction,
the entropy of the inner (O or Ne) shell frequently
increases and overtakes the outer shell(s), so
that such mergers are particularly prevalent shortly
before collapse as pointed out by \citet{collins_18},
who estimated that 40\% of stars between
$16\,M_\odot$ and $26\,M_\odot$ collapse during
an ongoing shell merger. Although such mergers occur in 1D models, they may occur more readily in 3D, and 3D
simulations are also necessary to capture the composition
inhomogeneities and nucleosynthesis during the dynamical merger phase.

Shell mergers have indeed been seen in several
recent 3D simulations.
\citet{mueller_16b} pointed out the breakout
of a thin O shell through an inert, non-convective O layer
into the active Ne burning zone in a $12.5\,M_\odot$
star in the last minute before collapse, which, however,
did not lead to a complete shell merger.
\citet{mocak_18} found a merger between 
the O and Ne shell in a $23\,M_\odot$ model,
and noted that the runaway entrainment leads
to a peculiar quasi-steady with two distinct
burning zones for O (at the base) and Ne (further out)
within the same convective shell.
 However, their simulation only
covered five turnover times and showed
the merger occurring during the initial transient
phase. \citet{yadav_19} simulated an 
O-Ne shell merger in an $18.88\,M_\odot$ over
15 turnover times, and were able to follow the evolution from
the pre-merger phase with a soft, but clearly
defined shell boundary and slow steady-state
entrainment through the
dynamical merger phase to a partially mixed
post-merger state at the onset of collapse..
They stressed the emergence of large-scale
asymmetries in the velocity field
(with extreme velocities of up to $1700\, \mathrm{km}\, \mathrm{s}^{-1}$) and
the composition during the merger, although
the compositional asymmetries are already washed
out somewhat at the point of collapse.

\paragraph{Impact on nucleosynthesis.}
With multi-D simulations of the
late-burning stages firmly established,
it is critical to identify observable fingerprints
of additional convective boundary mixing.
The nucleosynthesis yields may provide one
such fingerprint, which has already been
discussed by several studies, even though one
can only draw conclusions based on qualitative 
arguments and on 1D models with artificially
enhanced mixing so far.

\citet{davis_19} pointed out that the
assumptions for convective boundary mixing
can significantly affect the yields
of various $\alpha$-elements (C, O, Ne, Mg, Si),
simply as a consequence of the change in
shell structure. However, entrainment and
shell mergers may leave more specific
abundance patterns.
In their investigation of O-C shell mergers
in 1D \citet{ritter_19} found significant
overproduction of P, Cl, K, Sc, and possibly p-process
isotopes, and argue that the occurrence of
shell mergers may have important consequences
for galactic chemical evolution (GCE). 
More recently, \citet{cote_19} considered
a Si-C shell merger, for which they find
significant overproduction of ${}^{51}\mathrm{V}$
and
${}^{52}\mathrm{Cr}$, which allows them
to strongly constrain the rate of such
events based on observed Galactic stellar abundances.
This is related to the long-standing realization
that the ashes of hydrostatic silicon burning
under neutron-rich conditions cannot be ejected in large quantities
because of GCE constraints \citep{woosley_73,arnett_96}.

Supernova spectroscopy may also help constrain
additional convective boundary mixing and shell mergers
may via their nucleosynthetic fingerprints.
For example, the ejection of neutron-rich material
from the silicon shell that is mixed out by entrainment
before the explosion would lead to a supersolar Ni/Fe ratio
as observed in some supernovae \citep{jerkstrand_15b}.
Mixing of minimal amounts of Ca into O-rich zones
can also have significant repercussions
since only a mass fraction of a few $10^{-3}$ in
Ca is required for Ca to be the domnant coolant
during the nebular phase and quench O line mission
in a shell \citep{fransson_89,kozma_98b}.
This diagnostic potential of supernova spectroscopy
for convective boundary mixing needs to be explored
further in the future, but further (macroscopic) mixing 
during the explosion presents a major complication
as it is not straightforward to disentangle the
effect of mixing processes prior and during the 
explosion.

\paragraph{Secular effect on stellar evolution.}
Evaluating the observational consequences of the convective 
boundary mixing seen in 3D models is also difficult
because there is still no rigorous method for treating
these processes in 1D stellar evolution codes. 
A crude estimate for the shell growth by entrainment
can be formulated by noting that the
work required to entrain material with
 density contrast $\delta \rho/\rho$ against
 buoyancy  must be no larger
 than a fraction of the time-integrated convective energy flux \citep{spruit_15,mueller_16b}. During
 the late burning stage, the convective energy
 flux is set by the nuclear energy generation rate,
 and hence one can estimate that the
 entrained mass $\Delta M_\mathrm{entr}$
 over the lifetime of a shell with mass
 $M_\mathrm{shell}$ and radius $r$ is roughly
\begin{equation}
\label{eq:delta_m}
\frac{G M}{r}\frac{\delta \rho}{\rho}
\Delta M_\mathrm{entr}
\sim A M_\mathrm{shell} \Delta Q,
\end{equation}
where $A$ is the dimensionless
coefficient in Eq.~(\ref{eq:entrainment})
and $\Delta Q$ is the average $Q$-value for
a given  burning stage \citep{mueller_16b}.
Based on Eq.~(\ref{eq:delta_m}),
\citet{mueller_16b} estimates that O
shells could grow by tens of percent
in mass by entrainment; for Si shells one expects
a smaller effect, for C shells, the effect may be bigger.

How one can go beyond such simple estimates
by using improved recipes for convective boundary
mixing in stellar evolution codes is still
an unresolved question. A common approach, based on
the simulations of \emph{surface} convection
by \citet{freytag_96}, models entrainment
as diffusive overshooting with an exponential
decay of the MLT diffusion coefficient outside
the convective boundary. The length scale $\lambda_\mathrm{OV}$
for the exponential decay can be calibrated
against 3D simulations. This approach
has been followed by \citet{cote_19,davis_19} and
by many works on additional convective boundary mixing
in low-mass mass stars, but has several issues.
Entrainment is a very different process than
diffusive overshoot that operates in a distinct
physical regime (high P\'eclet number), and
hence one should not expect that it can be described
by the same formalism \citep{viallet_15}. It is also
unclear why diffusive overshoot should be applied
only for compositional mixing in the entrainment regime.
The common approach of expressing $\lambda_\mathrm{OV}$ by the pressure scale height is also open
to criticism because the relevant length scale should
be set by the convective velocities and the buoyancy
jump, so that one would rather expect
$\lambda_\mathrm{OV}\propto v_\mathrm{conv}^2/\Delta b$.

\citet{staritsin_13} proposed an alternative
approach of extending convectively mixed regions
with time following
the entrainment law (\ref{eq:entrainment}), which
better reflects the physics of the entrainment process.
However, this approach has not been applied to
the late neutrino-cooled burning stages yet (i.e., precisely
where entrainment should operate). It also has some
conceptual issues, for example, the entrainment law
(\ref{eq:entrainment}) obviously breaks down if there
is convection on both sides of a shell interface.
Yet another approach for extra mixing in
1D models was followed by \citet{young_05b}, who
handle mixing based on the local gradient Richardson number
$\mathrm{Ri}$ for shear flows, which is estimated using
an elliptic equation for the amplitudes of
waves excited by convective motions \citep{young_05a}.
This approach is physically motivated, but 
is still awaiting (and worthy of) a more quantitative comparison with 3D simulations beyond the qualitative
discussion in \citet{young_05a}

\paragraph{Flame propagation in low-mass progenitors.}
Around the minimum progenitor mass for supernova
explosions, multi-D effects can have a more
profound effect than merely changing shell masses
on a modest scale, and may decide about
the final fate of the star. This
regime is best exemplified by the electron-capture
supernova channel of super-asymptotic giant
branch (SAGB) stars (see \citealp{jones_13,doherty_17,nomoto_17,leung_19a}
for a broader overview and a discussion of uncertainties). In this evolutionary channel,
the star builds up a degenerate core composed
primarily of O and Ne. If this core grows
to $1.38\,M_\odot$, electron captures on
Ne and Mg trigger an O deflagration. Depending
on the interplay of deleptonization
(which decreases the degeneracy pressure)
and the nuclear energy release, the core
either contracts, collapses
to a neutron stars, and explodes as an electron
capture supernova, or the core does not collapse and explodes as a weak
thermonuclear supernova \citep{jones_16}.
Since the flame is turbulent, its propagation
needs to be modelled in multi-D, similar to deflagrations
in Type~Ia supernovae. Simulations of this
problem have been conducted by
\citet{jones_16,kirsebom_19} in 3D and  \citep{leung_19a,leung_19b} in 2D. Efforts
to improve the nuclear input physics and explore
the sensitivity to the ignition geometry,
general relativity, and flame physics are ongoing
\citep[e.g.,][]{kirsebom_19,leung_19b}.

For slightly more massive stars, one encounters
similar situations with convectively-bounded
flames after off-center ignition
of O or Si \citep{woosley_15}. Again, multi-D
effects may significantly affect the final
evolutionary phase before collapse in this regime,
but multi-D simulations of such supernova progenitors
are yet to be carried out (but see
\citealt{lecoanet_16} for idealized 3D simulations
relevant to this regime).

\subsection{Current and future issues}
Significant progress notwithstanding, multi-D simulations of 
the late stages of convective burning still face further 
challenges. For the evolution towards collapse,
models will eventually need to include a more 
sophisticated treatment of burning and deleptonization
in the QSE and NSE regime and forgo the current approach
of either using small networks or excising the Fe/Si
core.  Perhaps an even greater concern about the
conclusions of current multi-D simulations lies
with the time-scale problem, however. No 3D simulations
have yet been evolved sufficiently long to reach
the state of balanced power (or to reveal why
it would not be reached). This may have repercussions
for turbulent entrainment, which ultimately
taps the energy in turbulent motions and hence cannot be 
completely disconnected from the energy budget within a 
shell. 

Moreover, although a few attempts
have been made to simulate convection in rotating
shells  in 2D \citep{arnett_10,chatzopoulos_16}
and 3D \citep{kuhlen_03},
multi-D simulations have yet to investigate 
the angular momentum distribution and angular momentum
transport during the late pre-supernova stages in a
satisfactory manner. Three-dimensional
simulations are even more critical for this purpose
than in the non-rotating case since many relevant
phenomena such as (Rossby waves, Taylor-Proudman
columns)
cannot be modeled adequately in 2D. Simulations also need to
explore a larger parameter space since the convective
dynamics will depend on the Rossby number
$\mathrm{Ro}\sim v_\mathrm{conv}/(\Omega R)$
(where $\Omega$ is the rotational velocity).
Furthermore, the time scales become a bigger challenge
because models need to be run for several rotational
periods $T=2\pi \Omega^{-1}$ and several convective turnover
times $\tau_\mathrm{conv}$ (whichever
is longer), which is  problematic since
rotation in pre-supernova models is likely
slow \citep[e.g.,][]{heger_05} so that
 $\mathrm{Ro}\gg 1$  and $T\gg \tau_\mathrm{conv}$.
 On the bright side, multi-D simulations may
 reveal  much more interesting  differences to 
 1D stellar evolution models: Both \citet{kuhlen_03}
 and \citet{arnett_10} found pronounced differential
 rotation developing from a rigidly rotating initial
 state, and \citet{arnett_10} suggest that convective
 shells might adjust to a stratification with
 constant angular momentum as a function of radius
 rather than to uniform rotation as assumed
 in stellar evolution models. However, more simulations
 and more rigorous analysis is required to investigate these
 claims.

The problem of rotation can obviously not be solved
without including magnetic fields in the long run.
It is well known 
\citep[e.g.,][]{shu} that the criterion for the instability
of rotating flow becomes less restrictive in the
MHD case, and effects such as the
magnetorotational instability
\citep[MRI,][]{balbus_91}
or a small-scale dynamo may enforce a more uniform rotation
profile than hydrodynamic convection alone. But the
importance of magnetic fields is not confined to
the case rotating progenitors. Prompted by helioseismic
measurements that indicate smaller convective velocities
in the deeper layer of the solar convection zone \citep{gizon_12,hanasoge_12},
some simulations of magnetoconvection in the Sun
found a suppression of convective velocities
by up to 50\% compared
to hydrodynamic simulations due to
strong magnetic fields from a small-scale dynamo
that reach equipartition with the turbulent
kinetic energy \citep{hotta_15}. Magnetic fields
can also inhibit or enhance mixing in shear layers
\citep{brueggen_01} and may hence affect convective
boundary mixing. Thus, there is still plenty of
ground left to explore for simulations of the late burning 
stages.

\section{Core collapse and shock revival}
\label{sec:sn}
In the Introduction, we already outlined a variety of multi-D effects
than can play a role in reviving the stalled supernova shock as
a subsidiary agent to neutrino heating (i.e., neutrino-driven convection in the gain region, the SASI, and progenitor asphericities), or also as the main
drivers of the explosion (rotation and magnetic fields). Historically,
a number of works have also considered convection in the PNS interior as a 
means for precipitating explosions by enhancing the neutrino emission from the
PNS \citep{epstein_79,wilson_88,burrows_88b,wilson_93}, but these
hopes were not substantiated in subsequent decades. Nonetheless, convection
inside the PNS remains important for various aspects of the supernova
problem such as the neutrino and gravitational wave signals and the
nucleosynthesis conditions in the innermost ejecta.

Since each of these phenomena has proved rich and complex over the years,
it is no longer possible to treat them adequately within a chronological
narrative of the quest for the supernova explosion mechanism. Nevertheless,
ascertaining the explosion mechanism by means of first-principle simulations
remains the overriding concern in supernova theory,
and it is therefore still useful to recapitulate the progress
in supernova explosion modelling from the advent of the first 2D models with
a simplified treatment of neutrino heating and cooling in the 1990s
\citep{herant_92,shimizu_93,yamada_93,janka_93,herant_94,burrows_95,janka_95,janka_96}. A more detailed analysis
of the individual hydrodynamic phenomena beyond
this chronicle of simulations 
is then provided
in Sects.~\ref{sec:1d_struc}
to \ref{sec:pns_convection}.

\paragraph{Neutrino-driven explosions in 2D.}
Although the 2D simulations of the early and mid-1990s had shown multi-D effects
to be helpful for shock revival, these models did not utilize neutrino transport
on par with the best available methods for 1D simulations at the time.
In a first attempt to better model neutrino heating and cooling in 2D
by using the pre-computed neutrino radiation field from a 1D simulation
with multi-group flux limited diffusion, \citet{mezzacappa_98} were
unable to reproduce the successful explosions found in earlier 2D models.
This led to a resurgence of interest in accurate methods for neutrino transport,
culiminating in the development of Boltzmann solvers for relativistic
\citep{yamada_99,liebendoerfer_00,liebendoerfer_04}
and pseudo-Newtonian simulations \citep{rampp_00,rampp_02}.
The explosion problem was then revisited in 2D 
using various types of multi-group neutrino transport
from the mid-2000s onwards. Neutrino-driven explosions were obtained in many
of these 2D simulations for a wide range of progenitors
\citep{buras_06b,marek_09,mueller_12a,mueller_12b,mueller_13,janka_12,janka_12b,suwa_10,suwa_13,bruenn_13,bruenn_16,nakamura_15,burrows_18,pan_18,oconnor_18a}, though still with 
significant differences between the various simulation codes.

\paragraph{Challenges and successes in 3D.}
Following isolated earlier attempts at 3D modelling  using 
the ``light-bulb'' style
models of the 1990s \citep{shimizu_93,janka_96},
or gray flux-limited diffusion \citep{fryer_02},
the role of 3D effects in the explosion mechanism
was finally investigated vigorously in the last decade,
starting again with simple light-bulb models
\citep{nordhaus_10a,hanke_12,couch_12b,dolence_13}.
Except for spurious results in \citet{nordhaus_10a},
these light-bulb models indicated a similar
``explodability'' in 2D and 3D.
However, subsequent 3D models with rigorous neutrino transport
proved more reluctant to explode; indeed the first 3D models
of $11.2\,M_\odot$, $20\,M_\odot$, and $27\,M_\odot$ progenitors 
using multi-group, three-flavour neutrino transport 
did not explode at all \citep{hanke_13,tamborra_14a}.

Even though various groups have now obtained explosions
in 3D simulations, shock revival usually occurs later
than in 2D, and often requires additional (and sometimes hypothetical) 
ingredients to improve the heating conditions or a specific
progenitor structure. For low-mass
single-star \citep{melson_15a,mueller_16b,burrows_19}
and binary \citep{mueller_18} progenitors just above
 the iron-core formation limit, 3D simulations readily yield
 explosions since the steep drop of the density outside
 the iron core implies a rapid drop of the accretion rate onto
 the shock after bounce. For more massive stars the record
 is mixed. For standard, non-rotating progenitors 
 in the range between $11.2\,M_\odot$ and $27\,M_\odot$
and
unmodified, state-of-the-art  microphysics, no explosions
 were found in simulations using
 the \textsc{Vertex} code  \citep{hanke_13,tamborra_14a,melson_15b,summa_18}
 and the \textsc{Flash-M1} code \citep{oconnor_18b}.
 On the other hand, the Oak Ridge group 
 obtained an explosion for a $15\,M_\odot$ star \citep{lentz_15}
 with their \textsc{Chimera} code,
 and the Princeton group observed
 shock revival in   eleven out of fourteen models between $9\,M_\odot$
 and $60\,M_\odot $ \citep{vartanyan_19,burrows_19,radice_19,burrows_19b}
 with the \textsc{Fornax} code. In both
 cases the accuracy of the microphysics, the neutrino transport,
 and gravity treatment appears comparable to \textsc{Vertex}.
 Three-dimensional simulations using other codes (that are
 constantly evolving!) are more difficult to compare
 as they involve simplifications in the microphysics
 or transport compared to 
 \textsc{Vertex}, \textsc{Chimera}, and \textsc{Fornax}, although
 some of them compensate for this by higher resolution in real space
 and energy space and a better treatment of gravity. At any rate,
 results obtained with other codes 
such as \textsc{CoCoNuT-FMT}, \textsc{Fugra}, 
\textsc{Zelmani}, and \textsc{3DnSNe}
 add to the picture
 of simulations straddling the verge between successful shock revival 
 \citep{takiwaki_12,takiwaki_14,mueller_15b,roberts_16,chan_18,ott_18,kuroda_18} and  failure \citep{mueller_17,kuroda_18}
 for standard, non-rotating progenitors and standard or simplified
 microphysics.
 
These different results may simply indicate that the
neutrino-driven mechanism operates close
to the critical threshold for explosion in nature.
Observations of supernova progenitors indeed indicate that 
black hole formation occurs already at  relatively low masses
down to to $\mathord{\sim} 15\,M_\odot$ for single
stars \citep{smartt_09a,smartt_15}. Since the lack
of robust explosions in 3D persists to even
lower masses, and since strongly delayed explosions
in 3D may turn out too weak to be compatible with
observations, several groups have explored new avenues
towards more robust explosions. Some of the proposed
ideas invoke modifications or improvements to the 
microphysics that ultimately lead to improved
neutrino heating conditions, such as strangeness corrections to the  neutral-current scattering rate \citep{melson_15b} and   muonization \citep{bollig_17}. Other studies have
explored purely hydrodynamic effects. Among these, \citet{takiwaki_16,janka_16,summa_18} pointed out
that rapid progenitor rotation could be conducive
to shock revival even without invoking MHD effects.
Another idea posits that including seed perturbations
from the late convective burning stages can facilitate
shock revival. First studied by 
\citet{couch_13} and \citet{mueller_15a} using parametric
initial conditions,  this ``perturbation-aided'' mechanism
has subsequently been explored further using
pre-collapse perturbations from
3D models of the late burning stages, initially
with ambiguous results in the leakage simulations
of \citet{couch_15} and then in a number of 3D simulations
using multi-group neutrino transport
\citep{mueller_16b,mueller_17,mueller_19a}, where
it led to robust explosions over a wider mass range
from $11.8\,M_\odot$ to $18\,M_\odot$ for single stars.

Neutrino-driven explosion models have thus matured
considerably in recent years, but it would be
premature to declare the problem of shock revival
 solved. The discrepancies between the results
of different groups have yet to be sorted out, and
there is still no ``gold standard'' among
the simulations that combines the best
neutrino transport, the best microphysics, 3D progenitor
models, and general relativity. Moreover, phenomenological
models of neutrino-driven explosions \citep{ugliano_12,pejcha_15a,sukhbold_16,mueller_16a}
suggest that a different mechanism is still needed to
explain hypernova explosions with energies above
$2 \times 10^{51}\, \mathrm{erg}$. 

\paragraph{Magnetohydrodynamic simulations.}
The mechanism(s) behind hypernovae likely rely
on rapid rotation and strong magnetic fields \citep{akiyama_03,woosley_06b}, but
the importance of magnetic fields may not
end there. 
There may be a continuous transition from
neutrino-driven explosions to MHD-driven explosions
\citep{dessart_07a}, and strong magnetic fields may also a role in non-rotating progenitors 
as an important driving agent or as a subsidiary
to neutrino heating \citep{obergaulinger_14}.

Although the ideas of \citet{akiyama_03}
quickly triggered first 2D MHD core-collapse
supernova simulations
\citep[e.g.,][]{sawai_04,sawai_05,obergaulinger_06a,shibata_06}, there is still only a small
corpus of magnetorotational supernova explosion models, especially if we focus on models
of the entire collapse, accretion, and early explosion phase
using reasonably detailed microphysics
and disregard parameterized models of relativistic 
and non-relativistic jets and of collapsar disks.
\citet{dessart_07a} presented 2D simulations of magnetorotational explosions of a $15\,M_\odot$  progenitor
(later followed by MHD simulations of
accretion-induced by collapse in \citealt{dessart_07b}) with the
Newtonian radiation-MHD code
\textsc{Vulcan} and demonstrated the ready emergence
of jets powered by strong hoop stresses for sufficiently
strong initial fields. \citet{dessart_07a} made the important point that
these non-relativistic jets are a distinctly different phenomenon
from the relativistic jets seen in long gamma-ray bursts
(GRBs), which may be formed several seconds after
shock revival. 

Like most other subsequent 
simulations, these models relied on parameterized
initial conditions with artificially strong magnetic fields
to mimic the purported fast amplification
of much weaker fields in the progenitor by the
magnetorotational instability \citep{balbus_91,akiyama_03}.
They also imposed the progenitor rotation profile by hand.
The 2D studies of \citet{obergaulinger_17,obergaulinger_19,bugli_19} have
recently explored variations in the assumed
initial field strength and topology and the assumed
rotation profiles more thoroughly. While they find
considerable variation in the outcome of their models,
it is interesting to note that in some instances 
\citet{obergaulinger_19} even find magnetorotational explosions for the \emph{unmodified} rotation
profile and magnetic field strength
of two of the  $35\,M_\odot$ progenitor models from
\citet{woosley_06}, although it is not perfectly clear
what the precise geometry of the field in the stellar
evolution models ought to be.

The imposition of axisymmetry is an even bigger
concern in the case of magnetorotational supernovae
than for nuetrino-driven explosion models. Several
3D simulations of MHD-driven explosions have by now
been performed, but among these only
\citet{obergaulinger_19} included
multi-group neutrino transport, whereas the others
\citep{winteler_12,moesta_14b,moesta_18}
employed a leakage scheme. The prospects
for successful magnetorotational explosions in 
3D are still somewhat unclear. \citet{moesta_18}
reported the destruction of the emerging jets
by a kink instability, although the jet can
apparently be stabilised if the poloidal field strength
is comparable to the toroidal field strength
\citep{moesta_18}. Moreover, the explosion
dynamics already depends sensitively on the
assumed initial field geometry; strong
dipole fields appear to be required for
the most powerful explosions \citep{bugli_19}.
Given the vast uncertainties concerning
the initial rotation rates, field strengths,
and field geometries in the supernova progenitors,
considerably more work
is necessary before the magnetorotational
mechanism can be considered robust even for
a small sub-class of progenitors. We will therefore
focus only on the hydrodynamics of neutrino-driven
explosions in the subsequent discussion.

\begin{figure}
    \centering
    \includegraphics[width=0.474\linewidth]{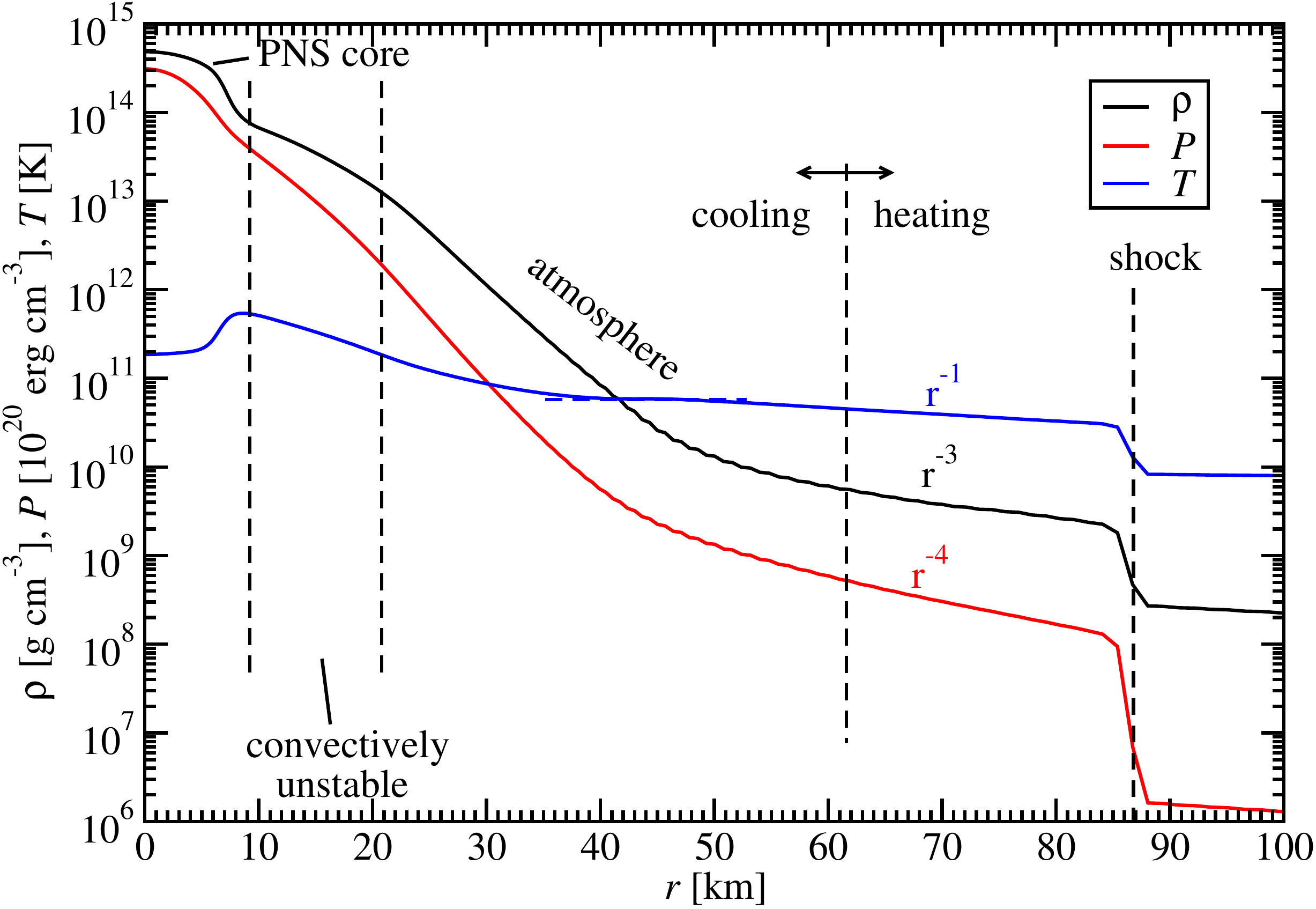}
    \hfill
    \includegraphics[width=0.5\linewidth]{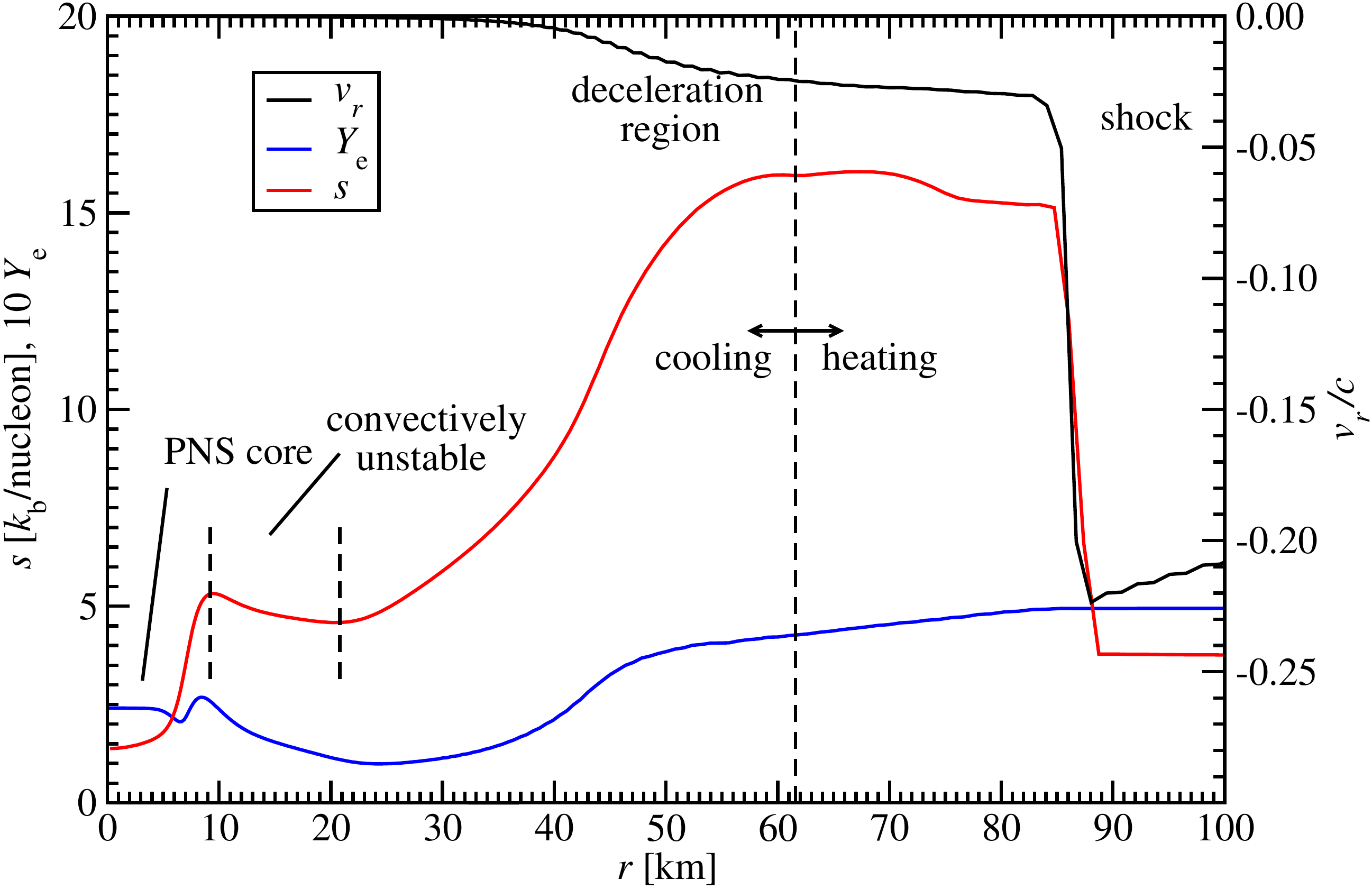}
    \caption{
    Schematic 1D structure of the supernova
    core after the formation of the gain region, illustrated by profiles of the density
    $\rho$, pressure $P$, temperature
    $T$ (left), radial velocity
    $v_r$, entropy $s$, and
    electron fraction $Y_\mathrm{e}$
    (right). The profiles are taken
    from a 1D radiation hydrodynamics
    simulation of
    the $20\,M_\odot$ progenitor
    of \citet{woosley_07} at a post-bounce
    time of $200\,\mathrm{ms}$. See text for details.}
    \label{fig:1d_struc}
\end{figure}

\subsection{Structure of the accretion flow and runaway conditions in spherical symmetry}
\label{sec:1d_struc}
Before analyzing the role of multi-D phenomena in 
core-collapse supernovae in greater depth, it is expedient to discuss
the structure of the supernova core that emerges
once the gain region has formed
a few tens of milliseconds after collapse in an idealized,
spherically-symmetric picture as shown in Fig.~\ref{fig:1d_struc}.
Our discussion closely follows the works of
\citet{janka_01,mueller_15a,mueller_16a} which may be consulted
for further details.

\paragraph{Structure of the accretion flow.}
At this stage, the PNS consists of an inner core of
about $0.5\,M_\odot$ (depending on the equation of state)
with low entropy, which is surrounded by an extended
mantle of about $1\,M_\odot$ that was heated to entropies
of about $6k_\mathrm{b}/\mathrm{nucleon}$ as the shock propagated
through the outer part of the collapsed iron core and
most of the Si shell. The mantle extends out to the
neutrinosphere at high subnuclear densities,
where neutrinos on average undergo
their last interaction before escape
 \citep[for details see][]{kotake_06,janka_17,mueller_19d,mezzacappa_19}. In the
atmosphere immediately outside the neutrinosphere
radius $R_\nu$, the pressure $P$ is 
dominated by non-relativistic baryons, and neutrino interactions
are still frequent enough to act as ``thermostat''
and maintain a roughly isothermal stratification, resulting
in an exponential density profile \citep{janka_01}:
\begin{equation}
    \rho = \rho_\nu \exp\left[-\frac{G M m_\mathrm{n}}{ r k_b T_\nu}\left (1-\frac{R_\nu}{r}\right)\right],
\end{equation}
where $M$ is the PNS mass, $R_\nu$, $T_\nu$, and $\rho_\nu$
are the neutrinosphere radius, temperature, and density,
and $m_\mathrm{n}$ is the neutron mass.
To maintain rough isothermality with the neutrinosphere,
the accreted matter must cool as it is advected through
the atmosphere. 
Below a density of about $10^{10}\, \mathrm{g}\, \mathrm{cm}^{-3}$,
the pressure is dominated by relativistic electron-positron pairs
and photons, and around this point neutrino heating
starts do dominate over neutrino cooling at the gain radius $R_\mathrm{g}$.\footnote{Properly speaking,
the EoS transition radius between
the baryon-dominated and the radiation-dominated regime
and the gain radius are  close, but
the gain radius is slightly larger \citep{janka_01}.
For many purposes it is not critical to distinguish them.}
Since the cooling and heating rate scale
with $T^6\propto P^{3/2}$ and $L_\nu E_\nu^2/r^2$ in terms
of the matter temperature $T$ and the electron-flavor neutrino luminosity $L_\nu$
and mean energy $E_\nu$ (appropriately
averaged over electron neutrinos and antineutrinos),
balance between heating and cooling defines an
effective thermal boundary condition
for the radiation-dominated gain region further out,
\begin{equation}
P^{3/2}\propto
\frac{L_\nu E_\nu^2}{R_\mathrm{g}^2}.
\end{equation}
Before shock revival, the stratification between the gain radius is
roughly adiabatic out to the shock\footnote{This is because neutrino heating
does not change the entropy appreciably as material traverses
the gain region as long as the heating conditions are
far from critical. Furthermore, mixing reduces
the entropy gradient in 3D once convection or SASI have developed.}, resulting in power-law profiles
$\rho\propto r^{-3}$ and $T\propto r^{-1}$ for the temperature
and density.
Ahead of the shock, the infalling
material moves with a radial
velocity of $|v_r|\approx \sqrt{GM/r}$ (i.e.,
a large fraction of the free-fall velocity),
and the
density is given in terms
of the mass acrretion rate $\dot{M}$ as $\rho= \dot{M}/(4\pi r^2 |v_r|)$. In a quasi-stationary situation, the stalled
accretion shock will adjust to a radius  $R_\mathrm{sh}$ such that
the jump conditions are fulfilled
and the post-shock pressure $P_\mathrm{sh}$ and
the pre-shock ram pressure $P_\mathrm{ram}=\rho v_r^2$ are
related by
\begin{equation}
\label{eq:shock}
P_\mathrm{sh}=
\frac{\beta}{\beta-1} P_\mathrm{ram}
=
\frac{\beta}{\beta-1} \frac{\dot{M}}{4\pi  R_\mathrm{sh}^2}\sqrt{\frac{GM}{ R_\mathrm{sh}}}
\propto \dot{M} R_\mathrm{sh}^{-5/2}.
\end{equation}
Here $\beta$ is the compression ratio in the shock, which 
varies from $\beta\approx 10$ early on, which is slightly
larger than the value of $\beta=7$ for an ideal gas
with adiabatic index $\gamma=4/3$ because of the dissociation
of nuclei in the shock, to $\beta\approx 4$ during the
explosion phase when there is a net release of energy
by burning in the shock.
Equation~(\ref{eq:shock}) immediately implies that
the  quasi-stationary accretion shock radius
increases with the post-shock pressure roughly as
$R_\mathrm{sh}\propto P_\mathrm{sh}^{2/5}$.
Recognizing that the heating
rate $\dot{q}_\mathrm{heat}\propto L_\nu T_\nu^2 r^{-2}$
and the cooling rate
$\dot{q}_\mathrm{cool}\propto T^6 \propto P^{3/2}$
balance at the gain radius, and using the adiabatic stratification
in the gain region and the jump conditions, one can
go further and derive that the shock radius scales 
as \citep{janka_12,mueller_15a}
\begin{equation}
\label{eq:rsh}
    R_\mathrm{sh}
    \propto \frac{(L_\nu T_\nu^2)^{4/9} R_\mathrm{g}^{16/9}}{\dot{M}^{2/3} M^{1/3}},
\end{equation}
in spherical symmetry in terms of $L_\nu$, $T_\nu$, $R_\mathrm{g}$,
$M$ and the mass accretion rate $\dot{M}$, which is
related to the density profile of the progenitor
\citep{woosley_12,mueller_16a}. 

\paragraph{Conditions for shock revival.}
So far, we have assumed a stationary accretion flow
in this picture. The problem of shock revival is,
however, related to the \emph{breakdown}
of stationary accretion solutions \citep{burrows_93,janka_01},
or more strictly speaking, to the development
of non-linearly unstable flow perturbations \citep{fernandez_12}. The transition
to runaway shock expansion can be understood
in terms of a competition of time scales, namely the advection or residence time
$\tau_\mathrm{adv}$ that the accreted material
spends in the gain region, and
the time scale $\tau_\mathrm{heat}$
for unbinding the material in the gain region
by neutrino heating \citep{thompson_00,janka_01,thompson_05,buras_06b,murphy_08b}. These can be computed in terms of
the binding energy $E_\mathrm{g}$ and mass $M_\mathrm{g}$ of the gain region, the
mass accretion rate $\dot{M}$, and the volume-integrated
heating rate $\dot{Q}_\nu$ as
\begin{align}
    \tau_\mathrm{adv}&=\frac{M_\mathrm{g}}{\dot M},
    \\
    \tau_\mathrm{heat}&=\frac{|E_\mathrm{g}|}{\dot Q_\nu}.
\end{align}
Transition to runaway expansion is expected
if $\tau_\mathrm{adv}/ \tau_\mathrm{heat}\gtrsim 1$,
which is borne out
 by 1D light-bulb simulations
 with a pre-defined neutrino luminosity \citep{fernandez_12}. Alternatively, the runaway condition can  be expressed in terms of
 a critical luminosity $L_\mathrm{crit}$  above which there are no  stationary 1D accretion solutions
 \citep{burrows_93}.  \citet{janka_12} and \citet{mueller_15a} have pointed out that
 these two descriptions are essentially equivalent
 by converting the time-scale criterion
 into a power law for the critical value of the
 ``heating functional'' 
 $\mathcal{L}=L_\nu E_\nu^2$,
\begin{equation}
\label{eq:lcrit1d}
\mathcal{L}_\mathrm{crit}=
(L_\nu E_\nu^2)_\mathrm{crit}
\propto (M \dot{M})^{3/5} R_\mathrm{g}^{-2/5}.
\end{equation}
Other largely equivalent ways to characterize
the onset of a runaway instability (at least
in spherical symmetry) are the notion that
the Bernoulli parameter reaches zero
somehwere in the gain region around
shock revival \citep{burrows_95,fernandez_12},
and the antesonic condition $c_\mathrm{s}^2>3/8 GM/r$
\citep{pejcha_12},
which is effectively a condition for
the flow enthalpy just like the Bernoulli parameter \citep{mueller_16b}.

\subsection{Impact of multi-dimensional effects
on shock revival}
\label{sec:dynamical_role}
\paragraph{Qualitative description.}
This simple picture is useful for qualitatively understanding  how 
multi-D effects modify the spherically-averaged bulk structure
of the post-shock flow and hence affect the conditions for shock 
revival. It is intuitive from Eq.~(\ref{eq:shock}) that
increasing the post-shock pressure (e.g., by turbulent 
heat transfer), or adding turbulent or magnetic stresses will
increase the shock radius and modify Eq.~(\ref{eq:rsh})
for the spherically symmetric case. This will 
then affect the time scales $\tau_\mathrm{adv}$ and 
$\tau_\mathrm{heat}$ and thereby modify the conditions for
runaway shock expansion driven by neutrinos. Moreover,
certain multi-D phenomena may also facilitate runaway shock 
expansion more directly by dumping extra energy into the gain region,  which may take the form of thermal
energy, turbulent kinetic energy, or magnetic energy.
This is, of course, only a coarse-grain interpretation
of the effect of multi-D effects, which needs to be
based on a more careful analysis of the underlying
hydrodynamic phenomena.

 The studies from the
1990s also outlined qualitative explanations for the beneficial
role of multi-D effects. \citet{herant_94} interpreted convection
as of an open-cycle heat engine that continuously pumps transfers
energy from the gain radius (where neutrino heating is strongest)
further out into the gain region, and \citet{janka_96} similarly
stress the importance of more effective heat transfer
from the gain region to the shock. 
\citet{herant_94} argued that large-scale mixing motions
are also advantageous because they continue to channel fresh matter
to the cooling region during the explosion phase so that
the neutrino heating is not quenched when the shock is revived.
Finally, \citet{burrows_95} pointed out what we now
subsume under the notion of turbulent stresses:
As the convective bubbles collide with
the shock surface with significant velocities (or in modern
parlance provide ``turbulent stresses'') and thereby deform and expand it.

\paragraph{Modified critical luminosity.}
Since then, the impact of multi-D effects has
been analyzed more quantitatively. Several studies
\citep{buras_06b,murphy_08b,hanke_12}
showed that the advection time scale
$\tau_\mathrm{adv}$ is systematically larger in multi-D,
while the onset of runaway shock expansion is still
determined by the criterion $\tau_\mathrm{adv}/\tau_\mathrm{heat}$. This suggests
that the runaway is still powered by neutrino heating
just as in 1D, and that
multi-D effects facilitate explosions facilitate
shock revival by somewhat expanding the stationary
shock, keeping a larger amount of mass in the
gain region, and thereby increasing the heating efficiency.\footnote{Note that this refers to a comparison
of multi-D and 1D models for a given set
of parameters of the accretion flow
($L_\nu$, $E_\nu$, $M$, $\dot{M}$, and $R_\mathrm{g}$.
When comparing multi-D and 1D models 
at the threshold to explosion
(with different $L_\nu$ and  $E_\nu$), the
heating efficiency can be lower in multi-D \citep{couch_14},
but this does not  mean that there is a different
runaway mechanism \citep{mueller_16b}.} To a lesser extent, mixing also reduces the binding energy of the gain region 
\citep{mueller_16b}, but this appears to be of secondary
importance for shock revival.

Building on the 1D picture from Sect.~\ref{sec:1d_struc}, the increase
of the quasi-stationary shock radius can be understood
as the consequence of additional ``turbulence''\footnote{It is important to stress that
``turbulence'' is something of a convenient
misnomer in this context and refers to any deviation from
quasi-stationary, spherically-symmetric flow. This should
be carefully distinguished from the usual
notion of turbulence in high-Reynolds number flow,
although the two concepts are frequently
conflated.} terms
that arise in a spherical Reynolds or Favre decomposition
of the flow, whose importance can be gauged
by the square of the turbulent Mach number $\mathrm{Ma}$
in the gain region \citep{mueller_12b}.
Using light-bulb simulations, \citet{murphy_12} demonstrated
quantitatively that the inclusion of Reynolds stresses
(``turbulent pressure'')
largely accounts for the higher shock radius in multi-D
models. The critical role of the turbulent pressure
was confirmed by \citet{couch_14} using a leakage
scheme and by \citet{mueller_15a} with multi-group neutrino transport.

The resulting effect on the critical
luminosity can be estimated by including
a turbulent pressure term\footnote{An alternative
approach to account for
the effect of the turbulent pressure is
to use an effective adiabatic index $\gamma>4/3$
in the gain region \citep{radice_15}.} in 
Eq.~(\ref{eq:shock}), which ultimately
leads to \citep{mueller_15a}
\begin{equation}
\label{eq:lcrit3d}
(L_\nu E_\nu^2)_\mathrm{crit}
\propto
(M \dot M )^{3/5} R_\mathrm{g}^{-2/5}
\left(1+\frac{4 \mathrm{Ma}^2}{3}\right)^{-3/5}
=
(L_\nu E_\nu^2)_\mathrm{crit,1D}
\left(1+\frac{4 \mathrm{Ma}^2}{3}\right)^{-3/5},
\end{equation}
where the  critical luminosity in 1D,
$(L_\nu E_\nu^2)_\mathrm{crit,1D}$, is
modified by a correction factor containing
the turbulent Mach number in the gain region.
Although based on a rather simple analytic model,
Eq.~(\ref{eq:lcrit3d}) describes shock
revival in 2D \citep{summa_16}
and 3D models \citep{janka_16} remarkably well. This suggests
that the critical parameter for increased explodability
in multi-D is indeed the turbulent Mach number,
although, as argued by \citet{mabanta_18},
the larger accretion shock radius may not
be due to turbulent pressure alone. Even if other effects such
as turbulent heat transport, turbulent dissipation
\citep{mabanta_18},
and even turbulent viscosity \citep{mueller_19b}
play a role, one expects a scaling law similar to 
Eq.~(\ref{eq:lcrit3d}) simply because
\emph{any} leading-order correction to the
1D jump condition (\ref{eq:shock}) from
a spherical Reynolds decomposition will scale
with $\mathrm{Ma}^2$, only with a slightly
different proportionality constant than
in Eq.~(\ref{eq:lcrit3d}).
The turbulent Mach number itself will be determined
by the growth and saturation mechanisms
of the non-radial instabilities in the gain region
as discussed in the following sections.

\subsection{Neutrino-driven convection in the gain region}
\label{sec:nu_conv}
Convection in the gain region develops because neutrino
heating establishes a negative entropy gradient. In many
respects, this ``hot-bubble convection'' resembles 
convection on top of a quasi-hydrostatic spherical background structure as familiar from the earlier phases of stellar evolution,
but there are subtle differences because the instability
occurs in an accrretion flow.

\paragraph{Condition for instability.}
Under hydrostatic conditions,  the Ledoux criterion for convective instability can be written as \citep{buras_06a},
\begin{equation}
C_\mathrm{L}
=
\frac{\pd \rho}{\pd r}-\frac{1}{c_\mathrm{s}^2}\frac{\pd P}{\pd r}
=
\left(\frac{\pd \rho}{\pd s}\right)_{P,Y_\mathrm{e}}
 \frac{\pd  s}{\pd r}
+ \left(\frac{\pd \rho}{\pd Y_\mathrm{e}}\right)_{P,s}
 \frac{\pd  Y_\mathrm{e}}{\pd r}
>0,
\end{equation}
in terms of the gradients of density, pressure,
entropy $s$ and electron fraction $Y_\mathrm{e}$.
Using a local stability analysis for a displaced
blob, one finds a growth rate $\mathrm{Im}\,\omega_\mathrm{BV}$,
where the Brunt-V\"ais\"al\"a frequency $\omega_\mathrm{BV}$
is defined as
\begin{equation}
    \omega_\mathrm{BV}^2=-\frac{g C_\mathrm{L}}{\rho}.
\end{equation}

In a stationary accretion flow, the radial derivatives
can be expressed in terms of the time derivatives
$\dot{s}$
and $\dot{Y}_\mathrm{e}$ and the advection velocity
$v_r$,
\begin{equation}
C_\mathrm{L}
=
\frac{1}{v_r}
\left[\left(\frac{\pd \rho}{\pd s}\right)_{P,Y_\mathrm{e}}
\dot{s}
+ \left(\frac{\pd \rho}{\pd Y_\mathrm{e}}\right)_{P,s}
 \dot{Y}_\mathrm{e}
 \right]
>0.
\end{equation}
Once the gain region forms around $80$--$100\, \mathrm{ms}$
after bounce, the electron fraction gradient plays a minor
role for stability, and the material in the gain region
can be well described as a radiation-dominated
gas with $P\propto (\rho s)^{4/3}$ so that
\begin{equation}
    \omega_\mathrm{BV}^2=-\frac{g \dot{q}_\mathrm{e}}
{v_r c_\mathrm{s}^2}.
\end{equation}
This has an important consequence: Different from
a hydrostatic background, the stability of a heated
accretion flow (or outflow) depends on the sign
of the advection velocity rather than on the profile
of the heating function. Using the aforementioned scaling
for the heating rate and assuming a linear velocity
profile behind the shock, we obtain an estimate
\begin{equation}
\omega_\mathrm{BV}^2
\propto
\frac{GM}{R_\mathrm{g}^2}\frac{\langle L_\nu E_\nu^2\rangle}{R_\mathrm{g}^2}
\left(\frac{\beta R_\mathrm{sh}}{G M}\right)^{1/2}
\left(\frac{R_\mathrm{sh}}{R_\mathrm{g}}\right)
.
\end{equation}

More importantly, however, advection can stabilize
the flow against convection because perturbations only
have a finite time to grow as they cross the gain region
as pointed out by \citet{foglizzo_06}, who demonstrated
that instability is regulated by a parameter
$\chi$,
\begin{equation}
\chi = \int\limits_{r_\mathrm{g}}^{r_\mathrm{sh}}\frac{\mathrm{Im}\,\omega_\mathrm{BV}}{|v_r\
|}\ud r .
\end{equation}
Instability
should only occur for $\chi \gtrsim 3$. This has indeed been
confirmed in a number of
 parameterized \citep{scheck_08,fernandez_14a,fernandez_15,couch_14b} and
 self-consistent simulations \citep{mueller_12b,hanke_13} in 2D and 3D.

\paragraph{Dominant eddy scale.}
Similar to the situation in convective shell burning,
the length scale of the most unstable linear mode is determined
by the width of the gain region according to
Eq.~(\ref{eq:ell_0}) \citep{foglizzo_06}. In 3D this remains the characteristic
length scale of convective eddies during the non-linear
saturation stage \citep[e.g.,][]{hanke_13}.
Since the ratio
of the shock and gain radius typically lies in the
range
$R_\mathrm{sh}/R_\mathrm{g}=1.5$--$2$
before the heating conditions become close
to critical, convection is characterized by
medium-scale eddies with angular wavenumbers
$\ell \approx 4$--$8$
during the accretion phase \citep{hanke_13,couch_14b}.
Around and after shock revival, large-scale modes
with $\ell=1$ and $\ell=2$ emerge.
By contrast, 2D simulations of
hot-bubble convection tend to develop large-scale 
($\ell=1$ and $\ell=2$) vortices during
the non-linear stage \citep{hanke_12,hanke_13,couch_12b,couch_14b} as a result of the
inverse turbulent cascade of
2D turbulence \citep{kraichnan_67}.

\paragraph{Non-linear saturation.}
The evolution towards shock revival typically proceeds over
sufficiently long time scales for hot-bubble convection 
to reach a quasi-stationary state.
Using 2D light-bulb simulations, \citet{murphy_12}
first demonstrated that this quasi-stationary
state closely mirrors
the situation in stellar convection
(cf.\ Sect.~\ref{sec:interior_flow}), i.e.,
neutrino heating, buoyant driving, and turbulent dissipation
balance each other \citep[see also][]{murphy_11}, and as a result the convective luminosity scales with the neutrino heating rate.
Alternatively, the quasi-stationary state can be characterized
by the notion of marginal stability; the flow adjusts itself
that such that the $\chi$-parameter for the spherically averaged
flow converges to $\langle \chi \rangle \approx 3$
\citep{fernandez_14a}. \citet{mueller_15a} showed that these
properties of the non-linear stage result in
a scaling law for the convective velocity that
is completely analogous to Eq.~(\ref{eq:vconv}).
In 2D simulations, the  velocity perturbations
$\delta v$ scale as
\begin{equation}
\label{eq:dv_conv2d}
    \delta v
    =(\dot{q}_\nu(R_\mathrm{sh}-R_\mathrm{g}))^{1/3},
\end{equation}
in terms of the average mass-specific neutrino  heating rate
$\dot{q}_\nu$ in the gain region. In 3D,
the convective velocities are slightly smaller \citep{mueller_16b},
\begin{equation}
\label{eq:dv_conv}
    \delta v
    =0.7[\dot{q}_\nu(R_\mathrm{sh}-R_\mathrm{g})]^{1/3}.
\end{equation}
The smaller proportionality constant in 3D can be motivated
by the tendency of the forward turbulent cascade to create
smaller structures, which decreases the dissipation length
and increases the dissipation rate of the flow.

\paragraph{Quantitative effect on shock revival.}
Based on these scaling laws for the convective
velocity, \citet{mueller_15a} determined that
convective motions should reach a characteristic
squared turbulent Mach number $\mathrm{Ma}\sim 0.3$--$0.45$ around the time of shock revival.
Using this value in 
Eq.~(\ref{eq:lcrit3d}) for the modified critical luminosity, they predict a reduction
of the critical luminosity by 15--25\%
due to convection, which is in the ballpark
of the numerical results
\citep{murphy_08b,hanke_12,couch_12b,dolence_13,fernandez_14a,fernandez_15}

One might
also be tempted to use
Eqs.~(\ref{eq:dv_conv2d})
and (\ref{eq:dv_conv})
to explain the lower explodability
of self-consistent 3D models compared to their
2D counterparts. The nature of the differences between 
2D and 3D is more complicated, however, since the critical luminosity for shock revival is roughly \emph{equal} in 2D and 3D light-bulb simulations.  Evidently, there are effects
that partly compensate for the smaller convective velocities
in 3D in some situations: The forward turbulent cascade \citep{melson_15a} and the different behavior 
of the  Kelvin--Helmholtz instability in 3D  \citep{mueller_15b}
affect the interaction between updrafts and downdrafts
and  can result in reduced cooling in 3D \citep{melson_15a}.
Moreover, compatible with earlier studies of
the Rayleigh--Taylor instability for planar
geometry \citep{yabe_91,hecht_95},
\citet{kazeroni_18} found a faster growth of
convective plumes and more
efficient mixing in a planer toy model of neutrino-driven
convection. Along similar lines,
\citet{handy_14} appealed to the higher volume-to-surface
ratio of convective plumes in 3D to explain the
reduced critical luminosity in 3D in their light-bulb 
simulations.
It is plausible that these factors establish 
a similar critical luminosity threshold for shock revival
in light-bulb simulations, but they do not explain the much
more decisive effect of dimensionality in self-consistent
simulations. One possible explanation lies in the fact
that explosions in self-consistent models usually occur
in a short non-stationary phase with a rapidly decreasing
mass accretion rate and neutrino luminosity around the infall of the Si/O shell interface; under these conditions the more
sluggish emergence of large-scale modes in 3D due to the
forward cascade may delay or inhibit shock revival
\citep{lentz_15}. Moreover, the more rapid response
of the mass accretion rate to shock expansion
in 3D \citep{melson_15a,mueller_15b} might be hurtful
around shock revival because this effect can reduce
the accretion luminosity and hence undercut neutrino
heating before a runaway situation can develop.

\paragraph{Resolution dependence and turbulence.}
Because of the turbulent nature of neutrino-driven
convection, the spectral properties of the flow
and the the resolution dependence in simulations
have received considerable attention
in the literature. Most self-consistent
models with multi-group neutrino transport
can only afford a limited resolution
(about $1.5$--$2^\circ$ in angle
and about 100 zones or less in the gain region)
and do not reach a fully developed turbulent
state, with \citet{handy_14} going so far
as to speak of ``perturbed laminar flow''
instead. Various authors \citep{abdikamalov_15,radice_15,radice_16} have argued
that considerably higher resolution is needed
to obtain clean turbulence spectra with
a developed inertial range and a Kolmogorov
spectrum and raised concerns
that a pile-up of kinetic energy
(``bottleneck effect'') at small scales might 
affect the overall dynamics.
However, the detailed spectral properties of the flow
are usually not critical, and integral
properties of the flow are more important for
the impact of convection on shock revival.
The resolution dependence nonetheless remains
a concern for the question of shock revival,
as  some 3D resolution studies \citep{hanke_12,abdikamalov_15,roberts_16} found a trend
towards decreasing explodability with increased
resolution. Recent work by \citet{melson_19}
resolved most of these concerns in a
resolution study using light-bulb simulations. They demonstrated
that the resolution dependence in \citet{hanke_12}
was a spurious effect connected to details
in their light-bulb scheme, and instead found
a trend towards increased explodability at
higher resolution. In rough agreement
with \citet{handy_14}, \citet{melson_19} found
that the overall flow dynamics converges
at an angular resolution of about $1^\circ$,
which is not far from what most self-consistent
simulations can afford (but one still needs
to bear in mind that the
resolution requirements depend on
the details of the numerical scheme,
cf.\ Sect.~\ref{sec:subsonic}). \citet{melson_19}
also pointed out that neutrino drag plays
a non-negligible role in the gain region,
so that merely increasing the resolution
does not add physical realism beyond numerical
Reynolds numbers of a few hundred unless
neutrino drag is also included as a non-ideal effect.
\citet{melson_19}  speculate that findings of decreased 
explodability with higher resolution in Cartesian 3D
models may be explained because the grid-induced
seed asphericities are lower at higher resolution.

\begin{figure}
    \centering
    \sbox0{\includegraphics{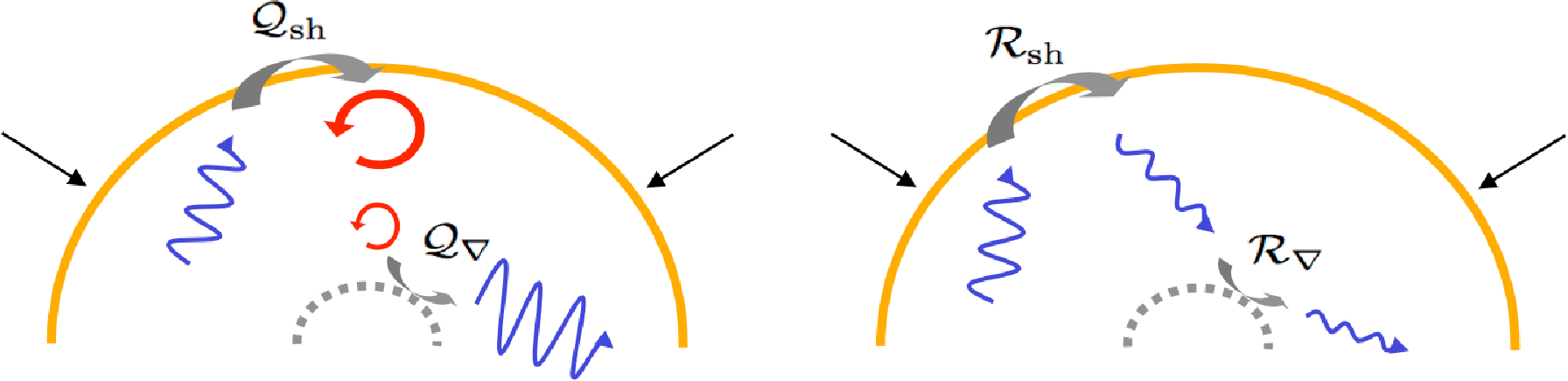}}
    \includegraphics[width=0.7 \linewidth,trim={0 0 {.5\wd0}  0},clip]{sasi_mech.pdf}
    \caption{The advective-acoustic mechanism
    for the standing accretion shock instability.
    Upward propagating acoustic waves
    (blue) generate vorticity perturbations
    (red) as they interact with the accretion
    shock (orange circle). The vorticity
    perturbations are advected downward
    with the accretion flow to the
    PNS surface where they generate
    acoustic waves due to advective-acoustic
    coupling in the steep density gradient.
    Instability for a given mode obtains
    if the product of the amplitude ratios 
    $\mathcal{Q}_\mathrm{sh}$
    and $\mathcal{Q}_\nabla$ for
   between ingoing and outgoing waves
   at the shock and PNS surface satisfies
   $\mathcal{Q}_\mathrm{sh} \mathcal{Q}_\nabla>1$.
    Image reproduced with permission from \citet{guilet_12}, 
    copyright by the authors.}
    \label{fig:sasi_mechanism}
\end{figure}

\subsection{The standing accretion shock instability}
\label{sec:sasi}
Using \emph{adiabatic} 2D simulations of spherical
accretion shocks, the seminal work of  \citet{blondin_03}
demonstrated that another instability,
dubbed ``SASI'' (standing accretion shock
instability), can  operate in the
supernova core even without a convectively unstable
gradient in the gain region. This instability
takes the form of large-scale ($\ell=1$ and sometimes $\ell=2$) oscillatory motions  of the shock,
and it was immediately realized that it can support
shock revival in a similar manner as convection.
In early 2D supernova simulations, the SASI was sometimes
confused with convection because the two phenomena
share superficial similarities like high-entropy
bubbles and low-entropy accretion downflows.
However, the SASI is set apart from convection by
dipolar (and sometimes quadrupolar) flow.
\citet{foglizzo_06} pointed out
that for the typical ratio between
the shock and gain radius in the pre-explosion
phase there are no unstable convective modes
with  $\ell=1,2$ in the gain region; instead
one finds $\ell \approx 4$--$8$ according
to Eq.~(\ref{eq:ell_0}), and for
$\chi<3$ the flow becomes stable against convection
altogether without strong perturbations
(see also Sect.~\ref{sec:nu_conv}).
This implies that a different instability mechanism
-- the one discovered by \citet{blondin_03} --
must be responsible for the $\ell=1$ and $\ell=2$
modes in supernova models with
small ratios $R_\mathrm{sh}/R_\mathrm{g}$.

\paragraph{Amplification mechanism.}
The stability of  accretion shocks
had in fact already been analyzed earlier in the
context of accretion onto compact objects
using linear perturbation theory
\citep{houck_92,foglizzo_01,foglizzo_02}, which provided
useful groundwork for identifying the physical
mechanism behind the SASI and explaining
the $\ell=1,2$ nature of the instability from
its dispersion relation. The accepted picture
is now that of a vortical-acoustic cycle
\citep{foglizzo_02,foglizzo_07}: Shock deformation
generates vorticity waves that are advected towards
the PNS surface, and due the deceleration
of the flow in the steep density gradient below
the gain region (see Fig.~\ref{fig:1d_struc}), these vorticity waves in turn
generate outgoing sound waves that again couple
to acoustic waves at the shock (Fig.~\ref{fig:sasi_mechanism}). Although
a purely acoustic amplification cycle has
been considered as well \citep{blondin_06}, analytical
\citep{laming_07,laming_08,yamasaki_07}
and numerical \citep{ohnishi_06,scheck_08,fernandez_09a,fernandez_09b} studies have on the whole supported
the advective-acoustic cycle, culminating
in the work of \citet{guilet_12} who sharpened
and summarized the arguments in favor of this amplification
mechanism.

Different from convection, the SASI is an oscillatory
instability with a periodicity 
$T_\mathrm{SASI}$ that is set by the sum of
the advective and acoustic crossing times
$\tau_\mathrm{adv}$
and $\tau_\mathrm{ac}$ between the shock
and the deceleration region
\citep{foglizzo_07},
\begin{equation}
\label{eq:sasi_timescale}
T_\mathrm{SASI}
=
\tau_\mathrm{adv}+\tau_\mathrm{ac}
=
\int_{r_\nabla}^{r_\mathrm{sh}} \frac{\ud r}{|v_r|}+
\int_{r_\nabla}^{r_\mathrm{sh}} \frac{\ud r}{c_s-|v_r|}.
\end{equation}
The advective time scale usually dominates, 
and neglecting a weak dependence on the PNS mass,
one can determine empirically that
the period of the $\ell=1$ mode of the SASI
roughly scales as \citep{mueller_14},
 \begin{equation}
\label{eq:sasi_period}
T_\mathrm{SASI} = 19 \, \mathrm{ms}
\left(\frac{R_\mathrm{sh}}{100 \, \mathrm{km}}\right)^{3/2}
\ln
\left(\frac{R_\mathrm{sh}}{R_\mathrm{PNS}} \right),
\end{equation}
where $R_\mathrm{PNS}$ is the PNS radius.
SASI-induced fluctuations in the neutrino emission
\citep{lund_10,tamborra_13,mueller_14,mueller_19a}
and gravitational waves \citep{kuroda_16b,andresen_17,kuroda_18}
could provide direct observational confirmation
for the SASI if this frequency can be identified
in spectrograms of the neutrino or gravitational wave signal.

The growth rate of the SASI is set both
by the period $T_\mathrm{SASI}$
and the quality factor $\mathcal{Q}$
of the amplification cycle \citep{foglizzo_06,foglizzo_07},
\begin{equation}
\omega_\mathrm{SASI}
=\frac{\ln |\mathcal{Q}|}{T_\mathrm{SASI}},
\end{equation}
where $\mathcal{Q}$ depends on the coupling 
between vortical and acoustic waves
at the shock and in the deceleration region,
and hence on the details of the density
profile and the thermodynamic stratification.
Nuclear dissociation and recombination
also affect the SASI growth rate 
and saturation amplitude
\citep{fernandez_09a,fernandez_09b}.

\paragraph{Interplay of SASI, convection, and neutrino heating.}
In reality, the SASI grows in an accretion flow
with neutrino heating, and in 2D,
it is not trivial at first glance to distinguish
SASI and convection in the non-linear phase where both 
instabilities lead to a  similar $\ell=1$ ``sloshing'' flow. Nonetheless, a clear distinction between
a SASI- and convection-dominated regime already
emerged in 2D models
using gray \citep{scheck_08} or multi-group
\citep{mueller_12b} neutrino transport, or
simpler light-bulb models
\citep{fernandez_14a}: Different from
convection-dominated models
SASI-dominated models clearly show an oscillatory
growth of the multipole coefficients of
the shock surface and coherent wave patterns
in the post-shock cavity in the linear regime, and
maintain a rather clear quasi-periodicity even
in the non-linear regime. 
The distinction
between the two different regimes tends to become
more blurred around shock revival, when large-scale
convective modes emerge and the periodicity of
the SASI oscillations is eventually broken.

The criterion
$\chi\approx 3$ roughly separates the two  regimes
even though unstable SASI modes can in principle exist
above this value. The reason is likely that
convection destroys the coherence of the
waves involved in the SASI amplification cycle \citep{guilet_10} if $\chi>3$. 
For $\chi<3$, high quality factors
$\ln |\mathcal{Q}|\sim 2$
can be reached and result in
rapid SASI growth. In terms of PNS parameters
and progenitor parameters, such low values of
$\chi<3$ are encountered in case of rapid PNS 
and shock contraction
\citep{scheck_08} and appear to occur preferentially
in high-mass progenitors with high mass accretion
rates \citep{mueller_12a}, although a detailed
survey of the progenitor-dependence of the $\chi$-parameter is still lacking.

Three-dimensional simulations 
with neutrino transport
\citep{hanke_13,tamborra_14b,kuroda_16b,mueller_17,ott_18,oconnor_18b}
as well as simplified leakage and light-bulb
models \citep{couch_14b,fernandez_15}
show an even  cleaner
distinction  between the SASI- and convection-dominated regimes for several reasons. The convective
eddies remain smaller in the non-linear stage than
in 2D because of the forward cascade, and without
the constraint of axisymmetry, the convective
flow is not prone to artificial oscillatory sloshing motions. The SASI, on the other hand, exhibits a cleaner
periodicity prior to shock revival in 3D,
and can develop a spiral mode that
is very distinct from convective flow 
(e.g., \citealp{blondin_07,fernandez_10,hanke_13};
see also 
Sect.~\ref{sec:compact_remnant} for possible
implications on neutron star birth periods).
Self-consistent models show that the post-shock
flow can transition back and forth between
the convection- and SASI-dominated regime
as the accretion rate and PNS parameters, and
hence the $\chi$-parameter change
\citep{hanke_13}.

\paragraph{Saturation mechanism.}
\citet{guilet_10} argued that parasitic
Kelvin--Helmholtz and Rayleigh--Taylor instabilities 
are responsible for the non-linear
saturation of the SASI, and showed
that this  mechanism
can explain the saturation amplitudes
in the adiabatic simulations of \citet{fernandez_09a}.
Assuming that the Kelvin--Helmholtz instability
is the dominant parasitic mode in 3D, one
can derive \citep{mueller_16b} a scaling law for
the turbulent velocity fluctuations $\delta v$
in the saturated state,
\begin{equation}
\label{eq:dv_sasi}
\delta v
\sim
\omega_\mathrm{SASI} (R_\mathrm{sh}-R_\mathrm{g})
\sim
\frac{\ln |\mathcal{Q}| (R_\mathrm{sh}-R_\mathrm{g})}
{\tau_\mathrm{adv}}
\sim 
\ln \mathcal{Q} |\langle v_r\rangle|,
\end{equation}
which is in good agreement with self-consistent
3D simulations. Interestingly,
this scaling results in similar
turbulent velocities as in the convection-dominated
regime for conditions typically encountered
in supernova core \citep{mueller_16b}.

The saturation of the SASI can also be understood
as a self-adjustment to marginal stability
\citep{fernandez_14a}, which is a closely related concept.
As the SASI grows in amplitude, the flow  is driven
towards $\langle \chi \rangle\approx 3$, but
stays slightly below this critical value \citep{fernandez_14a}.

\paragraph{Effect on shock revival.}
The SASI provides similar beneficial effects
as convection to increase the shock radius
and bring the accretion flow closer to
a neutrino-driven runaway, i.e., it generates
turbulent pressure, brings high-entropy bubbles
to large radii, channels cold matter towards
the PNS, and converts turbulent
kinetic energy thermal energy throughout
the gain region by turbulent dissipation.
Due to the different instability mechanism
(which feeds on the energy of the accretion flow
directly instead of the neutrino energy
deposition), and the different flow pattern
(which affects the rate of turbulent dissipation),
the quantitative effect on shock revival 
can be different from convection. Using light-bulb
simulations \citet{fernandez_15} indeed
found a significantly lower critical
luminosity in the SASI-dominated regime than
in the convection-dominated regime
and a lower critical luminosity
in 3D by $\sim 20\%$ compared to 2D, which 
he ascribed to the ability of the spiral mode
to store more kinetic energy than sloshing
modes in 2D. An even bigger difference
to convective models (albeit with a different
and very idealized setup) was found
by \citet{cardall_15}.
Self-consistent simulations, on the other hand,
have not found higher explodability in 3D
in the SASI-dominated regime \citep{melson_15b}.
The reason for this discrepancy could, e.g., lie
in the feedback of shock expansion on neutrino heating,
but is not fully understood at this stage.
\citet{cardall_15} also observed
considerably more stochastic variations
in shock revival in the SASI-dominated regime in their
idealized models (i.e., a smeared-out
critical luminosity threshold), but it again remains to be 
seen whether this is borne out by self-consistent
3D models, where the SASI oscillations tend to be
of smaller amplitude and shorter period than in
\citet{cardall_15}.

\subsection{Perturbation-aided explosions}
\label{sec:perturbations}
Progenitor asphericities from convective
shell burning can  aid shock revival by 
affecting  both the growth and  saturation of convection 
of the SASI. That a higher level of seed perturbations
leads to a faster growth of non-radial instabilities
behind the shock and thereby
fosters explosions (as in the early studies
of \citealt{couch_13,couch_15}) may be  intuitive, but
appears less important in practice. In self-consistent
simulations, shock revival typically occurs only once
convection or the SASI have already reached the
stage of non-linear saturation, and it is rather
the permanent ``forcing'' by infalling perturbations
that matters \citep{mueller_15a,mueller_17}. In either case, it is useful to separately
consider a) how the initial perturbations in the
porgenitor are translated to perturbations ahead
of the shock, and b) how the infalling perturbations
interact with the shock and the post-shock flow.

\paragraph{Initial state and infall phase.}
Typically, the Si and O shell (and sometimes
a Ne shell) are the only active convective shells
that can reach the shock at a sufficiently early
post-bounce time to affect shock revival.
As described in Sect.~\ref{sec:presn}, these
shells are characterized by Mach numbers
$\mathrm{Ma}_\mathrm{prog} \sim0.1$ with significant variations
between different shells and progenitors, and
can have a wide range of dominant angular wave
numbers $\ell$. Due to its subsonic nature,
the flow is almost solenoidal with $\nabla\cdot (\rho \mathbf{v})\approx 0$, and density perturbations
$\delta \rho/\rho \sim \mathrm{Ma}_\mathrm{prog}^2$ are small
within convective zones. Viewed as a superposition
of linear waves, the convective flow consists mostly
of vorticity and entropy waves with little contribution
from acoustic waves.

From analytic studies of perturbed Bondi accretion 
flows in the limit $r \rightarrow 0$
 in a broader context \citep{kovalenko_98,lai_00,foglizzo_00}, it
is known that such initial perturbations
are amplified during infall, and that acoustic waves are generated
from the vorticity and entropy perturbations.
Estimating the pre-shock perturbations for the
problem at hand \citep{takahashi_14,mueller_15a,abdikamalov_19}
involves some subtle differences, but the upshot
is rather simple: Advective-acoustic
coupling generates strong acoustic perturbations
ahead of the shock that scale \emph{linearly}
with the convective Mach number at the pre-collapse
stage \citep{mueller_15a,abdikamalov_19},
\begin{equation}
\delta P/P \sim \delta \rho/\rho\sim \mathrm{Ma}_\mathrm{prog}.
\end{equation}
According to simulations \citep{mueller_17}
and analytic theory \citep{abdikamalov_19}, this
scaling is roughly independent of the wave number $\ell$.\footnote{If strong acoustic perturbations were
present at the pre-collapse stage, these
modes with higher $\ell$ would grow faster
during the linear
stage \citep{takahashi_14}, but quickly undergo non-linear
damping \citep{mueller_15a}.}

\begin{figure}
    \centering
    \includegraphics[width=\textwidth]{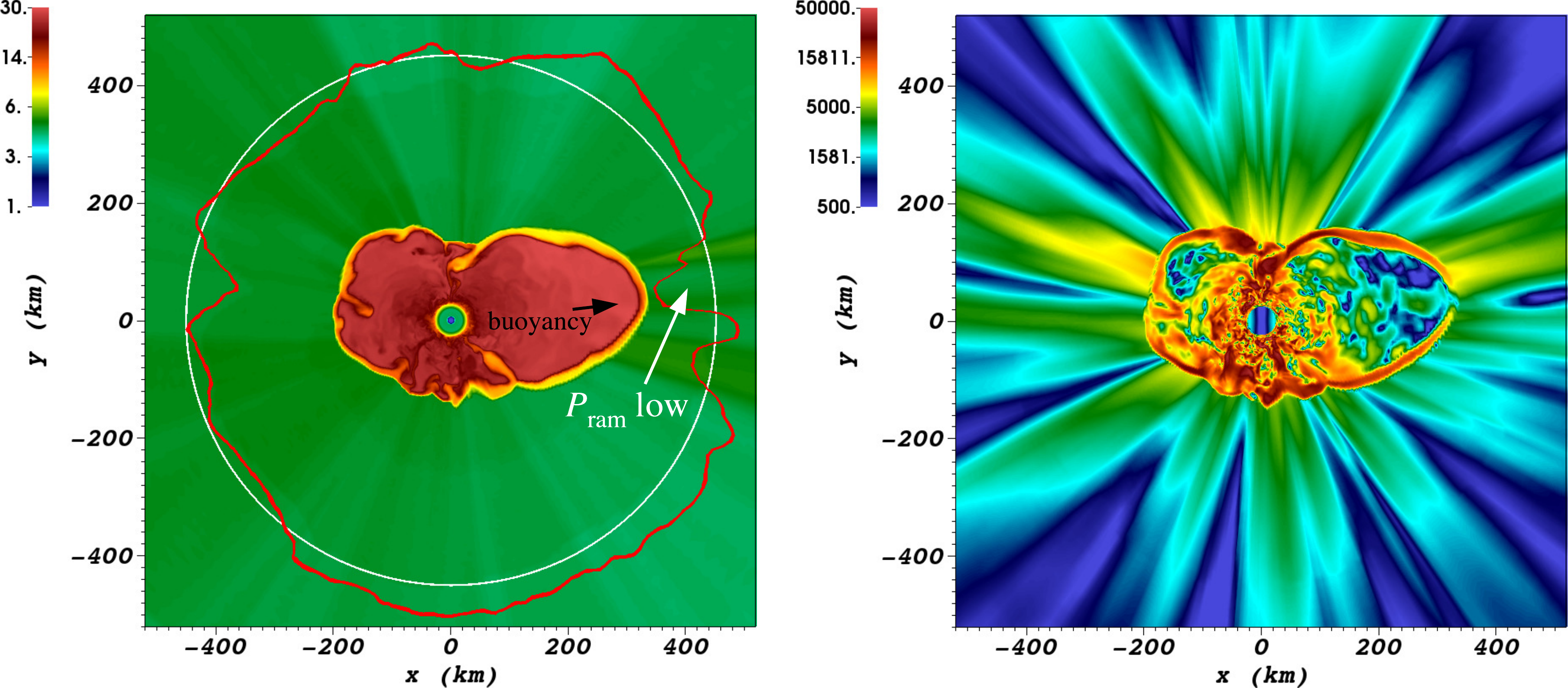}
    \caption{Interaction of infalling
    perturbations with the shock
    and the post-shock flow, illustrated
    by snapshots of the entropy
    (in units of $k_\mathrm{b}/\mathrm{nucleon}$,
    left panel) and the absolute value
    of the non-radial velocity (in units
    of $\mathrm{km}\, \mathrm{s}^{-1}$,
    right panel) in the $12.5\,M_\odot$
    model of \citet{mueller_19a} at a post-bounce
    time of $510 \, \mathrm{ms}$.
    The left panel also shows the deformation
    of the isodensity surface with
    $\rho = 7 \times 10^{6}\, \mathrm{g}\, \mathrm{cm}^{-6}$ (red curve).
    Due to the infalling  density perturbations, the pre-shock
    ram pressure is anisotropic
    and creates a protrusion of the shock.
    Additional
    energy is pumped into non-radial
    motions in the gain region both
    because of substantial lateral velocity perturbations
    ahead of the shock and because of the oblique
    infall of material through the deformed shock.
    }
    \label{fig:perturbations}
\end{figure}

\paragraph{Shock-turbulence interaction and forced shock deformation.}
The infalling perturbations affect the shock and the post-shock flow
in several ways \citep{mueller_15a,mueller_16c}. They provide a continuous
flux of acoustic and tranverse kinetic energy into the gain region,
and also create post-shock density perturbations that will be
converted into turbulent kinetic energy by buoyancy. Moreover,
the shock becomes deformed due to the anisotropic ram pressure
(Fig.~\ref{fig:perturbations}), which results in fast
lateral flow behind the shock, i.e., in the generation of additional
transverse kinetic energy. Thus, more violent turbulent
flow can be maintained in the gain region, which is conducive
to shock revival [cf.\ Eq.~(\ref{eq:lcrit3d})]. 
If the infalling perturbations are of large scale,
the deformation of the shock creates large and stable high-entropy bubbles.
This is also helpful for shock revival since runaway shock expansion
in multi-D appears to require the formation of such large bubbles
with sufficient buoyancy to rise and expand against the supersonic
drag of the infalling material \citep{fernandez_14a,fernandez_15}.

There is as yet no comprehensive quantitative theory for the interaction
of infalling perturbations with the shock and the instabilities
in the gain region, but several studies have investigated
aspects of the problem. Using order-of-magntitude estimates,
\citet{mueller_16c} argued that turbulent motions are primarily
boosted by the action of buoyancy on the injected post-shock
density perturbations. This hypothesis is supported by
controlled parameterized simulations of shock-turbulence
interactions in planar geometry \citep{kazeroni_19}.
\citet{mueller_16c} also attempted to derive a correction
for the saturation value of turbulent kinetic energy
depending on the convective Mach number $\mathrm{Ma}_\mathrm{prog}$
and wave number $\ell$ in the progenitor. They
predicted a reduction of the critical luminosity 
functional $\mathcal{L}_\mathrm{crit}=(L_\nu E_\nu^2)_\mathrm{crit}$ by
\begin{equation}
\label{eq:delta_lcrit}
\frac{\delta \mathcal{L}_\mathrm{crit}}{\mathcal{L}_\mathrm{crit}}
\approx 0.47 \frac{\mathrm{Ma}_\mathrm{prog}}{\ell \eta_\mathrm{heat} \eta_\mathrm{acc}}
\end{equation}
in terms of the heating efficiency $\eta_\mathrm{heat}$
and the accretion efficiency $\eta_\mathrm{heat}=L/(GM\dot{M}/R_\mathrm{g})$.
However, the analysis of \citet{mueller_16c}
did not account in detail for the interaction of the infalling
perturbations with the shock. This has been investigated using
linear perturbation theory \citep{takahashi_16,abdikamalov_16,abdikamalov_18,huete_18,huete_19}.
As a downside, this perturbative approach cannot easily capture the non-linear
interaction of the injected perturbations with fully developed
neutrino-driven convection and the SASI, but \citet{huete_18}
recently incorporated the effects of buoyancy downstream of
the shock. The more sophisticated treatment of \citet{huete_18}
predicts a similar effect size as Eq.~(\ref{eq:delta_lcrit}). 

\paragraph{Phenomenology of perturbation-aided
explosions.}
Whatever its theoretical
justification, Eq.~(\ref{eq:delta_lcrit}) successfully captures
trends seen in 2D and 3D simulations of perturbation-aided
explosions starting from parameterized initial
conditions or 3D progenitor models. Both high
Mach numbers $\gtrsim 0.1$
and large-scale convection with $\ell \lesssim 4$
are required for a significant beneficial effect
on the heating conditions \citep{mueller_15a}. In this
case, the perturbations can be the decisive 
factor for shock revival as in the $18\,M_\odot$ model
of \citet{mueller_17}. In leakage-based
models with high heating efficiency early on \citep{couch_13,couch_15},
the effect is smaller, especially if the pre-collapse
asphericities are restricted to medium-scale modes as
in octant simulations \citep{couch_15}. 

By now, there is a handful of exploding 
supernova models that use multi-group neutrino transport and 3D progenitor models
\citep{mueller_17,mueller_19a}. While this
is encouraging, more 3D simulations are needed
to determine to what extent convective
seed perturbations generally contribute
to robust explosions. At present, one
can nonetheless extrapolate the effect
size based on the properties of
convective shells in 1D stellar evolution
models using Eq.~(\ref{eq:delta_lcrit}). Analysing
over 2000 supernova progenitors computed with the \textsc{Kepler} code \citet{collins_18}
predict a substantial reduction of
the critical luminosity due to perturbation
by 10\% or more in the mass range
between $15\,M_\odot$ and $27\,M_\odot$,
and in isolated low-mass progenitors.
Below $15\,M_\odot$, the expected
reduction is usually 5\% or less, which
could still make the convective perturbations
one of several important ingredients
for robust explosions. In the vast majority
of progenitors, only asphericities from
oxygen shell burning are expected to
have an important dynamic effect.

\subsection{Outlook: Rotation and magnetic fields in neutrino-driven
explosions}
Earlier on, we already briefly touched
simulations of magnetorotational
explosion scenarios and the uncertainties
that still beset this mechanism. It
is noteworthy that rotation and magnetic
fields could also play a role within
the neutrino-driven paradigm.

\paragraph{Rotationally-supported explosions.}
Since early attempts to study the impact
of rotation on neutrino-driven explosions
either employed a simplified neutrino treatment \citep[e.g.,][]{kotake_03,fryer_03,nakamura_14,iwakami_14}
or were restricted to 2D in the case of
models with multi-group transport \citep{walder_05,marek_09,suwa_10},
more robust conclusions had to wait for  3D simulations
with multi-group neutrino transport \citep{takiwaki_16,janka_16,summa_18}.
The 3D simulations indicate that the overall
effect of rapid rotation is to support neutrino-driven
explosions. Centrifugal support reduces the infall velocities
and hence the average ram pressure at the shock
\citep{walder_05,janka_16,summa_18}. Moreover,
3D neutrino hydrodynamics simulations of rotating
models tend to develop
a strong spiral SASI
\citep{janka_16,summa_18}.
 This is in line with analytic theory 
\citep{yamasaki_08} and idealized simulations
\citep{iwakami_09,blondin_17,kazeroni_18}, which demonstrated
that rotation enhances the growth rate of the prograde
spiral mode and stabilises the retrograde mode.
For sufficiently
rapid rotation, an even more violent spiral corotation instability can
occur \citep{takiwaki_16}.
There is also a subdominant
adverse effect, since lower neutrino luminosities and mean energies at low latitudes close to the equatorial plane
are detrimental for shock revival \citep{walder_05,marek_09,summa_18}, which is particularly relevant since the explosion tends
to be aligned with the equatorial plane in the case of
rapid rotation \citep{nakamura_14}. 
\citet{summa_18} found that the overall combination
of these effects can be encapsulated by a further modification
of the critical luminosity,
\begin{equation}
\label{eq:lcrit_rot}
(L_\nu E_\nu^2)_\mathrm{crit}
=
(L_\nu E_\nu^2)_\mathrm{crit,1D}
\left(1+\frac{4 \mathrm{Ma}^2}{3}\right)^{-3/5}
\sqrt{1-\frac{j^2}{GM R_\mathrm{sh}}}.
\end{equation}
Here $j$ is the spherically-averaged angular momentum
of the shell currently falling through the shock.
The last factor accounts for the reduced pre-shock
velocities, and the effect of stronger non-radial
flow in the gain region is implicitly (but not
predictively) accounted for in the turbulent
Mach number.

\paragraph{Magnetic fields without rotation.}
Given the  expected pre-collapse
spin rates, rotation is unlikely to have a major
impact in the vast majority of supernova
explosions. It is harder to exclude 
a significant role of magnetic fields
\emph{a priori}. Even if the progenitor core
rotates slowly and does not have strong
magnetic fields, convection and the SASI
might furnish some kind of turbulent dynamo process 
that could generate dynamically relevant fields
in the gain region.
There could also be other processes to
provide dynamically relevant magnetic fields,
e.g., 
the accumulation of Alfv\'en waves 
at an Alfv\'en surface \citep{guilet_11}
or the injection of Alfv\'en waves generated
in the PNS convection zone \citep{suzuki_08}.

\begin{figure*}
  \includegraphics[width=0.49\textwidth]{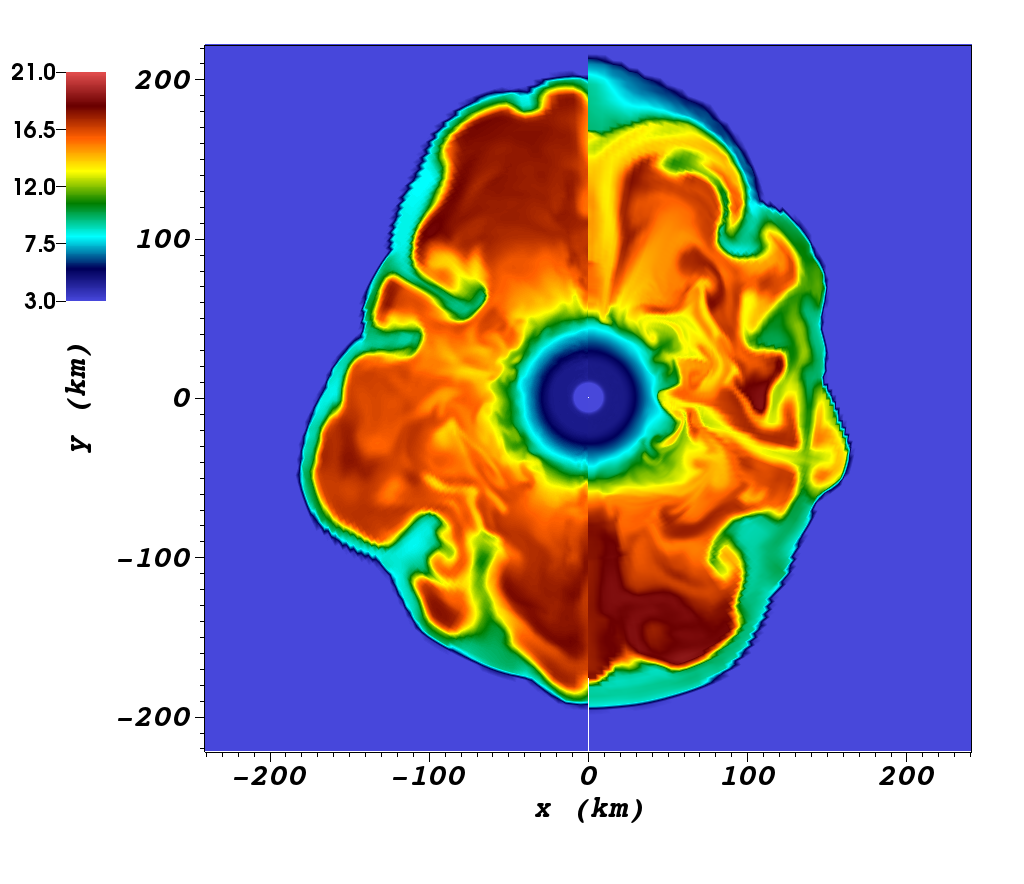}
\hfill
  \includegraphics[width=0.49\textwidth]{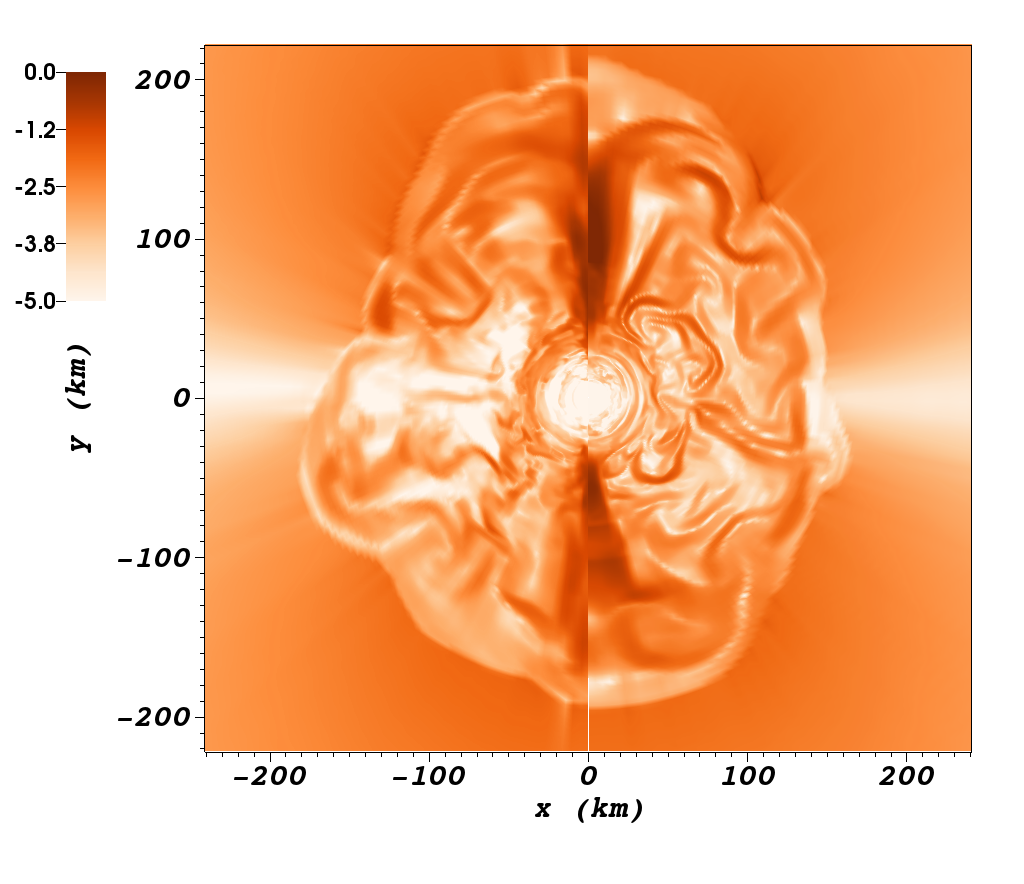}
\caption{Entropy $s$ in $k_\mathrm{b}/\mathrm{nucleon}$ (left panel)
and the logarithm $\log_{10} P_\mathrm{B}/P_\mathrm{gas}$
of the ratio between the magnetic pressure  $P_\mathrm{B}$
and the gas pressure $P_\mathrm{gas}$ (right panel) in a 3D simulation (left half
of panels) and a 2D simulation 
(right half of panels) of the slowly rotating progenitor
$15\,M_\odot$ progenitor m15b6 of \citet{heger_05}
with the \textsc{CoCoNuT-FMT} code.
The initial field is assumed to be combination of
a dipolar poloidal field and a toroidal field.
Outside convective zones, the field strength
is taken from the progenitor, inside convective
zones, the magnetic pressure is set
to a fraction of $10^{-4}$ of the thermal pressure.
The figures shows meridional
slices $140 \, \mathrm{ms}$ after bounce. Field amplification is
driven by convection. Strong fields are generated in regions
of strong shear, but these strong field are highly localized, and
the total magnetic energy in the gain region remains much smaller
than the turbulent kinetic energy and thermal energy.
}
\label{fig:mhd}      
\end{figure*}

The simulations available so far do not suggest
that sufficiently high field strengths can
be reached by a small-scale turbulent dynamo. In idealized 2D and 3D
simulations of \citet{endeve_10,endeve_12},
the SASI indeed drives a small-scale turbulent
dynamo, and strong field amplification
occurs locally up to equipartition and
super-equipartition field strengths, especially
when a strong spiral mode develops. On larger
scales, the magnetic field energy remains
well below equipartition, however, and
does not become dynamically important. The 
total magnetic energy in the gain region
remains one order of magnitude smaller
than the turbulent kinetic energy, and the
field does not organize itself into large-scale
structures. The situation is similar
in the 2D neutrino hydrodynamics simulations
of non-rotating progenitors of
\citet{obergaulinger_14} for initial
field strengths of up to $10^{11}\, \mathrm{G}$,
with even lower ratios between the
total magnetic and turbulent kinetic
energy in the gain region. Only for initial
field strengths of $\mathord{\sim}10^{12}\, \mathrm{G}$,
which yields magnetar-strength fields
after collapse,
do \citet{obergaulinger_14} find that magnetic
fields become dynamically important and accelerate
shock revival. If the fossil field hypothesis
for magnetars is correct and the fields of
the most strongly magnetized main-sequence stars
translate directly to supernova progenitor
and neutron star fields by flux conservation 
\citep{ferrario_06}, such conditions for
magnetically-aided explosions might be still realized
in nature in a substantial fraction of core-collapse
events.

Ultimately, a thorough exploration
of resolution effects and initial field configurations
in the convection- and SASI-dominated regime
will be required in 3D to confidently
exclude a major role of magnetic field in
weakly magnetized, slowly rotating progenitors.
First tentative results from
3D MHD simulations with neutrino transport
(Fig.~\ref{fig:mhd}) suggest a
picture of fibril flux concentrations with equipartition
field strengths, and sub-equipartition fields in most
of the volume, akin to the situation
in solar convection \citep[e.g.,][]{solanki_06}.

\begin{figure}
    \centering
    \includegraphics[width=0.48\textwidth]{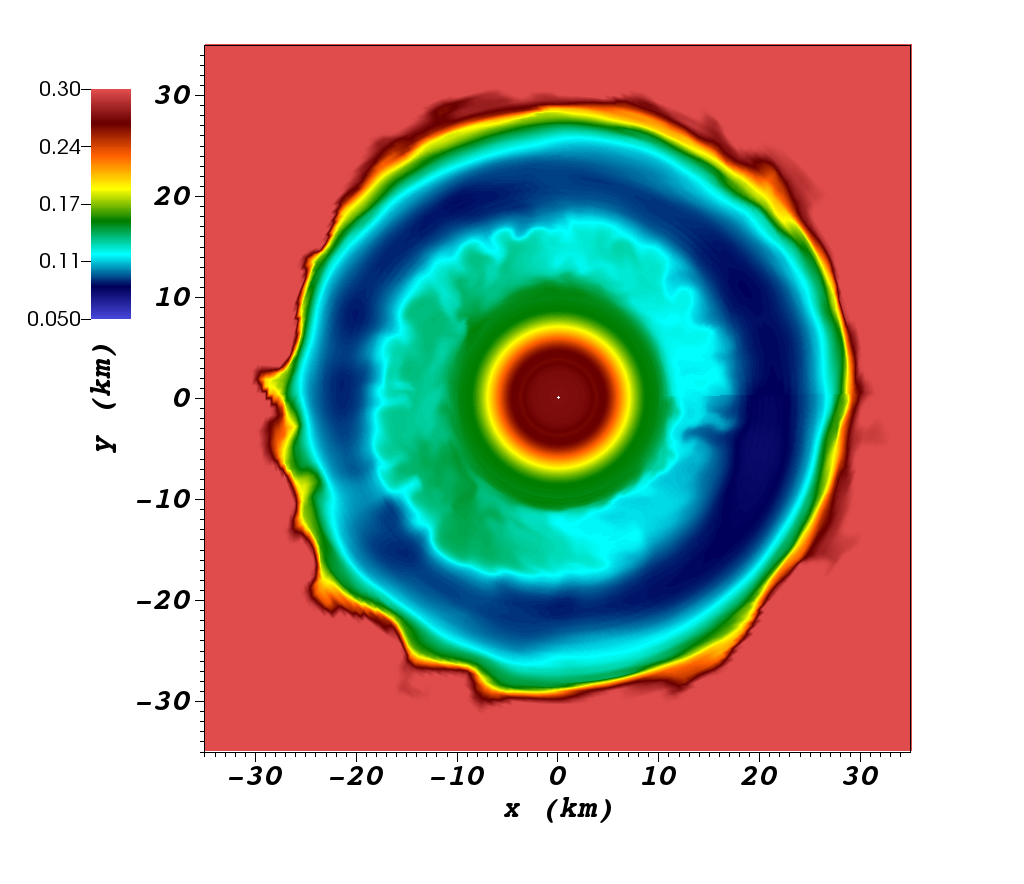}
    \hfill
\includegraphics[width=0.48\textwidth]{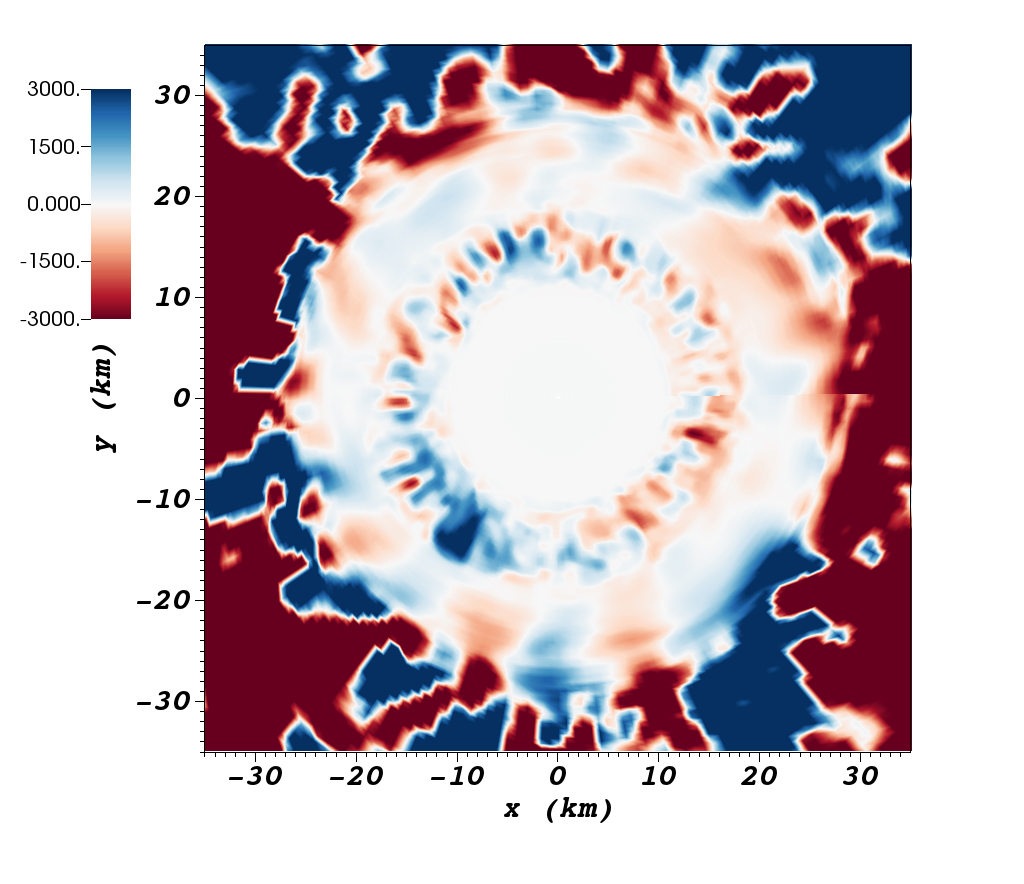}
\caption{LESA instability in a
simulation of an $18\,M_\odot$
star at time of $453 \, \mathrm{ms}$
after bounce, illustrated by
2D slices showing the electron fraction $Y_\mathrm{e}$
(left) and the radial velocity
in units of $\mathrm{km}\,\mathrm{s}^{-1}$
in the PNS convection zone. Note that the $Y_\mathrm{e}$
distribution in the PNS convection zone between
radii of $10\, \mathrm{km}$ and $20\, \mathrm{km}$
shows a clear dipolar asymmetry, whereas the
radial velocity field is dominated by small-scale modes
superimposed over a much weaker dipole mode.
Image repoduced with permission from \citet{powell_19},
copyright by the authors.
}
    \label{fig:lesa}
\end{figure}

\subsection{Proto-neutron star convection and LESA instability}
\label{sec:pns_convection}
\paragraph{Prompt convection.}
Convective instability also develops
inside the PNS. As already
recognized in the late 1980s \citep{bethe_87,bethe_90},
a first episode
of ``prompt convection'' occurs within milliseconds
after bounce around a mass coordinate
of $\sim 0.8\,M_\odot$ as the shock weakens and a negative
entropy gradient is established. The negative entropy
gradient is, however, quickly smoothed out, and
the convective overturn has no bearing on the
explosion mechanism, although it can leave
a prominent signal in gravitational waves
(see reviews on the subject; \citealp{ott_08b,kotake_13,kalogera_19}).

\paragraph{Proto-neutron star convection.}
Convection inside the PNS is triggered again
latter as neutrino cooling establishes
 unstable lepton number
\citep{epstein_79} and entropy gradients
(see profiles in Fig.~\ref{fig:1d_struc}).
PNS convection was investigated extensively
in the 1980s and 1990s as a possible means
of enhancing the neutrino emission from the PNS,
which would boost the neutrino heating and thereby
aid shock revival
\citep[e.g.,][]{burrows_87b,burrows_88b,wilson_88,wilson_93,janka_95,keil_96}. In particular, \citet{wilson_88,wilson_93}
assumed that PNS convection operates
as double-diffusive ``neutron finger'' instability that significantly increases the neutrino luminosity.

None of the modern studies of PNS convection since the mid-1990s
\citep{keil_96,buras_06b,dessart_06} found a sufficiently
strong effect of PNS convection on the neutrino emission
for a significant impact on shock revival.
PNS convection indeed increases the heavy flavor
neutrino luminosity by 
$\mathord{\sim} 20\%$ at post-bounce times
of $\gtrsim 150 \, \mathrm{ms}$, leaves the
electron neutrino luminosity about equal, but tends
to decrease the electron antineutrino luminosity,
and reduces the mean energy of all neutrino flavors \citep{buras_06b}. This can be explained by
the  effects of PNS convection on the bulk structure
of the PNS, namely a modest increase of the PNS
radius and a higher electron fraction (due to
mixing) close to the neutrinosphere of $\nu_e$
and $\bar{\nu}_e$ \citep{buras_06b}. Convective
instability appears to be governed by the
usual Ledoux criterion and does not develop
as a double-diffusive instability in the simulations.

\paragraph{LESA instability.}
Although PNS convection does not have a decisive
influence on shock revival, its indirect effect
on the gain region is quantitatively
important; effectively PNS convection changes
the inner boundary condition for the flow in the gain region.
PNS convection also has an important impact
on the neutrino signal from the PNS cooling
phase \citep[e.g.,][]{roberts_12b,mirizzi_16},
and may provide a sizable contribution
to the gravitational wave signal
\citep{marek_08,yakunin_10,mueller_13,andresen_17,morozova_18}.

Moreover, starting with \citep{tamborra_14a},
the dynamics of PNS convection has proved more
intricate upon closer inspection in recent years with potential repercussions on nucleosynthesis and gravitational wave emission.
In their 3D simulations, \citet{tamborra_14a}
noted that a pronounced $\ell=1$
asymmetry in the electron fraction develops
in the PNS convection zone, which leads to a sizable
anisotropy in the radiated lepton number
flux, i.e., the fluxes of electron neutrinos
and antineutrinos show a dipole asymmetry with opposite
directions. The direction of the dipole varies
remarkably slowly compared to the characteristic
time scales of the PNS. Curiously, no such pronounced dipole was
seen in the velocity field, which remained dominated
by small-scale eddies. This is very different
from convection in the gain region or the convective
burning in the progenitors, where the asymmetries
in the entropy and the composition are reflected
in the velocity field as well.
 This unusual phenomenon,  which
is illustrated in Fig.~\ref{fig:lesa}, 
has been christened LESA (``Lepton
Number Emission Self-Sustained Asymmetry''), and
could have
important repercussions on the composition of
the ejected matter, whose $Y_\mathrm{e}$ is sensitive
to the differences in electron neutrino and antineutrino
emission.

Since then this phenomenon has been reproduced by
many 3D simulations using very different methods
for neutrino transport
\citep{oconnor_18b,janka_16,glas_19,powell_19,vartanyan_19b},
and even in the 3D leakage models of \citet{couch_14b}
The dipolar $Y_\mathrm{e}$ asymmetry has also been seen in the 2D Boltzmann simulations of \citet{nagakura_19b}. This demonstrates
that the LESA is a robust phenomenon; claims
that it depends on the details of neutrino
transport \citep{nagakura_19b} are not convincing.

The nature of this instabiity is still not fully understood.
\citet{tamborra_14a} initially suggested a feedback
cycle between $\ell=1$ shock deformation, a dipolar asymmetry in the accretion flow,
and a dipolar asymmetry in the
lepton fraction in the
PNS convection zone. However, more
recent studies suggest that the LESA
does not depend on an \emph{external} feedback cycle between
asymmetries in the accretion flow and
asymmetries in the PNS convection zone.
\citet{glas_19} demonstrated that 
LESA can be even more pronounced
in exploding low-mass progenitor models
with low accretion rates, which suggests
that the mechanism behind LESA works
\emph{within} the PNS convection zone.

This, however, leaves the question 
why the flow \emph{within} the PNS convection zone
would organise itself to generate a dipolar
lepton fraction asymmetry.
Some papers have, however, formulated first
qualitative arguments to suggest that there
is an \emph{internal} mechanism for a dipole
asymmetry in the lepton fraction, and that
LESA may just be a very peculiar manifestation of
buoyancy-driven convection.
What appears
to play a role is that the lepton fraction gradient
becomes stabilizing against convection in the middle of the PNS
convection zone \citep{janka_16,powell_19}. 
\citet{janka_16} suggested that this can give rise
to a positive feedback loop because a hemispheric lepton asymmetry
will  attenuate or enhance the stabilizing effect
in the different hemispheres, thereby leading to more vigorous
convection in one hemisphere, which in turn maintains
the lepton asymmetry. On a different note, \citet{glas_19} sought
to explain the large-scale nature of the asymmetry by
applying the concept of a critical Rayleigh number
for thermally-driven convection \citep{chandrasekhar_61}.
However, one still needs to account for the fact that
the typical scales of the velocity and lepton number
perturbations appear remarkably different in the
PNS convection zone. This was confirmed
by the quantitative analysis of \citet{powell_19},
who found a very broad turbulent velocity
spectrum peaking around $\ell=20$, which conforms
neither to Kolmogorov or Bolgiano--Obukhov scaling for stratified
turbulence.
\citet{powell_19} suggested that this could
be explained by the scale-dependent effective buoyancy
experienced by eddies of different sizes as they
move across the partially stabilized central region
of the PNS convection zone. \citet{powell_19}
also remarked that the non-linear state
of PNS convection is characterized by a balance
between the convective and diffusive lepton number flux.
All these aspects suggest that the LESA could
be no more than a manifestation of PNS convection,
but that PNS convection is in fact quite dissimilar
from the high-P\'eclet number convection as familiar
from the gain region or the late convective burning stages.
A satisfactory explanation of the phenomenon likely
needs to go beyond concepts from linear stability
theory and the usual global balance arguments
behind MLT, and will have to take
into account scale-dependent forcing and dissipation, and the
``double-radiative'' nature of the instability.

Since the stabilising lepton number gradient in the
middle of the PNS convection zone figures prominently
in these attempts to understand LESA, one might justifiably
ask whether there is some role for double-diffusive
instabilities in the PNS after all.
Local stability analysis \citep{bruenn_95,bruenn_96,bruenn_04} in fact 
suggests that double-diffusive instabilities
(termed lepto-entropy fingers and lepto-entropy semiconvection)
should occur in the PNS. But why were such double-diffusive
instabilities never identified in multi-D simulations so far?
Further careful analysis and interpretation of the
simulation results and theory is in order to clarify this.
One possible interpretation could be that the characteristic step-like
lepton number profile established by LESA \citep{powell_19}
\emph{is} actually a manifestation of layer formation
in the subcritical regime as familiar from semiconvection
\citep{proctor_81,spruit_13,garaud_18}. However,
the slow, global turnover motions in LESA do
not readily fit into this picture.\footnote{That
the LESA is also seen in models without lateral diffusion
may not an obstacle for this interpretation. Lateral
diffusion is essential to obtain semiconvective overstability, but layer formation can occur below the threshold
for overstability \citep{proctor_81,spruit_13}.}
One should also beware premature conclusions because
PNS convection is an inherently difficult regime
for numerical simulations due to small convective
Mach numbers of order
$\mathord{\sim}0.01$ and the
importance of diffusive effects. The potential issues
go beyond the question of resolution and unphysically
high numrical Reynolds numbers (cf.\ \ref{sec:subsonic}),
and there are concrete reasons to investigate these
in more depth.
For example, although different codes agree qualitatively
concerning the region of instability and the qualitative features
of the convective flow, substantial
differences in the turbulent kinetic energy
density have been reported in a comparison
between the  \textsc{Alcar} and \textsc{Vertex} codes
in the PNS convection zone, even though the agreement
between the codes is otherwise excellent \citep{just_18}.
While it is unlikely that the uncertainties in models
PNS convection have any impact on the problem of shock revival,
they need to be addressed to obtain a better
understanding of LESA and reliable predictions of
gravitational wave signals and the nucleosynthesis
conditions in the neutrino-heated ejecta.

\section{The explosion phase}
\label{sec:expl_phase}
Regardless of whether the explosion is
driven by neutrinos or magnetic fields, there
is no abrupt transition to a quasi-spherical outflow
after shock revival. In this section, we
shall focus on the situation in neutrino-driven explosions,
which has already been quite thoroughly explored.

\subsection{The early explosion phase}
In typical neutrino-driven models, the multi-dimensional 
flow structure in the early explosion phase appears 
qualitatively similar to the pre-explosion phase at first
glance. Buoyancy-driven outflows and accretion downflows
persist for hundreds of milliseconds to seconds and
allow for simultaneous mass accretion and ejection.
Because of the ongoing accretion, high neutrino luminosities
and hence high heating rates can be maintained to
continually dump energy into the developing explosion.
As the shock radius slowly increases, large-scale
$\ell=1$ and $\ell=2$ modes start to dominate the
flow irrespective of whether medium-scale
convection or large-scale SASI modes dominated
prior to shock revival, even though
2D explosion models probably tended
to exaggerate this effect. The basic features of
this pictures have held since the 1990s
\citep{herant_92,shimizu_93,yamada_93,janka_93,herant_94,burrows_95,janka_95,janka_96}, and have proved
critical for explaining the energetics
of core-collapse supernovae.\footnote{Critiques
of the neutrino-driven mechanism have occasionally
overlooked \citep{papish_15} and then ultimately
rebranded the simultaneous outflows and downflows
as ``jittering jets'' \citep{soker_19}. In this
latest instalment, the alternative jittering-jet
scenario seems to have come down to little
more than a question of unconventional terminology
for well-established phenomena in neutrino-driven
explosions.}
Even in electron-capture supernova
progenitors, which explode even without
the help of multi-dimensional effects
\citep{kitaura_06}, there is a brief
phase of convective overturn after shock revival
\citep{wanajo_11}.

\begin{figure}
    \centering
    a)
    \includegraphics[width=0.46 \textwidth]{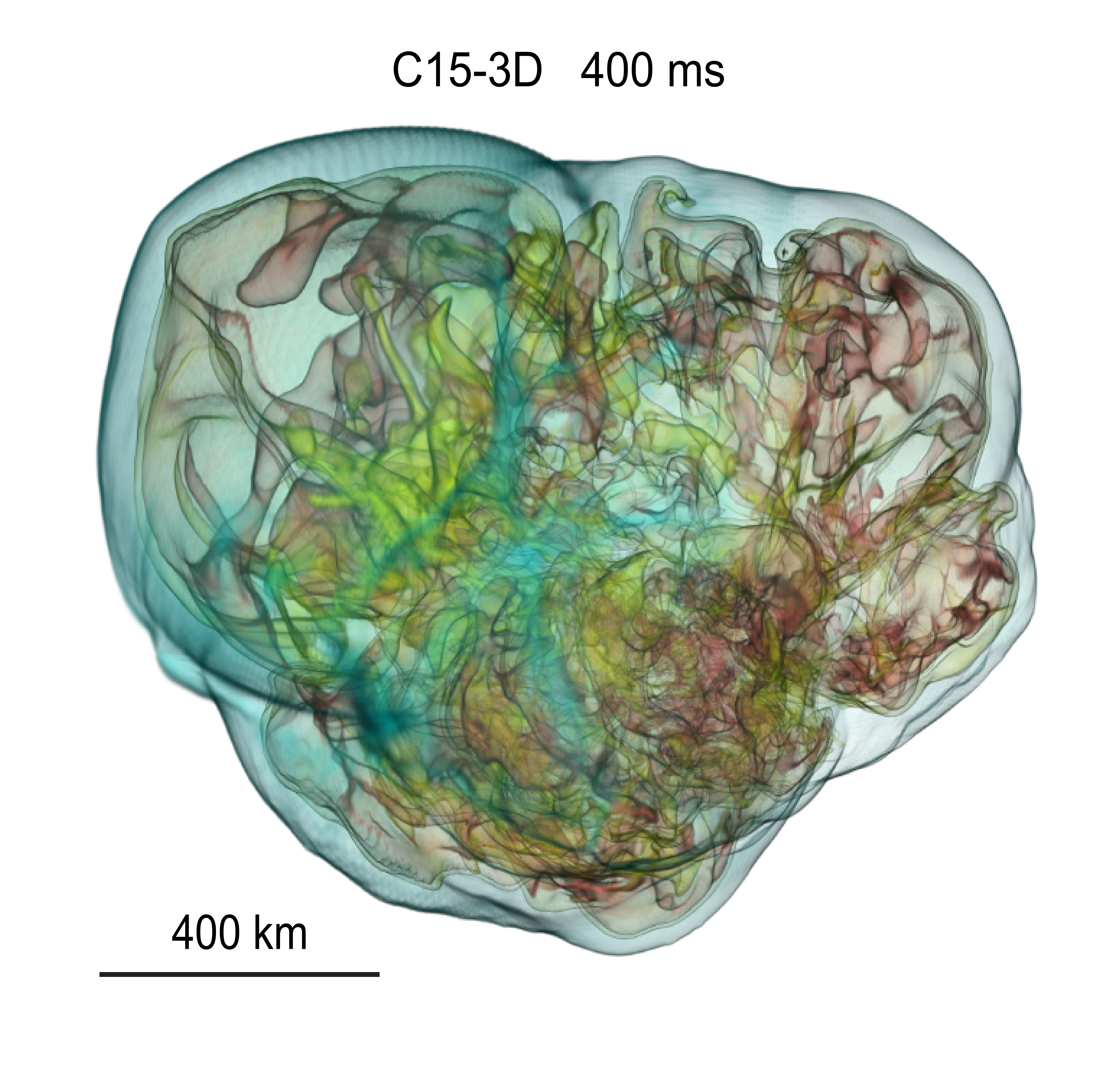}
    b)
    \hfill
    \includegraphics[width=0.46 \textwidth]{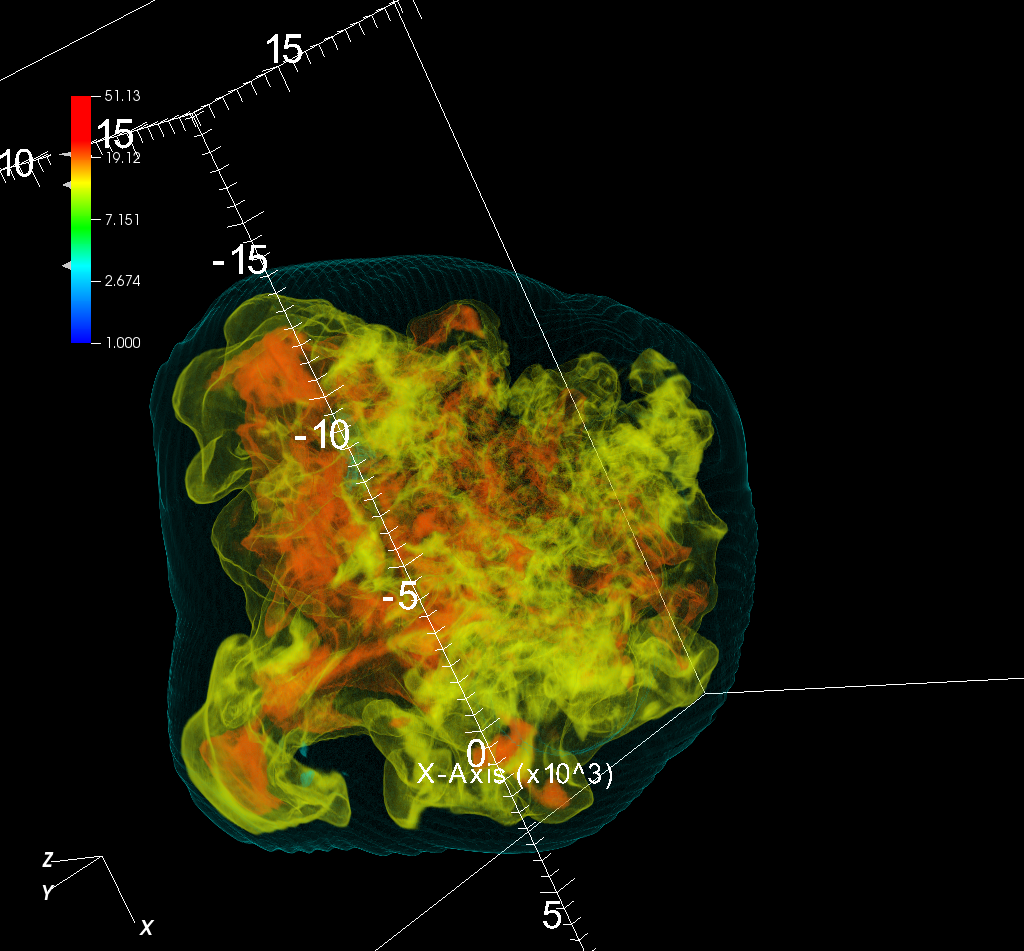}
    \\
    c)
    \includegraphics[width=0.46 \textwidth]{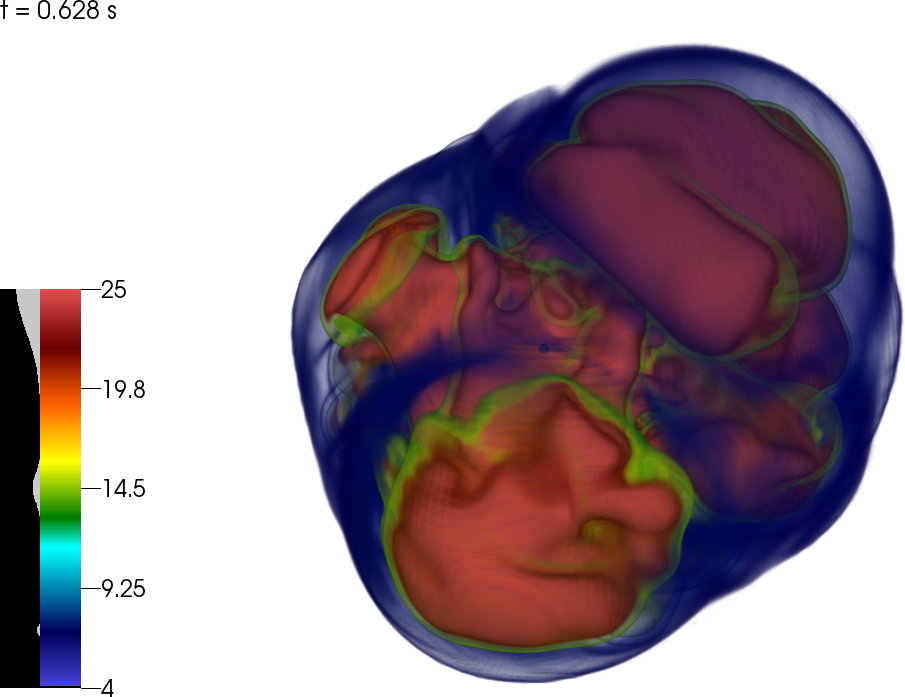}
    \hfill
    d)
    \includegraphics[width=0.46\textwidth]{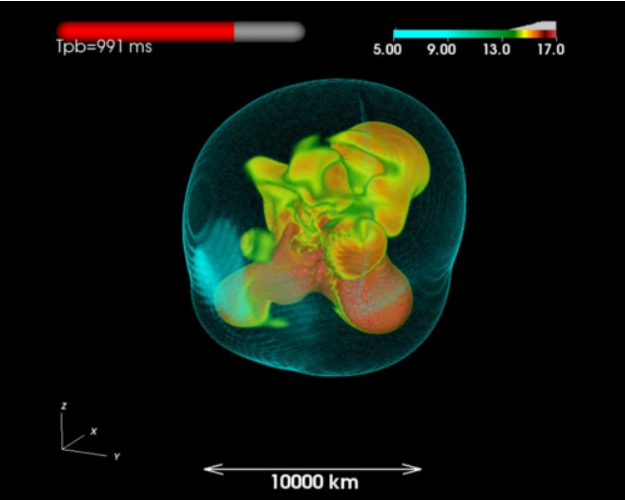}
    \caption{Volume renderings of the entropy in different
    3D supernova simulations showing the emergence of stable
    large-scale plumes around and after shock revival as
    a common phenomenon despite differences in resolution
    and in the neutrino transport treatment.
    The outer translucent surface is the shock, the structures inside
    are neutrino-heated high-entropy bubbles:
    \textbf{a)} $15\,M_\odot$ model of  \citet{lentz_15}
    with a unipolar explosion geometry
    at a post-bounce time of $400\, \mathrm{ms}$.
    \textbf{b})  $3\,M_\odot$ He star model of \citet{mueller_19a}
    at $1238 \, \mathrm{ms}$, with  two prominent plumes
    in the 7 o'clock  and
    11 o'clock directions and weaker
    shock expansion on the opposite side.
    \textbf{c)} $20\,M_\odot$
     model of \citet[][Fig.~8]{burrows_19b} with a more dipolar
     explosion geometry
     at $651 \, \mathrm{ms}$.
     \textbf{d}) $11.2\,M_\odot$
     model of \citet[][Fig.~6]{nakamura_19} at a time
     of $991 \, \mathrm{ms}$. 
     Images reproduced with permission from [a] \citet{lentz_15}, copyright by AAS; [c] from \citet{burrows_19b} and [d] from \citet{nakamura_19}, copyright by the authors.
    }
    \label{fig:expl_snapshots}
\end{figure}

More recent 3D explosion models using
multi-group transport
\citep{takiwaki_14,melson_15b,lentz_15,mueller_15b,mueller_17,mueller_19a,burrows_19b} have 
confirmed this picture, but paved the way
towards a more quantitative theory of the explosion
phase. In massive progenitors, shock expansion
is usually sufficiently slow for one or
two dominant bubbles of neutrino-heated ejecta
to form (Fig.~\ref{fig:expl_snapshots}).
Only at the low-mass end of the progenitor
spectrum \citep{melson_15a,gessner_18} do the convective
structures freeze out so quickly that the
neutrino-heated ejecta are organized in
medium-scale bubbles instead of a unipolar
or bipolar structure.

The detailed dynamics of the
outflows and downflows proved to be significantly different
in 3D compared to 2D, and that only
restricted insights on 
explosion and remnant properties and nucleosynthesis
can be gained from the impressive corpus 
of successful 2D simulations with multi-group transport
\citep{buras_06b,marek_09,mueller_12a,mueller_12b,mueller_13,janka_12,janka_12b,suwa_10,suwa_13,bruenn_13,bruenn_16,nakamura_15,burrows_18,pan_18,oconnor_18a}. Except at the lowest masses,
the 2D simulations are uniformly characterized by
almost unabated accretion through fast downflows
that reach directly to the bottom of the
gain region, by  outflows that are often weak and intermittent,
and by  a halting rise of explosion energies.
Long-time 2D simulations  showed
that this situation can persist out to more
than $10 \, \mathrm{s}$ \citep{mueller_15b},
and as a result, implausibly high
neutron star masses are reached.
The halting growth of explosion energies
in 2D can partly be explained by the topology
of the flow which lends itself to outflow
constriction by  equatorial downflows, but
the primary difference between 2D and 3D
lies in the velocity of the downflows.
\citet{melson_15a} already noticed that
the downflows appear to subside more quickly
in their 3D model of a low-mass
progenitor, which they ascribed
to the forward turbulent cascade in 3D;
this led to a slight enhancement
of the explosion energy by 10\% in 3D compared to 2D.
In more massive progenitors with stronger
accretion after shock revival stronger
braking of the downflows in 3D compared
to 2D is even more evident \citep{mueller_15b}. Instead
of crashing into a secondary accretion
shock at $\mathord{\sim} 100 \, \mathrm{km}$
at a sizable fraction of the free-fall velocity,
the downflows are gently decelerated, and
secondary shocks rarely form. \citet{mueller_15b}
ascribed this pathology of the 2D models to the behavior
of the Kelvin--Helmholtz instability between
the outflows and downflows, which
is stabilized at high Mach numbers in 2D,
but can always grow in 3D \citep{gerwin_68}.
Since the typical Mach number of the downflows
is higher during the explosion phase, the assumption
of 2D symmetry becomes even more problematic than during
the accretion phase.

\subsection{Explosion energetics}
\label{eq:energy}
\paragraph{Estimators for the explosion energy.}
Strictly speaking, the final demarcation between
ejected and accreted material\footnote{We
avoid the term ``mass cut'', which is
commonly used for describing artificial
1D explosion models. The boundary
of the ejecta region is not a
sphere, and does not correspond to
a unique mass shell under realistic conditions.} cannot
be determined before the explosion becomes
kinetically dominated after shock breakout,
and the same holds true for the final explosion energy $E_\mathrm{exp}$.
It is, however, customary and useful
to consider the \emph{diagnostic explosion
energy} $E_\mathrm{diag}$
(often shortened to ``diagnostic energy''
or ``explosion energy'' when there
is no ambiguity), which is
defined as the total energy of the material that is
nominally unbound at any given instance
\citep{buras_06b,mueller_12a,bruenn_13}.
By definition the diagnostic energy will eventually
asymptote to  $E_\mathrm{exp}$, but
might  do so only over considerably longer
time scales than can be simulated with neutrino
transport. In particular, the diagnostic energy
can in principle decrease as the shock sweeps up
bound material from the outer shells. To account for
this, one can correct 
$E_\mathrm{diag}$ for the binding energy
of the material ahead of the shock (``overburden'')
to obtain a more conservative estimate for
$E_\mathrm{exp}$ \citep{bruenn_16}. In practice,
$E_\mathrm{diag}$ usually rises
monotonically because energy continues to
be pumped into the ejecta over seconds, but
there are exceptions, most notably in cases
of early black hole formation \citep{chan_18}.
In most cases, one expects
that $E_\mathrm{diag}$ levels out after
a few seconds and then provides a good estimate for $E_\mathrm{exp}$.

\paragraph{Explosion energies from
self-consistent simulations.}
Unfortunately, $E_\mathrm{diag}$ has not
levelled off in
most of the available self-consistent
3D explosion models, though
the growth of the explosion energy has
already slowed down significantly
in some long-time simulations using
the \textsc{CoCoNuT-FMT} code \citep{mueller_17,mueller_18,mueller_19a}.
  Even in 2D, only
 some of the models of the Oakridge group appear
to have approached their final explosion energy
\citep{bruenn_16}.

This means that no final verdict on the
fidelity of the simulations can be pronounced
based on a comparison with observationally
inferred explosion energies. The models
of  \citet{mueller_17,mueller_18,mueller_19a}
and \citet{bruenn_16}, whose explosion
energies are admittedly on the high
side among modern simulations,  have demonstrated
that neutrino-driven explosions can reach
energies of up to $8 \times 10^{50}\, \mathrm{erg}$.
Similarly, plausible nickel masses
of several $0.01\,M_\odot$ appear within reach,
although no firm 
statements can be made for \textsc{CoCoNuT-FMT} models
due to uncertainties in the $Y_\mathrm{e}$ of the
ejecta from the approximative transport treatment,
and due to the highly simplified treatment of
nucleon recombination.
Explosion energies beyond $10^{51}\, \mathrm{erg}$
may simply be a matter of longer simulations,
different progenitor models, and slightly
improved physics; and there \emph{may} be no conflict
with the 
distribution of observationally inferred explosion energies
\citep{kasen_09,pejcha_15c,mueller_t_17} of Type~IIP supernovae. First attempts to extrapolate
the non-converged explosion
energies from simulations and compare them to observations 
using a rigorous statistical
framework \citep{murphy_19} indicate that
the predicted values are still somewhat too low,
but \citet{murphy_19} also point out that conclusions are premature
due to biases and uncertainties in the comparison.

\paragraph{Growth of the explosion energy.}
Even at this stage, the simulations
already elucidate how the energy of neutrino-driven
explosions is determined. Upon closer inspection,
the energy budget of the ejecta is quite complicated
and includes contributions from the
injection of neutrino-heated material from below,
form nucleon recombination and nuclear burning,
from the accumulation of bound material by the shock,
and from turbulent mixing with the downflows
\citep[for a broader discussion, see][]{marek_09,mueller_15b,bruenn_16}.
Nonetheless, a few key findings have emerged.
The most critical determinant for
the growth of $E_\mathrm{diag}$ is the mass
outflow rate $\dot{M}_\mathrm{out}$ of
neutrino-heated material. Neutrino heating
only marginally unbinds the material, and
the net contribution to $E_\mathrm{diag}$ 
comes from the energy
$\epsilon_\mathrm{rec}$
released by nucleon recombination, 
which occurs at a radius of about $300 \, \mathrm{km}$.
To first order, the resulting growth rate of the diagnostics
energy is
\citep{scheck_06,melson_15a,mueller_15b},
\begin{equation}
\dot{E}_\mathrm{diag}\approx
\dot{M}_\mathrm{out} \epsilon_\mathrm{rec}.
\end{equation}
In principle, $8.8\, \mathrm{MeV}$ per nucleon
can be released from full recombination
to the iron group, but
for the relevant entropies and expansion
time scales, recombination is incomplete
and does not convert all the neutrino-heated
ejecta to iron-group elements. Mixing between the outflows and downflows reduces the effective value of $\varepsilon_\mathrm{rec}$ further to about $5$--$6\, \mathrm{MeV}/\mathrm{nucleon}$. 

The mass outflow rate is roughly determined by
the volume-integrated neutrino heating
rate and the energy required to lift the
material out of the gravitational potential.
One can argue that the relevant energy scale
is the binding energy at the gain radius
$|e_\mathrm{gain}|$, so that
\begin{equation}
    \dot{M}_\mathrm{out} = 
    \eta_\mathrm{out}
    \frac{\dot{Q}_\nu}{|e_\mathrm{gain}|}.
\end{equation}
with some efficiency parameter $\eta_\mathrm{out}$
that accounts for the fact that only part of
the neutrino-heated matter is ejected. Initially,
one finds $\eta_\mathrm{out}<1$ as expected, but
\citet{mueller_17} pointed out that the
situation changes later after shock revival
because much of the ejected material never
makes it down to the gain radius and starts
is expansion significantly further out with lower
initial binding energy.
This  leads to efficiency parameters 
$\eta_\mathrm{out}>1$, and helps compensate
for the declining heating rates as the supply
of fresh material to the gain radius slowly
subsides.

The situation in 2D models is somewhat different \citep{mueller_15b}.
Here, the ejected material comes from close to
the gain radius, and the outflow
efficiency $\eta_\mathrm{out}$ is lower
than in 3D. Although the lack of turbulent mixing
results in a higher asymptotic specific
total energy and entropy in 2D, the net effect
is  a slower growth of the explosion energy
than in 3D. Moreover, the higher entropies in 2D
will result in reduced recombination to
the iron group and hence lower nickel masses.
Despite these  shortcomings,
2D simulations remain of some use because
they already allow extensive parameters
studies of explodability and explosion
and remnant properties
\citep{nakamura_15}.

\subsection{Compact remnant properties}
\label{sec:compact_remnant}
\paragraph{Accretion rates and remnant masses.} The
forward cascade and the stronger Kelvin--Helmholtz
instabilities between the outflows and downflows
in 3D imply that the accretion rate onto the
PNS drops more quickly than in 2D \citep{mueller_15b}.
As a result, some self-consistent
3D simulations have been able to determine
firm numbers for final neutron star
masses \citep{melson_15a,mueller_19a,burrows_19,burrows_19b},
barring the possibility of late-time fallback.
The predicted neutron star masses appear
roughly compatible with the range of observed
values \citep{oezel_16,antoniadis_16}, but
as with all other explosion and remnant properties,
a robust statistical comparison is not yet possible.

\paragraph{Neutron star kicks.}
Observations show that most neutron stars receive
a considerable ``kick'' velocity at birth.
The kick velocity is typically a few
hundred $\mathrm{km}\, \mathrm{s}^{-1}$,
but  there is a broad distribution ranging from very
low kicks up to more than $1000 \, \mathrm{km}\, \mathrm{s}^{-1}$ \citep[e.g.,][]{hobbs_05,faucher_06,ng_07}.
The large-scale ejecta asymmetries that emerge during
the explosion provide a possible explanation
for this phenomenon \citep[for an overview including other mechanisms such as aniostropic neutrino emission, see][]{lai_01,janka_17}.

The 2D simulations of the 1990s could not yet naturally
obtain the full range of observed kick velocities
by a hydrodynamic mechanism \citep{janka_94}, unless
unrealistically large seed asymmetries in the progenitor
were invoked \citep{burrows_96}. A plausible
range of kicks was first obtained in parameterized 2D simulations by \citet{scheck_04,scheck_06}, thanks
to more slowly developing explosions that
allowed the  $\ell=1$-mode of the SASI, or an
$\ell=1$ convective mode to emerge. 
The work of \citet{scheck_04,scheck_06} revealed
that the kick velocity can grow for well over
a second in models with slower shock expansion.
They concluded that the kick arises primarily from
the asymmetric gravitational pull of over- and
underdense regions in the ejecta 
(later termed ``gravitational tugboat mechanism''
\footnote{The term ``tugboat mechanism'' 
was in fact suggested later by
Jeremiah Murphy and introduced
into the literature in \citet{wongwathanarat_13}.}
) rather than
from pressure force and hydrodyanmic momentum fluxes
onto the PNS; anisotropic neutrino emission was found
to play only a minor role. Subsequent simulations have not
fundamentally changed this analysis. Although
various studies showed that
the momentum flux onto the PNS can be comparable
to the gravitational force onto the PNS
\citep{nordhaus_10b,nordhaus_12,mueller_17},
this does not invalidate 
the tugboat mechanism. Effectively, the
contribution of each parcel of accreted material to the
PNS momentum via the hydrodynamic flux and
the gravitational tug almost cancel, and the
net acceleration of the PNS is due to
the gravitational pull of the material that is
actually ejected.

Three-dimensional simulations have not altered
this picture substantially. Even though 2D
simulations tend to obtain higher kicks,
values of several hundred $\mathrm{km} \, \mathrm{s}^{-1}$
were already obtained in the parameterized
3D simulations of
\citet{wongwathanarat_10b,wongwathanarat_13}.
Recently,  \citet{mueller_17,mueller_19a} performed
sufficiently long 3D simulations with multi-group
neutrino transport to extrapolate the final kick velocities, which  fall nicely within the observed range of up to $1000 \, \mathrm{km}\, \mathrm{s}^{-1}$.

Based on the physics of the kick mechanism, various 
authors have posited a correlation between the kick and
the ejecta mass \citep{bray_16} or, using
more refined arguments, on the explosion energy
\citep{janka_17,vigna_gomez_18}. Tentative
support for a loose correlation comes from the small
kicks obtained
in simulations of low energy, ultra-stripped supernovae
\citep{suwa_15,mueller_17} and electron-capture
supernovae \citep{gessner_18}, and from more
recent 3D simulations over a larger range
of progenitor masses \citep{mueller_19a}.

\paragraph{Neutron star spins.} If the downflows
hit the PNS surface with a finite impact
parameter, they also impart angular momentum
onto the PNS. While this was realized already
by \citet{spruit_98}, 3D simulations are needed
 for quantitative predictions of PNS spin-up
by asymmetric accretion. The predicted spin-up
in 3D models of non-rotating varies. Parameterized
 simulations \citep{wongwathanarat_10b,rantsiou_11,wongwathanarat_13} tend to find longer neutron star spin periods
 of hundreds of milliseconds to seconds (but
 extending down to $100 \, \mathrm{ms}$  in
 \citealp{wongwathanarat_13}). Recent
 3D simulations using multi-group
 transport \citep{mueller_17,mueller_19a}
 obtain spin periods between $20\, \mathrm{ms}$ and $2.7\, \mathrm{s}$, which roughly coincides with the range
 of observationally inferred birth periods
  \citep{faucher_06,perna_08,popov_12,noutsos_13}.
  Assuming the core angular momentum is conserved
  after the collapse to a PNS and 
  not changed by angular momentum transport
  during the explosion, current stellar evolution models
  computed with the Tayler--Spruit dynamo,
  predict spin periods in the same range \citep{heger_05}, which makes
  it difficult to draw inferences on the
  explosion mechanism or the progenitor rotation from the observed spin  periods. None of the simulations
  can as yet explain the spin-kick alignment that
  is suggested by observations \citep{johnston_05,ng_07,noutsos_13}.
  Proposed mechanisms for natural spin-kick alignment 
  by purely hydrodyanmic processes \citep{spruit_98,janka_17} have not been borne out
  by the models. However, the possibility of
  natural spin-kick alignment in rotating progenitors
  has yet to be investigated.

\paragraph{Role of the spiral SASI mode.}
In the first idealized simulations of the
spiral mode of the SASI, \citet{blondin_07}
noted a significant flux of angular momentum
into the ``neutron star'' (modeled by an
inner boundary condition) that would lead
to rapid neutron star rotation in the opposite
direction to the SASI flow
with angular frequencies
of the order of $100 \, \mathrm{rad}\, \mathrm{s}^{-1}$
even in the case of non-rotating progenitors.
This idea has been explored further in
several numerical \citep{hanke_13,kazeroni_16,kazeroni_17}
and analytical \citep{fernandez_14b} studies. The 
potential for spin-up of non-rotating progenitors
may be more modest than initially thought;
both numerical and analytical results suggest
that the angular momentum imparted onto the
PNS is only a few $10^{46}\, \mathrm{erg}\, \mathrm{s}$,
corresponding to spin periods of hundreds
of milliseconds \citep{hanke_13,fernandez_14b}. Moreover, part of the angular momentum
contained in the spiral mode may be accreted after
shock revival, negating the separation of angular
momentum previously achieved by the SASI. Spin-up
and spin-down by SASI in the case of rotating
progenitors still merits further investigation; the idealized simulations of \citet{kazeroni_17} suggest
different regimes of random spin-up and spin-down
for slow progenitor rotation, systematic spin-down for
intermediate rotation, and weaker spin-down
for high rotation rates in the regime of the corotation
instability. The possibility of magnetic field
amplification due to the induced shear in the PNS
surface region in the case of
significant spin-up or spin-down by the SASI also needs to be explored.

\begin{figure}
    \centering
    a)
    \includegraphics[width=0.63 \textwidth]{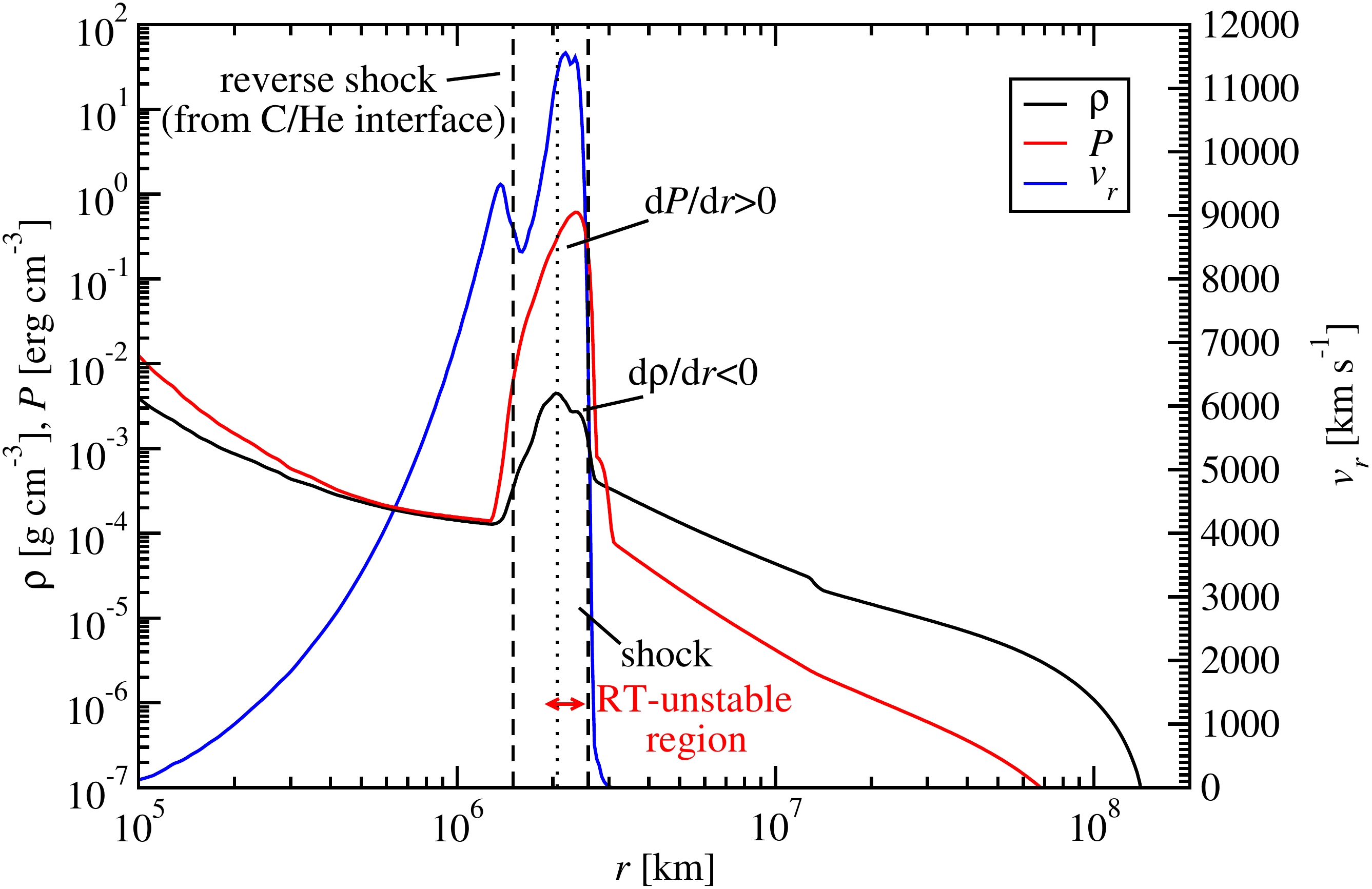}
    \hfill
    b)
    \includegraphics[width=0.3 \textwidth]{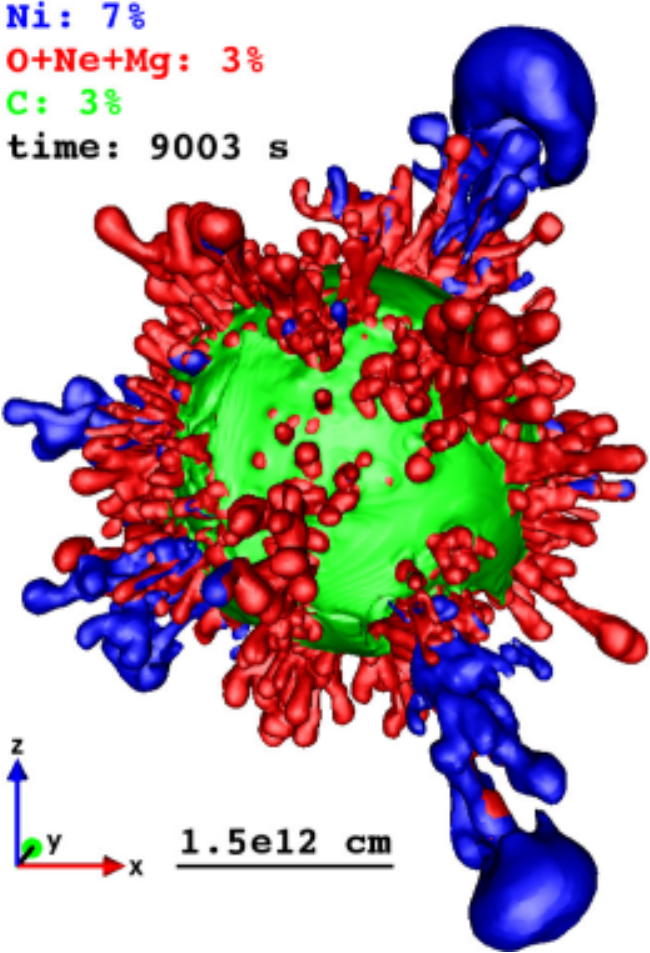}
    \caption{\textbf{a)}
    Emergence of Rayleigh--Taylor instability during
    the propagation of the shock through the envelope, illustrated by spherically averaged profiles
    of density $\rho$, pressure $P$, and
    radial velocity $v_r$ from a 2D long-time simulation of a $9.6\,M_\odot$ star based on the
    explosion model of \citet{mueller_13}. At this stage ($140 \,\mathrm{s}$
    after the onset of the explosion), the
    shock has reached the He shell,
    and a reverse shock has formed.
    Behind the forward shock, the 
    pressure gradient is positive and
    decelerates the expansion of the ejecta.
    Rayleigh--Taylor (RT) instability grows
    in a region with $\ud \rho/\ud r<0$ behind the shock. Note that the
    structure of the blast wave can
    be more complicated in general
    with several unstable regions and
    reverse shocks that interact with each
    other.
    \textbf{b)}  Mass fraction isocontours in a 3D model
    of mixing in SN~1987A. Note that while the biggest Ni-rich
    Rayleigh--Taylor clumps are seeded by large-scale asymmetries
    from the early explosion phase, these develop into the
    finger-like structures characteristic of the Rayleigh--Taylor
    instability, and there is also considerable growth of small-scale
    plumes.
    Image reproduced with permission from \citet{wongwathanarat_15},
    copyright by ESO.
}
    \label{fig:mixing}
\end{figure}

\subsection{Mixing instabilities
in the envelope}
\label{sec:mix}
\paragraph{Structure of the flow in the
later explosion phase.}
As the propagating shock scoops up matter
and as the explosion energy levels off,
the structure of the post-shock region 
changes (Fig.~\ref{fig:mixing}a).
Early on, the post-shock expansion velocities
are subsonic and the outflows are accelerated
by a positive pressure gradient, but eventually
the post-shock flow enters a Sedov-like
regime where a positive pressure gradient
is established and matter is decelerated behind
the shock \citep{chevalier_76}. Generally, the
shock velocity $v_\mathrm{sh}$ also decreases\footnote{Note
that deceleration of the post-shock matter
and deceleration of the shock do not
always coincide, though they are closely
related phenomena. One can have $\dot{v}_r<0$
and $\dot{v}_\mathrm{sh}>0$.} as
the mass $M_\mathrm{ej}$ of the shock ejecta grows;
it roughly scales as $v_\mathrm{sh}\propto (E_\mathrm{exp}/M_\mathrm{ej})^{1/2}$.
The shock can, however, transiently accelerate when
it encounters density gradients 
steeper than $\rho \propto r^{-3}$ at shell
interfaces \citep{sedov_59}. Both effects can
be captured by the formula of \citet{matzner_99},
\begin{equation}
\label{eq:vshock}
v_\mathrm{sh}
\approx 
0.794
\left(\frac{E_\mathrm{exp}}{M_\mathrm{ej}}\right)^{1/2}
\left(\frac{M_\mathrm{ej}}{\rho r^3}\right)^{0.19}.
\end{equation}

The post-shock pressure and density profiles adjust to variations in shock velocity to establish
something of a ``quasi-hydrostatic'' stratification
behind the shock with an effective gravity that
is directed \emph{outward}.
However, once the post-shock velocities become supersonic,
the post-shock pressure profile can no longer
globablly adjust to changing shock velocities,
and reverse shocks are formed. A first
reverse shock forms typically forms at a few
$1000\, \mathrm{km}$ as the developing neutrino-heated
wind crashes into more slowly moving ejecta.
Later on, further reverse shocks emerge after the shock encounters
various shell interfaces. Their strength depends
on the density jump at the shell interface. In
hydrogen-rich progenitors, the reverse shock from
the H/He interface is particularly strong (especially
in red supergiant) and therefore sometimes referred
to simply as \emph{the} reverse shock. 

\paragraph{Rayleigh--Taylor instability.}
The non-monotonic variations in $v_\mathrm{sh}$
result in mon-monotonic post-shock entropy and density
profiles, and some layers become Rayleigh--Taylor
unstable \citep{chevalier_76,mueller_91,fryxell_91}.
In the relevant highly compressible regime, the growth rate for the Rayleigh--Taylor instability from
a local stability analysis is given by
\citep{bandiera_84,benz_90,mueller_91}
\begin{equation}
\label{eq:omrt}
\omega_\mathrm{RT}=
\frac{c_\mathrm{s}}{\Gamma}
\sqrt{\frac{\pd \ln P}{\pd r}
\left (
\frac{\pd \ln P}{\pd r}-
\Gamma \frac{\pd \ln \rho}{\pd r}
\right)}
=
\sqrt{ g_\mathrm{eff}
\left (
\frac{1}{\Gamma}\frac{\pd \ln P}{\pd r}-
 \frac{\pd \ln \rho}{\pd r}
\right)},
\end{equation}
where $g_\mathrm{eff}=\rho^{-1} \pd P/\pd r$
is the effective gravity. The second form
elucidates that stability is determined by
the sub- or superadiabaticity of the density
gradient as for buoyancy-driven convection. In the relevant radiation-dominated regime,
composition gradients have a minor impact on stability,
and the entropy gradient is the deciding factor.
However, since the effective gravity points
outwards, positive entropy gradients are now
unstable. Such positive entropy gradients arise
when the shock accelerates at shell interfaces.
One should note that the actual growth rate of perturbations
depends on their length scale \citep{zhou_17a}, and
Eq.~(\ref{eq:omrt}) roughly applies to the
fastest growing modes with a wavelength comparable
to the width of the unstable region. Since the
unstable regions tend to be narrow, Rayleigh--Taylor
mixing tends to produce smaller, clumpy structures, but
large-scale asymmetries can also grow considerably
for sufficiently strong seed perturbations.

Intriguingly, it has been suggested that
the overall effect of  Rayleigh--Taylor mixing can roughly be 
captured in 1D by an appropriate turbulence
model \citep{duffell_16b,paxton_17}.
The key idea here is to incorporate the proper
growth rate, saturation behavior, and a velocity-dependent
mixing length \citep{duffell_16b}. While a first
comparison  of this model with 3D results from
\citet{wongwathanarat_15} proved encouraging
\citep{paxton_17}, a few caveats remain. 
A detailed analysis of Rayleigh--Taylor
mixing in a 3D model of a stripped-envelope
progenitor by \citet{mueller_17} unearthed some
basis for a phenomenological 1D description
of Rayleigh--Taylor mixing (most notably
buoyancy-drag balance in the non-linear
stage) and suggested some improvements to
the model of \citet{paxton_17}, but cast doubt
on the use of local gradient to estimate the density
and composition contrasts of the Rayleigh--Taylor
plumes. In particular, the Rayleigh--Taylor instability
sometimes produces partial inversions of the
initial composition profiles, which cannot
be modeled by diffusive mixing in 1D.

In addition to the Rayleigh--Taylor instability,
the Richtmyer--Meshkov instability \citep[see][for
details of the instability mechanism]{richtmyer_60,zhou_17a} can develop
because the shock is generally aspherical
and hits the shell interfaces obliquely.
The literature on mixing instabilities in supernovae
is extensive, and we can only provide a very condensed
summary of extant numerical studies. We will exclusively
focus on the optically thick phase of
the explosion and not consider the remnant phase.

\paragraph{Simulations of mixing in SN~1987A.}
After a few earlier numerical experiments, 
significant interest in mixing instabilities 
was prompted by observations of SN~1987A
that pointed to strong early mixing of
nickel \citep[see][and references
therein]{arnett_89,mccray_93}. 
Two-dimensional simulations of mixing instabilities 
\citep{arnett_89b,mueller_91,fryxell_91,hachisu_90,benz_90,benz_92} in the wake of SN~1987A took a first
step towards explaining the observed mixing.
The typical  picture revealed by these
models is that of a strong Rayleigh--Taylor
instability at the H/He-interface with linear
growth factors of thousands \citep{mueller_91},
Some models \citep{mueller_91,fryxell_91}
also showed a second strongly unstable region at the
He/C-interface that eventually
merges with the mixed region further outside.
Mixing was dominated by many small-scale plumes
in these first-generation simulations.
However, the maximum velocities of nickel
plumes still fell short by about a factor
of two compared to the observed velocities
of up to $\mathord{\sim} 4000\, \mathrm{km} \, \mathrm{s}^{-1}$.

Many subsequent studies have investigated stronger,
large-scale initial seed perturbations as a possible
explanations for the strong mixing in SN~1987A
and other observed Type~II supernovae. Such seed perturbations
are naturally expected in the neutrino-driven
paradigm from the SASI and low-$\ell$ convective modes,
and in magnetorotational explosions with
veritable jets.
Most simulations have explored the effect of
large-scale seed perturbations by specifying
them \emph{ad hoc}
\citep[e.g.,][]{nagataki_98,nagataki_00,hungerford_03,couch_09,ono_13,mao_15,ellinger_12}. One must therefore
be cautious in drawing conclusions on the role of ``jets'' 
in explaining the observed mixing. In fact, simulations
with artificially injected jets rather serve to
rule out kinetically-dominated jets in
Type~IIP supernovae based on early
spectropolarimetry, though thermally-dominated
jets are not excluded in principle \citep{couch_09}.

In their 2D AMR simulations, \citet{kifonidis_00,kifonidis_03,kifonidis_06}
followed a more consistent approach by starting
from light-bulb simulations of neutrino-driven
explosions. While the seed asymmetries from
the early explosion phase initially led
to high nickel clump velocities in 
the Type~IIP model of \citet{kifonidis_00,kifonidis_03},
the final velocities were still too small
because the clumps were caught behind the reverse
shock and underwent fast deceleration by supersonic
drag after crossing it. \citet{kifonidis_00}
found that this can be avoided with a more
slowly developing and more aspherical explosion.
In this case, they found less clump deceleration
because the clumps make it beyond the H/He interface before the reverse shock develops, and also found strong downward
mixing of hydrogen with the help of the Richtmyer--Meshkov
instability.

A very convincing picture of mixing in SN~1987A
has emerged since the advent of 3D simulations.
Simulations of single-mode perturbations by
\citet{kane_00} already suggested a faster
growth of the Rayleigh--Taylor instability in 3D.
A first 3D simulation of mixing
in SN~1987A based on 
a 3D explosion model using gray neutrino transport
was conducted by \citet{hammer_10}, who were
able to obtain realistic mixing of nickel,
hydrogen, and other elements even without
the need for strong initial shock deformation
and a strong Richtmyer--Meshkov instability.
\citet{hammer_10} explain the reduced deceleration
of plumes as a result of a more favorable
volume-to-surface ratio of the clumps in 3D
compared to 2D, where the clumps are actually toroidal.
Stronger mixing in 3D was also confirmed
in a study not specifically focused
on SN~1987A \citep{joggerst_10}.
The attainable nickel clump velocities are, however,
quite sensitive to the progenitor structure
\citep{wongwathanarat_15}.
Interestingly, the origin of the
fastest and biggest clumps in \citet{hammer_10}
and in some of the other subsequent 3D
simulations could be traced back
to the most prominent convective bubbles that formed
around shock revival, i.e.\ the late-time
instabilities still contain traces of the
early asymmetries imprinted by the neutrino-driven
engine. As a next step towards model validation,
synthetic light curves based on the
3D models of \citet{wongwathanarat_15} were
computed by \citet{utrobin_15}, and the results
are encouraging. While the fit to the observed
light curve is still not perfect, the discrepancies
likely indicate uncertainties in the progenitor
structure and the precise initial conditions
after shock revival and not a problem of
the neutrino-driven explosion scenario for SN~1987A.

\paragraph{Stripped-envelope supernovae.}
Mixing instabilities are also highly relevant
in the context of stripped-envelope supernovae.
Due to the lack of a H envelope (or
a small mass of the H envelope in the case
of Type~IIb events), the early asymmetries are
not shredded as strongly by Rayleigh--Taylor
mixing as in Type~IIP supernovae, so that
spectroscopy and spectropolarimetry
offer a more direct glimpse on global
asymmetries generated by the engine (see \citealp{wang_08} for observational diagnostics of asymmetries).
Moreover, the presence or absence of
He lines in Ib/c supernovae is sensitive
to the mixing of nickel
\citep{dessart_12,dessart_15,hachinger_12}, 
and so is the detailed shape of the light curve
\citep{shigeyama_90,yoon_19}.

Two-dimensional simulations of Rayleigh--Taylor
mixing in Ib/c supernovae were first conducted by
\citet{hachisu_91,hachisu_94}. These simulations
were based on helium star models, which are
a viable approximation for progenitors that
lost their envelope due to Case~B/Case~C mass transfer,
but used artificially triggered explosions.
\citet{hachisu_91,hachisu_94} found indications
of stronger mixing in less massive helium stars.
\citet{baron_92} interpreted this as pointing
towards an association of Ib and Ic supernovae
with low- and high-mass helium stars, respectively.
\citet{kifonidis_00,kifonidis_03} triggered
the explosion somewhat more realistically using
a light-bulb scheme, but constructed their
Ib supernova progenitor by artificially
removing the hydrogen envelope at collapse (implying
an inconsistent envelope structure); their
finding on the mixing of nickel were qualitatively 
similar to \citet{hachisu_91,hachisu_94}.

\citet{wongwathanarat_17} took an ambitious step
towards comparing stripped-envelope models with
observations by performing a 3D simulation
of a neutrino-driven explosion that matches
the global asymmetries in the distribution
of $^{44}\mathrm{Ti}$ and $^{56}\mathrm{Ni}$
and the neutron star kick 
in Cas~A to an astonishing degree. The required
progenitor for a Type~IIb (i.e., partially
stripped) supernova was again
constructed by manually removing part of the envelope
at collapse, but in terms of simulation
fidelity this is likely less of an issue than the
fact that the neutrino-driven explosion was still
tuned to match the desired explosion energy.

A first 3D simulation of mixing in an
ultra-stripped progenitor starting from a self-consistent
explosion model was conducted by \citet{mueller_18}
with a view to observations of fast and faint
Ic supernovae \citep{drout_13,de_18}. The model
showed mixing of substantial amounts of nickel
in a few narrow dense plumes out to about half way through the He envelope. These findings are, however,
difficult to extrapolate to other, less extreme
stripped-envelope supernova progenitors.
A more thorough exploration of mixing 
in Ib/c supernovae and a quantitative comparison
of 3D models of mixing instabilities with
the spectropolarimtery of observed Ib/c  events
is called for.

\paragraph{Mixing and fallback.} 
Mixing instabilities have also been studied
as a possible explanation of abundances in
extremely metal-poor stars.
\citet{umeda_03} suggested that the high
[C/Fe] in some of these stars can be explained
by invoking a combination of Rayleigh--Taylor mixing and 
fallback in the supernovae that supposedly contributed
to their initial composition.\footnote{Jet-driven
explosions could provide an alternative mechanism
to explain the observed abundance patterns
\citep[e.g.,][]{maeda_03,nomoto_06}.}
\citet{joggerst_09} conducted 2D simulations
of this scenario using the \textsc{Flash} code.
Their simulations indeed showed enhanced fallback
in low- and zero-metallicity progenitors,
and hence a possible mechanism for low iron-group
yields in metal-poor environments, but Rayleigh--Taylor 
mixing was not sufficient for ejecting the
required amount of iron-group and intermediate-mass elements
to match observed abundances. In a follow-up
study that surveyed a broader range of
rotating progenitors with two different metallicities
($Z=0$ and $Z=10^{-4}\,Z_\odot$)
using the \textsc{Castro} code, \citet{joggerst_10b}
were able to find better matches of the supernova
yields to abundance patterns from ultra-metal poor stars.
A similar study was conducted by \citet{chen_17}
to explain the abundances in the most
iron-poor star to date (SMSS J031300.36-670839.3, 
\citealp{keller_14})  by fallback in an explosion with 
modest energy. However, all of these simulations
were restricted to 2D and imposed seed perturbations
by hand in spherically symmetric models. A first
3D simulation of a fallback supernova from collapse
to shock revival by the neutrino-driven
mechanism, through black hole formation, and on to
shock breakout was performed by \citet{chan_18}.
In their model, fallback proceeds in a qualitatively
different manner than in previous studies; 
the iron-group material is accreted early, the
post-shock flow involves global asymmetries during
the first tens of seconds (which could potentially
generate substantial black hole kicks and spins), but no mixing instabilities occur later on. While the
work of \citet{chan_18} has demonstrated the feasibility
of a forward-modelling approach to fallback
supernovae, they explored only a single progenitor,
and a broader investigation is necessary to
understand the phenomenology of fallback in 
three dimensions.

\begin{acknowledgements}
I wish to thank
S.~Bruenn, A.~Burrows, C.~Collins, J.~Guilet, T.~Foglizzo, H.-Th.~Janka, K.~Kotake, E.~Lentz, A.~Mezzacappa, E.~M\"uller, J.~Powell, A.~Skinner, T.~Takiwaki, and A.~Wongwathanarat
for their kind permission to reproduce figures from their papers. I am grateful to A.~Heger for providing 1D progenitor models at various evolutionary stages.
This work was supported by the Australian Research Council (ARC) through
Future Fellowship FT160100035. The author acknowledges support
from the ARC Centre of
Excellence for Gravitational Wave Discovery (OzGrav) as associate investigator.
This research was undertaken with the assistance of
resources obtained via NCMAS and ASTAC  from the National Computational Infrastructure (NCI), which
is supported by the Australian Government and was supported by
resources provided by the Pawsey Supercomputing Centre with funding
from the Australian Government and the Government of Western
Australia. 
\end{acknowledgements}

\bibliographystyle{spbasic}    

\bibliography{paper}

\end{document}